\documentclass[twocolumn,final,epj]{svjour}
\pdfoutput=1
\usepackage[utf8]{inputenc}
\usepackage[T1]{fontenc} 
\usepackage{graphicx}  
\usepackage{amssymb}
\usepackage{amsmath}
\usepackage{hyperref}
\hypersetup{
    colorlinks=true,       
    linkcolor=blue,          
    citecolor=blue,        
    filecolor=magenta,      
    urlcolor=cyan           
}
\usepackage{color}

\def\p{{\bf p}}
\def\k{{\bf k}}
\def\r{{\bf r}}
\def\e{{\bf e}}
\def\bnabla{{\bf \nabla}}

\def\a{{\bf a}}
\def\A{{\cal A}}
\def\one{{\bf 1}}
\def\eps{\varepsilon}

\newcommand{\bra}[1]{\langle #1 |}
\newcommand{\ket}[1]{| #1 \rangle}

\begin{document}

\title{Cooper approach to pair formation 
in a tight-binding model of La-based cuprate superconductors}

\author{
Klaus\,M. Frahm\inst{1} 
\and
Dima L. Shepelyansky\inst{1}}

\institute{
Laboratoire de Physique Th\'eorique, 
Universit\'e de Toulouse, CNRS, UPS, 31062 Toulouse, France
}

\titlerunning{Cooper approach to pair formation in a  model of cuprate superconductors}
\authorrunning{K.M.~Frahm and D.L.~Shepelyansky}

\date{Dated: 19 September 2022}

\abstract{We study numerically, in the framework of the Cooper approach 
  from 1956, 
  mechanisms of pair formation in a model of La-based cuprate superconductors
  with longer-ranged hopping parameters reported in the literature 
  at different values of center of mass momentum.
  An efficient numerical method allows to study lattices with more than a 
  million sites.
  We consider the cases of attractive Hubbard and d-wave type interactions
  and a repulsive Coulomb interaction. The approach based on a frozen Fermi 
  sea leads to a complex structure of accessible relative momentum 
  states which is very sensitive to the total pair momentum 
  of static or mobile pairs.
  It is found that interactions with attraction of approximately
  half of an electronvolt give a satisfactory agreement
  with experimentally reported results for the  critical 
  superconducting temperature and its dependence on hole doping.
  Ground states exhibit d-wave symmetries for both attractive Hubbard and 
  d-wave interactions which is essentially due to the particular 
  Fermi surface structure and not entirely to an eventual d-wave 
  symmetry of the interaction. We also find pair states created by
  Coulomb repulsion at excited energies above the Fermi energy 
  and determine the different mechanisms of their formation. In particular, we 
  identify such pairs in a region of negative mass at rather modest excitation 
  energies which is due to a particular band structure.
}


\maketitle

\section{Introduction}
\label{sec1}

The properties and features of high temperature superconductivity (HTC), 
discovered in \cite{hts1986},
are still lacking a complete physical understanding 
as admitted by various experts of the field
(see e.g. \cite{dagotto,kivelson,proust}). 
The complexity of the phase diagram and 
strong interactions between electrons (or holes)
creates significant difficulties for the theoretical and numerical analysis. 
As a simplified, but still a generic model, 
it was proposed to use a one-body Hamiltonian with nearest-neighbor hopping on 
a two-dimensional (2D) square lattice formed by Cu ions \cite{anderson}.
In this framework the interactions between charges are considered as
the 2D Hubbard interaction resulting from a screened Coulomb interaction 
\cite{anderson}. Starting from  
\cite{emery1,emery2,varma,loktev} other models were developed and extended on 
the basis of extensive computations with various numerical methods of quantum chemistry
(see e.g. \cite{markiewicz,bansil,fresard} and Refs. therein). 
They showed the importance
of next-nearest one particle hoppings and allowed to determine 
longer-ranged tight-binding parameters. 

In this work, we extend the Cooper approach \cite{cooper} considering
two interacting particles (holes or electrons)
in a vicinity of a frozen Fermi surface
using the 2D longer-ranged tight-binding parameters reported 
in  \cite{fresard} for the one particle model
of La-based cuprate superconductors.
In contrast to the Cooper case \cite{cooper}
with a spherical (3D) or circle (2D) Fermi surface,
we show that for the above model with the parameters taken from \cite{fresard}
(called HTC model)
the frozen Fermi surface has a significantly more complex structure
due to the band structure of the lattice.
The complexity of the Fermi surface 
becomes really amazing for the case of
mobile pairs with nonzero total momentum (or twice 
the center of mass momentum)
of a pair (usually the total pair momentum 
is considered to be zero in the Cooper approach \cite{cooper}).
For comparison, we also present some data for the case of only
nearest-neighbor hoppings (called NN model).

We consider three types of interactions between particles:
attractive Hubbard interaction, a specific type
of attractive d-wave interaction discussed in \cite{bansil}
and a repulsive Coulomb interaction on the lattice
studied recently in \cite{prr2020,htcepjb}.
The physical origins and reasons of such model interactions
are not discussed in this work.
We note that in \cite{prr2020,htcepjb} it was shown
that pair formation can take place even for a Coulomb repulsion 
due to the appearance of an effective narrow or flat band
for mobile pairs with certain values of nonzero total momentum of a pair.
However, it is important to analyze the proximity of
such Coulomb pair states with respect to the Fermi surface
that was not done in \cite{prr2020,htcepjb}
and is performed here in the framework of the Cooper approach
for a pair in the vicinity of a frozen Fermi sea.

For the cases of attractive Hubbard and d-wave interactions, 
we find the appearance of a gaped coupled pair 
state below the Fermi surface and investigate the gap
dependence on interaction strength $U$
and hole doping. The obtained results
are compatible with the experimental findings
for $La_{2-x}Sr_xCuO_4$ (LSCO) 
(see  \cite{markiewicz})
at the attraction strength $U \approx -0.5\,$eV.
We also determine the gap dependence
on total momentum of mobile pairs.
An efficient numerical method allows to study
lattices with about million sites
providing results in the limit of infinite
lattice size.

For the case of Coulomb repulsion
the formation of pairs takes place only
for pair energies above the Fermi surface.
We establish three different mechanisms
of such Coulomb pair formation
and discuss their possible relations with
the pseudogap phenomenon.

Section~\ref{sec2} describes the basic features of 
the tight-binding model for typical HTC materials with a model 
of 5 different hopping matrix elements and other details concerning 
the Cooper pair approach with a frozen Fermi sea at given filling $n$. 
In particular, the effective sector Hamiltonian in relative momentum space 
for two interacting particles (holes) above (below) the Fermi energy 
for a given conserved value of the total momentum $\p_+$ is defined for three 
different types of interactions being the attractive Hubbard interaction, 
a similar attractive interaction with d-wave symmetry and a repulsive 
Coulomb interaction. 
In Sections~\ref{sec3} (for $\p_+=0$; static pairs) and \ref{sec4} 
(for $\p_+\neq 0$; mobile pairs) results for various ground state 
properties of electron pairs for the attractive Hubbard and d-wave 
interaction are presented. 
Sections~\ref{sec5} (for $\p_+=0$) and \ref{sec6} (for $\p_+\neq 0$) 
concentrate on pairs of hole excitations, and in particular 
in Section~\ref{sec5}, we present numerical results for the gap as a 
function of hole-doping which can be compared to experimental data. 
In Section~\ref{sec7}, we discuss excited pair states for two particular 
examples in the presence of repulsive Coulomb interaction and we identify 
three mechanisms of pair formation. 
The final discussion is presented in Section~\ref{sec8}.

Additional Figures S1-S17 are given in Supporting Material (SupMat).

\section{Generalized tight-binding model on a 2D lattice and sector Hamiltonian}
\label{sec2}

In the NN and HTC models, each electron moves on a square lattice 
of size $N\times N$ with periodic boundary conditions. 
The one-particle tight-binding Hamiltonian reads:
\begin{equation}
\label{eq_H1p}
H_{1p}=-\sum_{\r}\sum_{\a\in\A} t_\a\,\bigl(\ket{\r}\bra{\r+\a}+
\ket{\r+\a}\bra{\r}\bigr) \;\; .
\end{equation}
Here the first sum is over all discrete lattice points $\r$ (measured 
in units of the lattice constant) and $\a$ belongs to a certain set of 
{\em neighbor vectors} $\A$ such that 
for each lattice state $\ket{\r}$ there are non-vanishing hopping matrix 
elements $t_\a$ with $\ket{\r+\a}$ and $\ket{\r-\a}$ for $\a\in\A$.
The same model was used in \cite{htcepjb}
and we repeat here its description for convenience,
keeping the same notations. The hopping parameters of the HTC model
are taken from \cite{fresard}.
The set $\A$  contains all neighbor vectors $\a=(a_x,a_y)$
in one half plane with either $a_x>0$ or $a_y>0$ if $a_x=0$ such that 
$\A'=\A\cup(-\A)$ is the {\em full set} of all neighbor vectors. 
For each vector $\a$ of the full set $\A'$ any other vector $\tilde\a$ 
that can be obtained from $\a$ by a reflection at either the $x$-axis, 
$y$-axis or the $x$-$y$ diagonal also belongs to the full set $\A'$ and 
has the same hopping amplitude $t_\a=t_{\tilde\a}$. 

For the usual nearest neighbor tight-binding model (NN model),
considered in \cite{prr2020}, we have 
the set $\A_{\rm NN}=\{(1,0),(0,1)\}$ with $t_{(1,0)}=t_{(0,1)}=t=1$. 
A part of the numerical results is presented for the NN model 
(for illustration and comparison) but the main studies are done
for a longer-ranged tight-binding lattice 
\cite{fresard} denoted as the HTC model.
For this case the set of neighbor vectors is 
$\A_{\rm HTC}=\{(1,0),(0,1),(2,0),(0,2),(1,\pm 2),(2,\pm 1), (1,\pm 1)$, 
$(2,\pm 2)\}$ and the hopping amplitudes are: 
$t=t_{(1,0)}=1$, $t'=t_{(1,1)}=-0.136$, $t{''}=t_{(2,0)}=0.068$, 
$t{'''}=t_{(2,1})=0.061$ and $t^{(4)}=t_{(2,2)}=-0.017$ 
corresponding to the values given in Table 2 of \cite{fresard} (all energies 
are measured in units of the hopping amplitude $t=t_{(1,0)}=t_{(0,1)}$ 
which is  set to unity here; see also Fig.~6a of \cite{fresard}
for the neighbor vectors of the different hopping amplitudes).
The hopping amplitudes for other vectors such as $(0,1)$, $(1,-1)$, $(2,1)$, 
$(1,-2)$ etc. are obtained from the above amplitudes by the 
appropriate symmetry transformations, e.g. $t_{(1,-1)}=t_{(1,1)}=t'=-0.136$ 
etc. For comparison with experimental results
in LSCO we use the physical value of hopping $t=0.43$eV from
\cite{markiewicz}. We also put the Planck constant to unity, $\hbar=1$, 
thus using particle momentum $p_x, p_y$ and related
wave vectors $k_x, k_y$ to be the same.

The one-particle eigenstates of $H_{1p}$ (\ref{eq_H1p}) are simple plane waves:
$\ket{\p}=\sum_\r\,e^{i\p\cdot\r}\,\ket{\r}/N$
with energy eigenvalues:
\begin{equation}
\label{eq_en1p}
E_{1p}(\p)=-2\sum_{\a\in\A} t_\a\cos(\p\cdot \a)
\end{equation}
and momenta $\p=(p_x,p_y)$ such that $p_x$ and $p_y$ are integer multiples 
of $2\pi/N$ (i.e. $p_\alpha=2\pi l_\alpha/N$, $l_\alpha=0,\ldots,N-1$, 
$\alpha=x,y$). For the HTC model the energy dispersion reads:
\begin{equation}
\begin{aligned}
E_{1p}&(p_x,p_y) = - 2\left[\cos(p_x)+\cos(p_y)\right] \\
&-4t' \cos(p_x) \cos(p_y) - 2t{''} \left[\cos(2p_x)+\cos(2p_y)\right] \\
&-4t{'''} \left[\cos(2p_x) \cos(p_y) + \cos(2p_y) \cos(p_x)\right] \\
&-4t^{(4)} \cos(2p_x) \cos(2p_y)
\end{aligned}
\label{2dlattice}
\end{equation}
which corresponds to Eq. (30) of \cite{fresard} (assuming 
$t=1$ and $t^{(5)}=t^{(6)}=t^{(7)}=0$).

\begin{figure}
\begin{center}
\includegraphics[width=0.95\columnwidth]{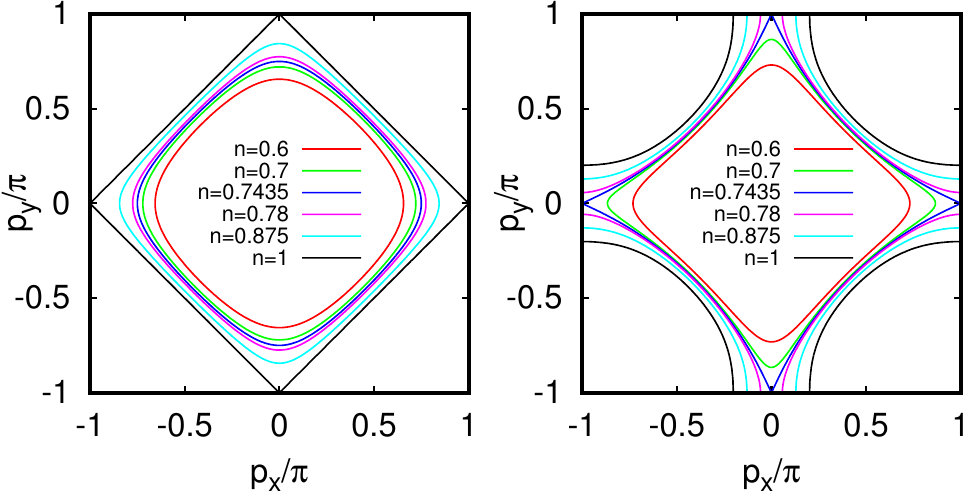}%
\end{center}
\caption{\label{fig1}
\label{fig_band}
Fermi surface for different filling factors $n$ for the NN model (left panel) 
and the HTC model (right panel). The value $n=0.7435$ is close 
to the separatrix value $n=0.743465958$ for the HTC model and 
$n=1$ is the separatrix value for the NN model.
}
\end{figure}

The energy Fermi surface of one particle is determined by
the dispersion relation (\ref{2dlattice}) and depends on the electron filing
factor $n$ and related Fermi energy $E_1=E_F$.
Examples of the Fermi surface at various
fillings $n$ are shown in Fig.~\ref{fig1}.
We note that the separatrix case corresponds
to the filling $n=0.74346...$ for the HTC model
and $n=1$ for the NN model. The separatrix
separate bounded and unbounded curves of
fixed energy on an infinite plane $(p_x,p_y)$
(rotation from libration as for a pendulum).
The filling $n$ corresponds to the electron filling
while the hole filling is $n_h = 1- n$.
The dependencies of one-particle density $\rho(E_1)$
of states on energy $E_1$ and filling factor $n$
are given in Fig.~S1 of SupMat. The density is strongly
peaked at $n=1$ (NN model) and $n=0.74346...$ (HTC model)
corresponding to the separatrix (and related Van Hove singularity).
Indeed, on a separatrix the frequency of motion $\omega_s$
becomes zero and thus  $\rho(E_1) \propto 1/\omega_s$
becomes singular.

The quantum Hamiltonian of the model with two interacting 
particles (TIP) has the form:
\begin{equation}
\label{eq_quant_Ham}
H=H_{1p}^{(1)}\otimes \one^{(2)}+\one^{(1)}\otimes H_{1p}^{(2)}+
\sum_{\r_1,\r_2}\bar U(\r_2-\r_1)\ket{\r_1,\r_2}\bra{\r_1,\r_2}
\end{equation}
where $H_{1p}^{(j)}$ is the one-particle Hamiltonian (\ref{eq_H1p}) 
of particle $j=1,2$ with positional coordinate $\r_j=(x_j,y_j)$ and 
$\one^{(j)}$ is the unit operator of particle $j$. 
The last term in (\ref{eq_quant_Ham}) represents, for the moment, 
a generic interaction to be specified below. 

In absence of interaction ($\bar U(\r_2-\r_1)=0$) 
the energy eigenvalues of the 
two electron Hamiltonian (\ref{eq_quant_Ham}) with  
given momenta $\p_1$ and $\p_2$ are:
\begin{equation}
\begin{aligned}
\label{eq_Ec}
E_c(\p_1,\p_2)&=E_{1p}(\p_1)+E_{1p}(\p_2)\\
&=-4\sum_{\a\in\A} t_\a\cos(\p_+\cdot \a/2)\cos(\Delta\p\cdot \a)
\end{aligned}
\end{equation}
where $\p_+=\p_1+\p_2$ is the total momentum 
(or $\p_+/2=(\p_1+\p_2)/2$ is the center of mass momentum) 
and $\Delta \p=(\p_2-\p_1)/2$ is the momentum associated to the relative 
coordinate $\Delta\r=\r_2-\r_1$. Note that the possible values of 
the components $\Delta p_\alpha$ ($\alpha=x,y$) are either integer 
or half-integer multiples of $2\pi/N$ depending on the center of 
mass momentum component $p_{+,\alpha}/2$ being an integer or half-integer 
multiple of $2\pi/N$. For the NN model Eq. (\ref{eq_Ec}) becomes 
$E_c(\p_1,\p_2)=-4\sum_{\alpha=x,y} \cos(p_{+\alpha}/2)\cos(\Delta p_\alpha)$. 

Due to the translational invariance of the interaction, it couples 
only pair momentum 
states $\ket{\p_1,\p_2}$ and $\ket{\p'_1,\p'_2}$ 
with identical conserved total momentum $\p_+=\p'_+$, i.~e.:
\begin{align}
\label{eq_Umat_elem}
\bra{\p'_1,\p'_2}\bar U\ket{\p_1,\p_2}&=
\delta_{\p'_+,\p_+}\,U_p(\Delta\p'-\Delta\p)\ ,\\
\label{eq_UFT}
U_p(\Delta\p'-\Delta\p)&=
\frac{1}{N_2}\sum_{\Delta\r} e^{-i(\Delta\p'-\Delta\p)\cdot\Delta\r}
\bar U(\Delta\r)
\end{align}
with $N_2=N^2$ being the size of the square $N\times N$ lattice 
and $U_p(\Delta\p'-\Delta \p)$ being (proportional to) the 
discrete Fourier transform 
of $\hat U(\r)$. Therefore, the two-particle Ha\-miltonian
(\ref{eq_quant_Ham}) 
can be diagonalized separately for each sector corresponding to 
a particular value of total momentum $\p_+$. 

In \cite{htcepjb}, the quantum time evolution inside such sectors was 
computed (for the repulsive Coulomb interaction; see below) 
using sector eigenstates in $\Delta\r$-representation with 
periodic (or anti-periodic) boundary conditions for the case 
of integer (half-integer) values of $Np_{+,\alpha}/(4\pi)$ ($\alpha=x,y$).
In this work, we compute the eigenstates in $\Delta\p$-representation, 
using the diagonal energies (\ref{eq_Ec}) (minus two times 
the Fermi energy; see below) in absence of interaction plus the 
interaction coupling matrix elements (\ref{eq_Umat_elem}). We have 
verified that the resulting eigenstates coincide (in absence of 
a frozen Fermi sea; see below) up to numerical precision 
with those of \cite{htcepjb} once the proper transformation between 
$\Delta\p$- and $\Delta\r$-representations are applied (the half-integer 
case corresponds now to periodic boundary conditions in 
$\Delta\p$-representation but the possible values of 
$\Delta p_{x,y}$ are half-integer 
multiples of $2\pi/N$). Furthermore, as explained in \cite{htcepjb}, 
we consider symmetric wavefunctions with respect to 
particle exchange, i.e. with respect to the parity symmetry in the 
relative momentum $\Delta\p\to -\Delta\p$. This case corresponds 
to an antisymmetric spin-singlet state.

Concerning the choice of the interaction, we consider three cases here~:
 
(i) As in \cite{htcepjb}, we use a (regularized) repulsive 
Coulomb type long-range interaction (see Section~\ref{sec7}) 
$\bar U(\r_2-\r_1)=U/[1+r(\r_2-\r_1)]$ 
with amplitude $U>0$ and the effective distance 
$r(\r_2-\r_1)=\sqrt{\Delta\bar x^2+\Delta\bar y^2}$ between the 
two electrons on the lattice with periodic boundary conditions. 
(Here $\Delta\bar x = \min(\Delta x,N-\Delta x)$;  
$\Delta\bar y = \min(\Delta y,N-\Delta y)$;
$\Delta x = x_2-x_1$; $\Delta y = y_2-y_1$ and the latter differences 
are taken modulo $N$, i.e. $\Delta x=N+x_2-x_1$ if $x_2-x_1<0$ 
and similarly for $\Delta y$). For the purpose of numerical diagonalization 
in $\Delta\p$-representation, we compute the discrete Fourier transform 
of this interaction numerically by (\ref{eq_UFT}) and we do not use 
any analytical approximation in this context. We mention that the 
numerical Fourier transform gives the approximate behavior 
$U_p(\k)\sim 1/|\k|^{3/2}$ for large $|\k|$ (with $\k=\Delta\p'-\Delta\p$) 
while the analytic 2D-Fourier 
transform of the (non-regularized) Coulomb interaction (in infinite 
continuous space) behaves as 
$U_p(\k)\sim 1/|\k|$.

(ii) We also consider the case of an attractive Hubbard interaction
$\bar U(\r_2-\r_1) = U \delta_{\r_1,\r_2}$ ($U<0$) 
with $U_p(\Delta\p'-\Delta\p)=U/N_2=-|U|/N_2$ being constant 
as it was the case for the Cooper problem \cite{cooper}.

(iii) We also analyze the case of an attractive interaction with d-wave 
symmetry and interaction coupling matrix elements being
$U_p(\Delta\p',\Delta\p)=(U/N_2) g_{\Delta\p'}\, g_{\Delta\p}$ ($U<0$) with 
$g_{\Delta\p}=(\cos\Delta p_x-\cos\Delta p_y)/2$. This interaction cannot 
simply be obtained from some interaction potential $\bar U(\r)$ since the 
matrix elements do not depend on the difference $\Delta\p'-\Delta\p$. 
It corresponds to an effective interaction in the context of the 
Bardeen-Cooper-Schrieffer (BCS) formalism assuming that the 
superconducting gap obeys the d-wave symmetry $\Delta_{\k}\sim g_{\k}$
(see for example Section 4.2 of \cite{bansil}). In particular, using this 
kind of interaction (in the sector $\p_+=0$, i.e. $\p_2=-\p_1=\Delta\p$), 
it is easy to verify that the classical BCS variational ansatz indeed produces 
the gap dependence $\Delta_{\k}=g_{\k}\bar\Delta$ where the universal 
parameter $\bar\Delta$ is determined by some implicit equation.
As with the classical BCS approach, one can argue that this interaction 
represents {\em certain relevant contributions} of the global interaction which 
is more complicated. We do not claim here that this ``d-wave'' interaction 
is ``really'' present as such in typical HTC-superconductors and our aim 
is more to compare its influence on pair eigenstates and ground state 
energies with the attractive Hubbard interaction where no d-wave symmetry 
is ``injected'' in the interaction itself.

In the following, we consider a model of two interacting electrons (or holes) 
with momenta $\p_1=\p_+/2-\Delta\p$, $\p_2=\p_+/2+\Delta\p$
which are excitations of a {\em frozen Fermi sea} where 
momentum states below the Fermi energy $E_F$, corresponding to a certain 
filling value $n$, are occupied. In this case, only values of $\Delta\p$ are
accessible such that both $E_{1p}(\p_+/2\pm\Delta \p)>E_F$ 
(or $<E_F$ for the hole case). As we will see later, depending on the value 
of $\p_+$, the structure of available states in the $\Delta\p$-plane 
is potentially quite complicated and very interesting. The choice of $\p_+$ 
itself is actually quite arbitrary, as long as the set of accessible 
$\Delta\p$ values is not empty. We may choose $\p_+=0$ for static pairs 
or $\p_+\neq 0$ for mobile pairs. Occasionally, we will use the notion 
of a ``virtual filling'' $n_v$ if the center of mass $\p_+/2$ of a 
pair lies on the Fermi surface at filling $n_v$ which may be different 
from the actually filling $n$ which is used to determine the frozen Fermi 
sea. 

The effective Hamiltonian (for accessible values of $\Delta\p$), 
for each sector $\p_+$, also called {\em sector Hamiltonian}, 
has diagonal matrix elements given by 
$\pm[E_c(\p_1,\p_2)-2E_F]$ (with ``$+$'' for electrons and ``$-$'' 
for holes) which are coupled by the interaction matrix elements 
$U_p(\Delta\p',\Delta\p)$ according to the different types of interactions 
we consider. Depending on the interaction, we either use full 
numerical diagonalization of the effective Hamiltonian (for the 
case of the Coulomb interaction; see Section~\ref{sec7}) or we 
compute by an efficient method, described in 
Appendix~\ref{appA1}, the ground state and its energy 
(for the cases of attractive Hubbard or d-wave interaction; see 
Sections~\ref{sec3}-\ref{sec6}) based on the ideas of Cooper \cite{cooper} 
and exploiting the rank-1 structure of the interaction matrix elements. 
As a consequence the energy eigenvalues can be obtained from the numerical 
solution of an implicit equation of the form of a sum over all
two-particle momentum states with each particle being above the frozen 
Fermi sea (Cooper considered the case of an infinite system where 
the sum is reduced to an integral \cite{cooper}) 
and the corresponding eigenstates are 
obtained from an explicit formula once the energy eigenvalues are known 
(see Appendix~\ref{appA1} for details). This method allows to significantly
reduce the numerical effort and to find the ground state
of a Cooper pair for lattices with more than a million sites. 

In the remainder of this work, when we speak of eigenstate energies etc. 
we refer to the eigenvalues of the sector Hamiltonian introduced above, i.e.
taking into account a shift with ``$-2E_F$'' and an additional minus sign 
for the hole case concerning the diagonal matrix elements of this Hamiltonian. 
Therefore, the ground state energy $E_{\rm min}$ of such a sector Hamiltonian 
is typically close to zero (corresponding to the Fermi energy) except for the 
cases where we have a strong gap $\Delta=-E_{\rm min}/2$ with possible 
negative values of $E_{\rm min}$ and other eigenvalues are positive.

\section{Properties of static Cooper pairs}
\label{sec3}

\begin{figure}
\begin{center}
\includegraphics[width=0.95\columnwidth]{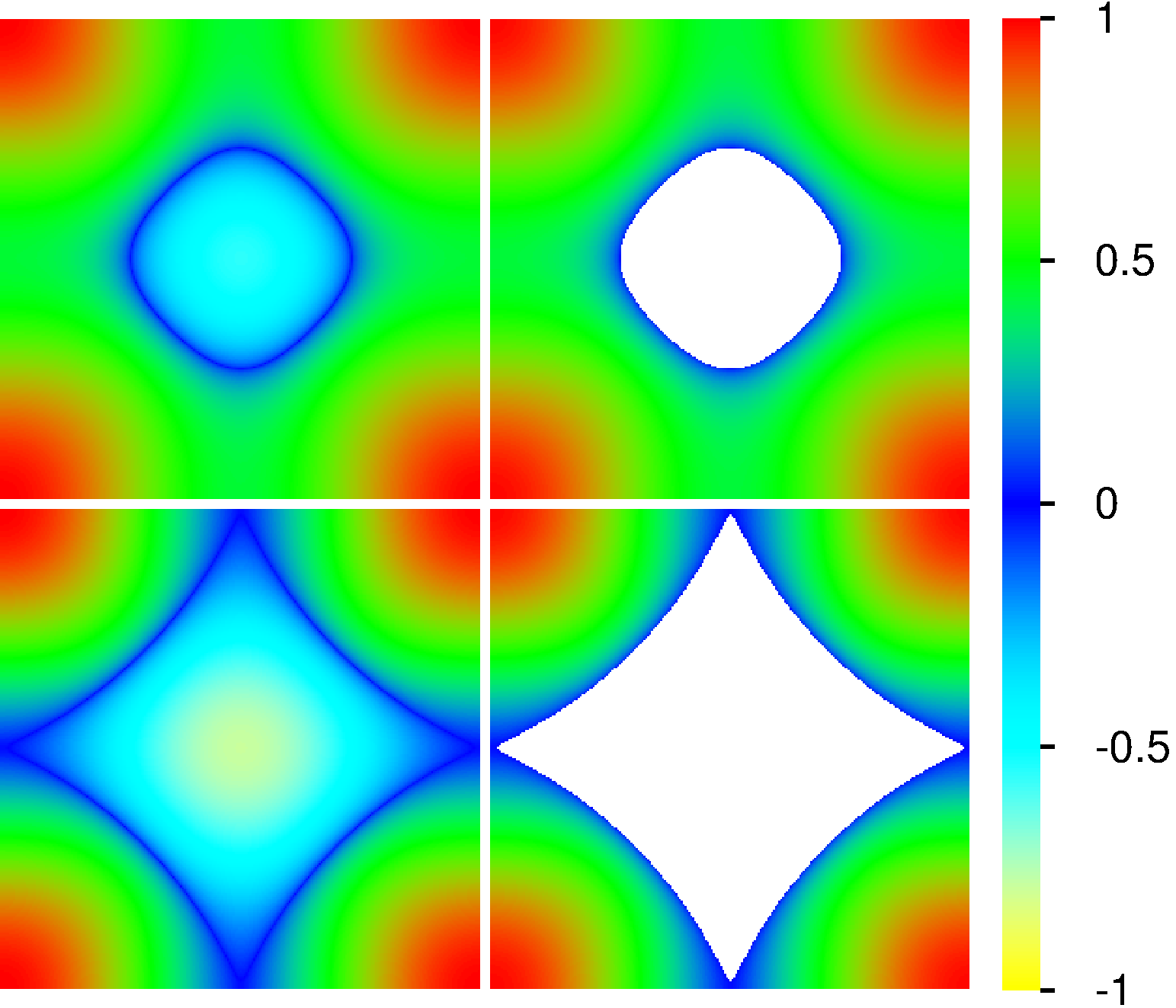}%
\end{center}
\caption{\label{fig2}
\label{fig_energy_P0}
Landscape of kinetic pair energy of particles 
in the $\Delta \p$-plane. Left panels show color plots 
of $E_c(\p_+/2-\Delta \p,\p_+/2+\Delta \p)-2E_F$ 
for the HTC model in the $\Delta p_x$-$\Delta p_y$ plane for 
$-\pi\le \Delta p_{x,y}<\pi$ in the sector $\p_+=0$. 
The Fermi energy $E_F$ corresponds to the filling factor 
$n=0.3$ ($n=0.74$) for top (bottom) panels.
The colors red (green) correspond to 
positive maximum (intermediate), blue to zero value
and yellow (cyan) to strongest (intermediate) negative values 
(the shown color bar applies to this and all subsequent color density 
plot figures of the same style, eventually with nonlinear 
rescaling to increase the visibility of small value regions). 
Right panels are as left panels but the forbidden 
zones of $\Delta \p$ such that each one-particle energy is below 
the Fermi energy $E_F$, i.e. 
$E_{1p}(\p_+/2-\Delta \p)<E_F$ and 
$E_{1p}(\p_+/2+\Delta \p)<E_F$,  
are replaced by white color. 
For $\p_+=0$ the white zones simply 
correspond to the colors yellow (cyan) for negative values in the left panel. 
However, for different sectors with $\p_+\neq 0$ shown 
in later figures this simple correspondence is no longer true 
and the structure of white zones is more complicated.
}
\end{figure}

We first consider static Cooper pairs of electrons
created by the Hubbard attraction
when the total pair momentum is $\p_+=0$.
The dependence of the quantity $E_c-2E_F$ with $E_c$ 
given by (\ref{eq_Ec}) on the relative momentum in the $\Delta\p$-plane
is shown in Fig.~\ref{fig2}
(left column) for two filling factors $n=0.3; 0.74$.
The region of the frozen Fermi sea is also shown by white color 
in Fig.~\ref{fig2} (right column). (In the following, we will refer to 
this type of figures as ``energy landscape'' figures.)
Thus in the quantum case all transitions
induced by interaction between TIP states
take place only outside the white zone
corresponding to the Cooper approach \cite{cooper}.

\begin{figure}
\begin{center}
\includegraphics[width=0.95\columnwidth]{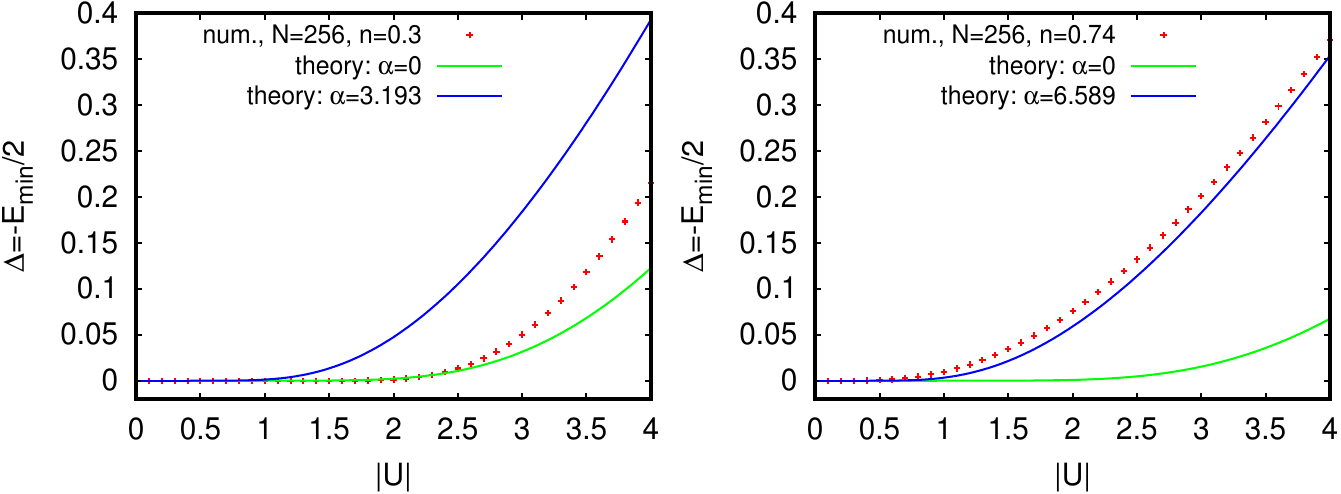}%
\end{center}
\caption{\label{fig3}
\label{fig_coopertheory}
Left (right) panel shows the ground state energy gap
$\Delta = -E_{\rm min}/2$ versus absolute Hubbard interaction strength 
$|U|$ (case of attractive interaction with $U<0$) for 
$\p_+=0$, for $N=256$, $n=0.3$ ($n=0.74$). 
(Here $E_{\rm min}$ is the Hubbard ground state energy of the sector 
Hamiltonian at $\p_+=0$.)
The red data points have been obtained by a numerical solution 
of the implicit eigenvalue equation and coincide up to numerical precision 
with a full numerical diagonalization.
Examples of ground states at specific $U$ values are shown in 
Fig.~S2 of SupMat.
The full curves correspond to the analytical approximation 
assuming a two-particle sector-density of states of the form 
$\rho_2(\eps)=\rho_2(0)/[1+\alpha(\eps/\eps_{\rm max})]$ where 
$\eps_{\rm max}$ is the energy bandwidth in the sector and 
the parameter $\alpha$ is either zero 
(green curve; case of constant DOS case) or obtained 
by a numerical fit (blue curve) from the integrated density of states
(see Fig.~S4 of SupMat). }
\end{figure}

We compute numerically, by the method of Appendix \ref{appA1}, the 
ground state and its energy for the attractive Hubbard interaction 
at different values of the interaction strength. 
The numerically obtained dependence of the gap
$\Delta = - E_{\rm min}/2$ on the Hubbard attraction
$U$ between electrons
(excitation energy above the frozen Fermi sea of electrons)
is shown in Fig.~\ref{fig3}
for fillings $n=0.3$ and $n=0.74$ where $E_{\rm min}$ 
is the ground state energy of the effective sector Hamiltonian at 
$\p_+=0$ (with diagonal matrix elements being $E_c(\p_1,\p_2) - 2E_F$, 
$\p_{2,1}=\p_+/2\pm\Delta\p$ as explained above and interaction 
coupling matrix elements $U_p(\Delta\p',\Delta\p)$ 
for $\Delta\p,\Delta\p'$ outside the forbidden zone due to the 
frozen Fermi sea).
As for the Cooper case \cite{cooper}
the gap sharply drops for
small interactions $|U|<1$
and grows strongly for large $|U| > 1$. 

Examples of related ground states at specific
$U$ values are shown in Fig.~S2 of SupMat
for $U=-2.5$ ($n=0.3$) and $U=-1$ ($n=0.74$).
In the coordinate space the ground state represents
a compact pair state with a size $|\Delta \r| \sim 2$
and in the momentum space ($\Delta\p$-plane) 
the probability of the ground state
is concentrated near the Fermi surface shown in Fig.~\ref{fig2}.
(We also show similar ground states
for the case of d-wave interaction at same fillings
in Fig.~S3 of SupMat).

Fig.~\ref{fig3} also shows 
the analytical result (\ref{eq_dosfit2}) of Appendix~\ref{appA1} 
(blue curve) based on the fit ansatz (\ref{eq_dosfit1}) 
for the density of states (of diagonal energies of the sector 
Hamiltonian) assuming a power law decay with exponent $-1$ for large energies. 
The green curve corresponds to the analytical expression 
(\ref{eq_Eminresult1}) assuming a constant density of states and 
which is essentially Cooper's well known result \cite{cooper} 
(with different notations/parameters). Further details and analytical 
expressions of $E_{\rm min}$ for small and strong interactions values 
are given in Appendix~\ref{appA1}. 

For the Cooper case \cite{cooper} the gap $\Delta$ was determined by the
attraction strength and the density of states near the Fermi surface since 
only a small interval corresponding to the Debye energy 
contributes to the pair formation.
In our case all energies above the Fermi sea contribute
to the formation of pairs. Thus the approximation
of a constant density of states does not work well, especially 
for $n=0.74$ which is close to the separatrix and the van Hove singularity. 
For this case the fit ansatz (\ref{eq_dosfit1}) works very well 
as can be seen in Fig.~S4 of SupMat (showing the integrated density 
of states) and indeed the blue curve in (the right panel of) Fig.~\ref{fig3} 
coincides very well with the numerical data points for the gap energy. 

For $n=0.3$, the situation is different and here the green curve 
in (the left panel of) Fig.~\ref{fig3} 
is for modest interaction values ($|U|\lesssim 2.5$) very close to the 
numerical data points while the blue curve is significantly higher. 
The reason is that in this case, the density of states is initially, 
for smaller energies (lower 20\%), 
quite constant (integrated density of states 
close to a linear function; see red data points and green curve in the 
left panel of Fig.~S4 of SupMat) 
thus that Cooper's original expression 
works very well. However, for larger interaction values in the region 
$|U|\approx 8$ (not shown in Fig.~\ref{fig3}) the blue curve is actually 
closer to the numerical data points and the reason is that here all energies, 
also outside the initial region of linear integrated density of states, 
contribute. Fig.~S4 of SupMat shows indeed that also for $n=0.3$ 
the fit ansatz is more accurate for larger energies (above 20\%). 

\begin{figure}
\begin{center}
\includegraphics[width=0.95\columnwidth]{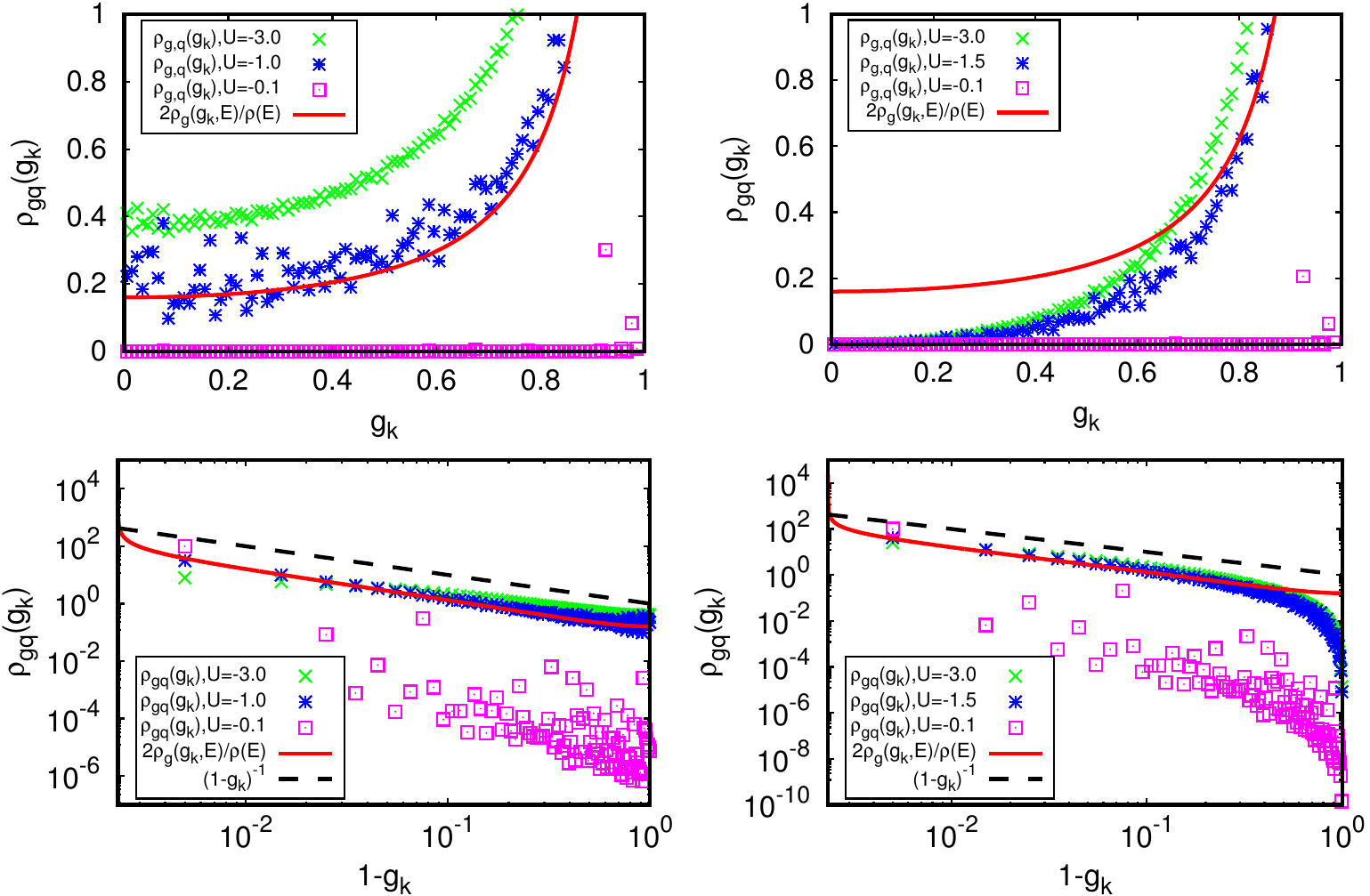}%
\end{center}
\caption{\label{fig4}
\label{fig_gkquantdens}
Comparison of the quantum distribution $\rho_{gq}(g_k)$
(see text), obtained for three interaction values $U$ 
(green, blue and pink data points), 
with the renormalized classical density $2\rho_g(g_k)/\rho(E)$
(red line) for the HTC model. Left (right) panels correspond 
to the attractive Hubbard (d-wave) interaction for $N=1024$, sector 
with $\p_+=0$ (for the quantum densities) and filling 
factor $n=0.74$ (for both quantum and classical densities). 
The relative factor $2/\rho(E)$ applied to the classical density 
ensures the proper 
normalization with respect to integration in the range $0\le g_k\le 1$
since $\int_{-1}^1 \rho_g(g_k,E)\,dg_k=\rho(E)$ where $\rho(E)=0.712$ is 
the classical density of states at $n=0.74$ (factor $2$ due to $\pm g_k$ 
symmetry). 
Top panels show a normal representation with a limited maximal value 
for the $y$-axis and lower panels show a double logarithmic representation 
using $1-g_k$ for the $x$-axis and the full range of density values. 
The black dashed line in lower panels shows the power law $(1-g_k)^{-1}$ 
for comparison (see also Appendix \ref{appA2} and Fig.~S5 of SupMat). }
\end{figure}

From Fig.~\ref{fig1} it follows that the angle resolved local density 
$\rho_\varphi(\varphi,E)$ on the energy Fermi surface
should significantly depend on the phase angle $\varphi$ 
of the vector
$\Delta \p = (k_x,k_y) \propto (\cos \varphi, \sin \varphi)$. 
This angle resolved density is proportional
to the area between two Fermi curves in Fig.~\ref{fig1}
taken at two close filling factors $n$ and $n+\delta n$ 
and between two close angles $\varphi$ and $\varphi+\delta\varphi$.
(See Appendix \ref{appA2} for the precise definition, computation 
and limiting behavior close to the separatrix of 
$\rho_\varphi(\varphi,E)$.)

For $n=0.3$ the Fermi curve is close to a circle
and the density $\rho_\varphi(\varphi,E)$ is rather constant. However, 
for $n=0.74$ the Fermi surface is drastically different from a circle
and we expect that
$\rho_\varphi(\varphi,E)$ is minimal for the symmetric case
$k_x=k_y$ or $\varphi\approx \pi/4$ (known as node in ARPES experiments
with HTC superconductors \cite{kivelson,proust,vishik1,vishik2})
and it is maximal for the 
asymmetric case $k_x \approx 0$ or $k_y \approx 0$, i.e. 
$\varphi\approx \pi/2$ or $\varphi\approx 0$
(known as antinode in ARPES).

In the ARPES experiment \cite{vishik1,vishik2}
the d-wave form is typically presented via
the parameter $g_k= (\cos k_x - \cos k_y)/2$, 
which can also be used to characterize a certain point on a given Fermi surface 
instead of $\varphi$, in particular we have $g_k\approx 1$ ($0$, $-1$) for 
$\varphi\approx \pi/2$ ($\pi/4$, $0$) for Fermi curves close to the 
separatrix curve. Therefore, 
we prefer to use the $g_k$-local density of states on
the Fermi surface given by 
$\rho_g (g_k,E) = \rho_\varphi (\varphi, E)/(dg_k/d \varphi )$.
(See Appendix \ref{appA2} for the details of the precise definition, 
computation and an analytical approximation of $\rho_g (g_k,E)$ 
for $E$ being close to the separatrix.)

Fig.~S5 of SupMat shows this density for the NN- and HTC-model and 
at different fillings. For the separatrix case, we have a 
power law $\rho_g (g_k,E)\approx C_1/(1-|g_k|)$ with a constant 
$C_1$ that can be computed analytically (as a function of the 
band-structure parameters) and with values 
$C_1\approx 0.025$ ($C_1\approx 0.052$) for the NN- (HTC-) model. 
For Fermi curves close but different from the separatrix curve the 
density is close to this power law but there is a cutoff 
at some maximal value $g_{\rm max}<1$ (with a square root singularity 
close to the cutoff; see Appendix \ref{appA2} for more details). 
The value of $g_{\rm max}$ corresponds to the case where either 
$k_x=0$ and $|k_y|$ maximal but typically smaller than $\pi$ 
(except for the separatrix case) or $k_y=0$ and $|k_x|$ maximal. 

We have also computed the quantum probability density 
$\rho_{gq} (g_k)$ for certain ground states (states similar 
as in Figs.~S2, S3 of SupMat) for the cases of the Hubbard and 
d-wave interaction, at certain interaction strengths, filling 
$n=0.74$, $N=1024$ and $\p_+=0$. 
This quantum distribution can be obtained from the interacting ground 
state $\psi(k)$, with $k=\Delta\p$ being the momentum in the relative 
coordinate, 
from a $g_k$-histogram by summing all probabilities 
$|\psi(k)|^2$ for those $k$-values such that $g_k$ falls in the same 
histogram bin with bin-width $\Delta g_k=0.01$. 
To ensure proper normalization with 
respect to integration in the range $0\le g_k\le 1$ an additional 
factor $1/\Delta g_k$ has been applied to the histogram values 
to obtain a properly integration normalized distribution $\rho_{gq}(g_k)$. 
Note that this quantity represents a pure $g_k$-distribution, a priori for all 
possible energies, while the classical local density $\rho_g(g_k,E)$ 
is specific to a certain classical energy $E$. 
Both quantities are shown and compared for the two cases 
of the attractive Hubbard and d-wave interaction in Fig.~\ref{fig4} 
(with a properly corrected normalization of $\rho_g(g_k,E)$ as explained 
in the figure caption of Fig.~\ref{fig4}). 

For small $|U|=0.1$ the quantum density is strongly inhomogeneous, essentially 
with one single peak at $g_k=0.995$ with about $99.5$\% of probability 
(only visible in the lower panels with logarithmic representation). The reason 
is that in this case the ground state is a small perturbation from 
the pure momentum state 
with $k$ closest to the Fermi surface. The fact that for this $k$ value 
we have $g_k\approx 1$ is a coincidence (but still with a strongly enhanced 
probability due to the nearly singular classical density at $g_k\approx 1$). 
For other parameters (fillings $n$, etc.) other $k$- and $g_k$-values for 
these peaks are in principle possible (a similar situation was discussed
for eigenstates of rough billiards \cite{roughbil}).

For moderate $U = -1; -1.5$ the quantum distribution $\rho_{gq} (g_k)$
is close to the (renormalized) classical distribution 
$2\rho_g (g_k,E)/\rho(E)$
in the case of Hubbard interaction but for the d-wave interaction
there are still significant differences. To explain this, we 
remind the expression (\ref{eq_psicooper2}) of Appendix \ref{appA1}, 
showing that the eigenstate amplitudes are given by 
the analytical formula~: $\psi(k)\sim a_k/(2\Delta+\eps_k)$ where $\eps_k$ 
represents a diagonal energy matrix element of the effective sector 
Hamiltonian. The factor $a_k$ is either $a_k=1$ for the Hubbard interaction 
or $a_k=g_k$ for the d-wave interaction. 

At very small interactions (e.g. $|U|=0.1$) 
in the perturbative regime, we also have according to (\ref{eq_limitsmallU}) 
a very small gap $\Delta\sim |U|/N_2$ such that only one single $k$-value 
satisfies the condition $\eps_k<2\Delta$ providing an isolated peak of 
the ground state in $\Delta\p$-representation. At modest interaction $U=-1$, 
the gap is significantly larger but still small in comparison to classical 
energy scales. Therefore, the eigenstate (for the Hubbard case with $a_k=1$) 
is concentrated at $k$- (or $\Delta\p$-) values close to the Fermi 
surface with an effective energy width $\approx 2\Delta$ which is perfectly 
confirmed by Fig.~S2 of SupMat. However, the width of this region 
around the Fermi surface 
in $k$-space is not uniform, it is enhanced for $k$ values with 
$|g_k|\approx 1$ 
and reduced for $|g_k|\approx 0$. Actually, a closer study of 
Fig.~S2 of SupMat shows that in the region $|g_k|\approx 0$ (i.e. 
$k_x\approx k_y$) there is still a peak-structure which is due to the finite 
grid for $N=256$ or $N=1024$. 

The reason of this is simply that the distance 
between the two Fermi curves at $E_F$ and $E_F+2\Delta$ is quite large 
at the region close to the separatrix point (with maximal $|g_k|$) 
and quite small at $k_x\approx k_y$ (with $|g_k|\approx 0$) 
in accordance with the nearly 
singular behavior of the classical density $\rho_g(g_k,E)$ for 
$|g_k|\approx 1$.
When computing the quantum distribution $\rho_{gq}(g_k)$, we consider 
a priori {\em all} $k$-values but the analytical expression of the 
amplitudes $\psi(k)\sim 1/(2\Delta+\eps_k)$ selects automatically 
the energies closest to the Fermi surface. This explains that 
(for the Hubbard) interaction the blue data points for $U=-1$ coincide quite 
well with the red curve for the classical (properly renormalized) density 
in the left panels of Fig.~\ref{fig4}. However, the blue data points 
still show some fluctuations (at $g_k<0.8$) which are due to the finite grid
structure of the possible $\eps_k$ values. 

For the stronger interaction $U=-3$ the green data points deviate 
significantly from the classical curve, also for the Hubbard case. 
The reason is that here the 
gap is significantly larger than for $U=-1$ (see Fig.~\ref{fig3}) 
and the quantum distribution corresponds actually to an energy average 
of the classical distribution over 
a quite large energy width of size $2\Delta$ 
which changes the shape of the distribution 
(reduction of the singular part at $|g_k|\approx 1$, increase of the density 
at modest values $|g_k|<0.9$).

Concerning the d-wave interaction (right panels of Fig. \ref{fig4}), we have 
the additional factor $g_k$ applied to the eigenstate amplitude $\psi(k)$ 
which provides an {\em additional} reduction of the density at $|g_k|\approx 0$ 
(and {\em additional} enhancement of the density at $|g_k|\approx 1$) which 
is clearly visible both in Fig.~S3 of SupMat and the right panels 
of Fig.~\ref{fig4}.

In conclusion, Fig.~\ref{fig4} and also Figs.~S2, S3 of SupMat show, that there 
are two ``d-wave'' effects: (i) enhancement of the 
$g_k$-density and wave function 
amplitudes at $|g_k|\approx 1$ simply due the HTC band structure, providing 
an increased number/area of momentum or $k$ values between two close 
Fermi curves if $|g_k|\approx 1$, (ii) an additional enhancement if the 
d-wave factor $g_k$ is artificially injected in the interaction (case 
of d-wave interaction).

\begin{figure}
\begin{center}
\includegraphics[width=0.95\columnwidth]{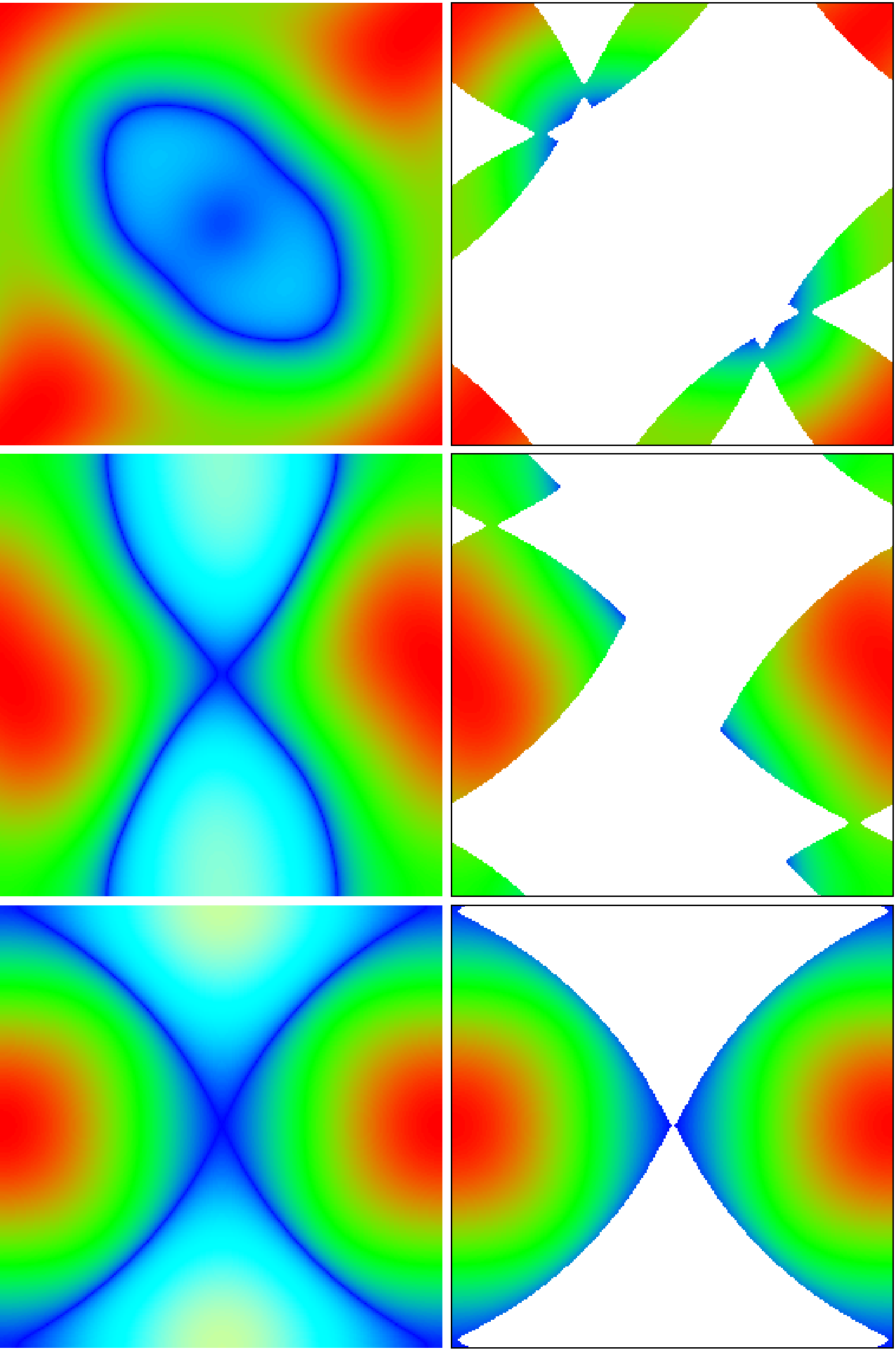}%
\end{center}
\caption{\label{fig5}
\label{fig_energy_P0p74}
Energy landscape for mobile Cooper pairs.
Left panels show color plots 
of $E_c(\p_+/2-\Delta \p,\p_+/2+\Delta \p)-2E_F$ 
for the HTC model in the $\Delta p_x$-$\Delta p_y$ plane for 
$-\pi\le \Delta p_{x,y}<\pi$. The Fermi energy $E_F$ 
corresponds to the filling factor $n=0.74$ (in all panels). 
Top (center, bottom) panel corresponds to 
the sector $\p_+=2\pi(103,103)/256$ of node case
($\p_+=2\pi(46,172)/256$ intermediate case,
$\p_+=2\pi(0,248)/256$ antinode case).
The three values of $\p_+$ are chosen such that the center of 
mass momentum $\p_+/2$ is very close to the Fermi surface 
of virtual filling factor $n_v=0.74$ with three cases 
of $p_{+x}=p_{+y}$, $p_{+x}\approx p_{+y}/4$ and $p_{+x}$ ($p_{+y}$) 
minimal (maximal). 
The choice of discrete values is motivated by subsequent 
quantum computations at $N=256$ with these exact identical parameters.
The colors red (green) correspond to 
positive maximum (intermediate), blue to zero value
and yellow (cyan) to strongest (intermediate) negative values 
(color bar as in Fig.~\ref{fig2}). 
Right panels are as the left panels but the forbidden 
zones of $\Delta \p$ (for particle excitations) 
such that each one-particle energy is below 
the Fermi energy, i.e. 
$E_{1p}(\p_+/2-\Delta \p)<E_F$ and 
$E_{1p}(\p_+/2+\Delta \p)<E_F$,  
are replaced by white color. Note that here 
the white zones include not only 
the negative value zones (yellow/cyan) in left panels but 
also additional zones of positive values due to $\p_{+}\neq 0$ and 
the more complicated selection rule using individual 
one particle energies.
}
\end{figure}

The issue of quantum ergodicity on the Fermi surface, 
eventually with a peak structure due to a finite grid at modest 
values of $N$ in the region $k_x\approx k_y$, is actually quite similar 
to the problem of rough billiards in the regime of quantum chaos 
\cite{roughbil}. 
Even for the cases where the quantum density $\rho_{gq} (g_k)$ differs 
from the classical density $\rho_g(g_k,E)$, 
the general tendency from classical ergodicity
remains valid:  $\rho_{gq} (g_k)$ is small for small $g_k$ values
(near node) and large for large values of $g_k$ (antinode).
It is interesting to note that the global dependence
of  $\rho_{gq} (g_k)$ at moderate interactions
is similar to the experimentally found gap dependence $\Delta(g_k)$,
see for example Fig. 3 in \cite{vishik1} for LSCO
where $\Delta$ is small for small $g_k$ and larger for $g_k > 0.5$.
It is important to stress that a somewhat similar dependence of  
$\rho_{gq} (g_k)$ is already visible for the Hubbard interaction 
which corresponds actually to an s-wave interaction.
Thus on this basis, we argue that the d-wave features
of HTC superconductors
can appear already for s-wave interactions due to the 
absence of s-wave symmetry for the Fermi surface and 
the particular band structure of HTC superconductors (point (i) above).
We think that this is an important message of this work.

\begin{figure}
\begin{center}
\includegraphics[width=0.95\columnwidth]{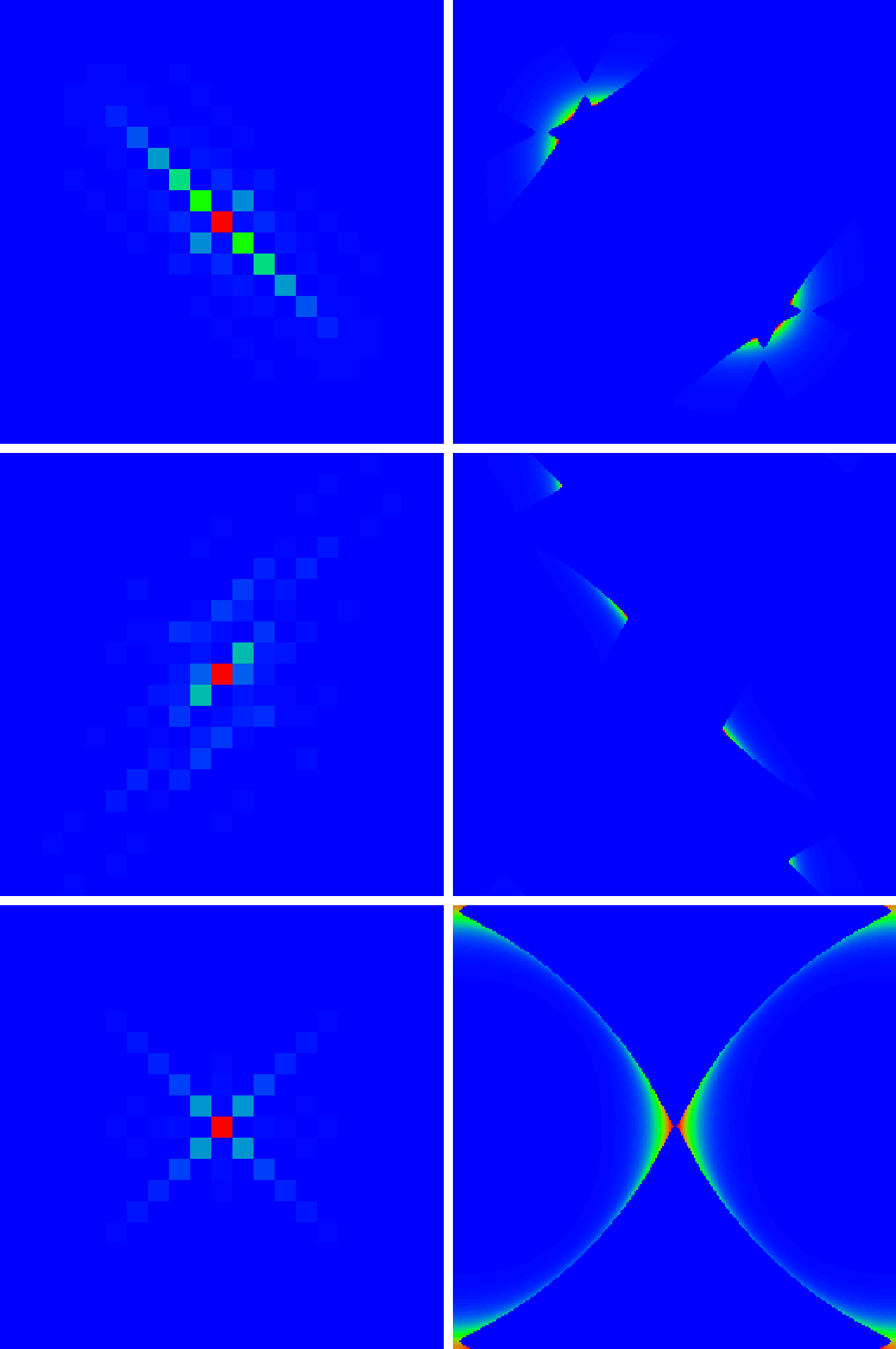}%
\end{center}
\caption{\label{fig6}
\label{fig_states_P0p74}
Ground state density plots for the Hubbard interaction, system size $N=256$, 
particle excitations, 
filling factor $n=0.74$ and three sectors $\p_+\neq 0$ (same 
values as in Fig.~\ref{fig5}). Top (center, bottom) panels 
correspond to $U=-4.5$, $\p_+=2\pi(103,103)/256$ 
($U=-7$, $\p_+=2\pi(46,172)/256$; $U=-3$, $\p_+=2\pi(0,248)/256$).
Left panels show the ground state 
in $\Delta \r$-representation in a zoomed region with 
$-10\le \Delta x,\Delta y\le 10$ 
(color values outside the zoomed regions are blue) 
and right panels show the state 
in $\Delta \p$-representation (with $-\pi\le \Delta p_{x,y}<\pi$).
The two particle ground state energies $E_{\rm min}$ 
in units of the basic hopping matrix element $t$ are $-0.1360$ ($-0.1427$, 
$-0.3823$) for top (center, bottom) panels.
}
\end{figure}

\section{Properties of mobile Cooper pairs}
\label{sec4}

In the previous Section, we discussed the ground state properties
for static Cooper pairs of electrons with zero total momentum  $\p_+=0$.
However, it is interesting to consider also the case of mobile pairs
with   $\p_+ \neq 0$. Indeed, such mobile pairs
can be related to the formation of stripes
observed in HTC superconductors (see e.g. \cite{stripe1,stripe2} and 
Refs. therein).
For particles with a quadratic dependence
of kinetic energy on momentum, considered by Cooper \cite{cooper},
the kinetic energy of a pair is the sum of
its internal motion energy and the center of mass motion energy.
Thus the kinetic energy of center of mass simply
adds a constant and plays therefore no role 
in the pair formation in a continuous media.
The situation is drastically different
for LSCO with a rather complex dispersion law
for each particle (\ref{2dlattice}). In this case, at 
$\p_+\neq 0$, the conditions $E_{1p}(\p_+/2\pm\Delta \p)>E_F$ 
for allowed transitions above
the frozen Fermi sea provide a nontrivial structure 
for the space of available $\Delta\p$ values. 

Fig.~\ref{fig5}
shows examples of the energy landscape of pair energy 
in the $\Delta\p$-plane within a fixed $\p_+$ sector without
Fermi restrictions (left column) and with restrictions
imposed by the frozen Fermi sea (right column)
at the filling factor $n=0.74$. 
The restrictions induced by the frozen Fermi sea create a 
very complex structure of the accessible $\Delta\p$-space, with ``tongues'' 
and multiple complicated borders, and it depends in a nontrivial
manner on the particular choice of $\p_+$. In 
Fig.~\ref{fig5}, we have chosen three examples of $\p_+$ 
such that the center of mass momentum $\p_+/2$ is very close to the 
Fermi surface 
of virtual filling factor $n_v=0.74$ with $p_{+x}=p_{+y}$, 
$p_{+x}\approx p_{+y}/4$ and $p_{+x}$ ($p_{+y}$) minimal (maximal). 

\begin{figure}
\begin{center}
\includegraphics[width=0.95\columnwidth]{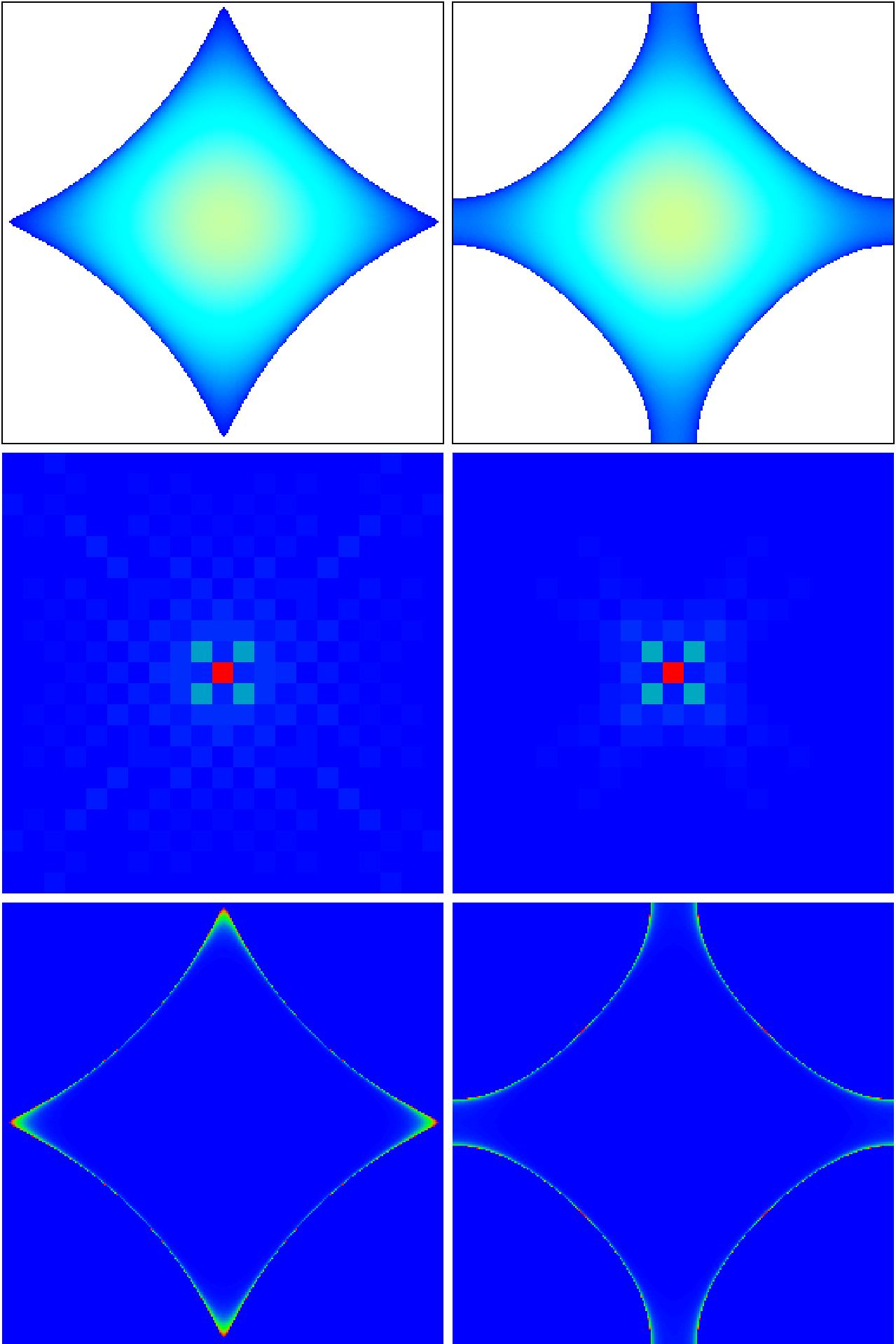}%
\end{center}
\caption{\label{fig7}
  \label{fig_hubbard_Holes0}
Energy landscape $E_c$ and ground states of static hole pairs 
in the HTC model.
Top panels show color plots 
of  $E_c(\p_+/2-\Delta \p,\p_+/2+\Delta \p)-2E_F$ 
on  $\Delta p_x$-$\Delta p_y$ with
$-\pi\le \Delta p_{x,y}<\pi$ in the sector $\p_+=0$. 
The forbidden zones for holes of $\Delta \p$ such that each one-particle 
energy is above the Fermi energy, i.e. 
$E_{1p}(\p_+/2-\Delta \p)>E_F$ and 
$E_{1p}(\p_+/2+\Delta \p)>E_F$,  
are replaced by white color. The Fermi energy $E_F$ 
corresponds to the filling $n=0.74$ (left panels) 
or $n=0.84$ (right panels). 
Center and bottom panels show ground state density plots for 
hole excitations, Hubbard interaction with $U=-1.5$, 
system size $N=256$ and sector $\p_+=0$.
Center panels show the ground state 
in $\Delta \r$-representation in a zoomed region with 
$-10\le \Delta x,\Delta y\le 10$ 
(color values outside the zoomed regions are blue) 
and bottom panels show the state 
in $\Delta \p$-representation (with $-\pi\le \Delta p_{x,y}<\pi$).
The two particle ground state energies $E_{\rm min}$ 
(of the effective sector Hamiltonian) 
in units of the basic hopping matrix element are $-0.03734$ ($-0.04222$)
for  $n=0.74$, ($n=0.84$). 
The colors red (green) correspond to 
positive maximum (intermediate), blue to zero value
and yellow (cyan) to strongest (intermediate) negative values. }
\end{figure}

In spite of the complexity of the energy landscape the implicit method 
for the computation of ground state properties
(see Appendix \ref{appA1})
still works perfectly that allows us to obtain results for
lattices with a large number of sites.
The ground states for the mobile Cooper pairs
with Hubbard attraction are shown in Fig.~\ref{fig6}
for parameters of Fig.~\ref{fig5} and interactions values 
between $U=-3$ and $U=-7$. 
We see that the ground states correspond
to compact pairs in $\Delta\r$-representation (left column)
and their densities in $\Delta\p$-representation (right column) 
are concentrated at certain borders of the frozen 
Fermi sea (``blue'' Fermi sea borders with small excitations energies; see 
right column of Fig.~\ref{fig5}). 
However, to find such nice pairs, it is necessary to considerably 
increase the value of $|U|$ as compared to static Cooper pairs 
(at $\p_+=0$). For smaller values of $|U|$ (not shown 
in Fig.~\ref{fig6}), the ground states are perturbative 
with isolated points in $\Delta\p$-representation 
and quite extended in $\Delta\r$-representation. The reason for the 
required larger values of $|U|$ is that the sector density of states 
close to the Fermi surface (number of available states at the blue Fermi sea 
border regions) is quite reduced as compared to the static case. 

Results similar to those of Figs.~\ref{fig5},~\ref{fig6}
are presented for another filling factor $n=0.84$ (and 
virtual filling $n_v=0.84$ for the choice of $\p_+/2$) 
in Figs.~S6, S7 of SupMat.

We discuss more features of mobile Cooper pair in the next Sections.

\section{Gap dependence on hole doping in LSCO for static pairs}
\label{sec5}

Up to now, we discussed the properties of Cooper pairs of electrons
at fixed electron doping $n$.
However, for LSCO the superconducting phase is formed by
doping of holes. This feature can be easily incorporated
in the framework of the Cooper approach considering hole excitation 
of the frozen Fermi sea at fixed hole doping $n_h = 1 -n$. 
Mathematically, one applies two fermionic hole creation operators 
(being two electron annihilation operators) to the frozen Fermi sea 
and as usual in the context of particle-hole transformation the 
one-body matrix elements between such hole-pair states acquire an additional 
negative sign while two-body matrix elements due to interactions 
are not changed. 

In particular, now the set of accessible $\Delta\p$ values must satisfy 
the condition of both electrons, associated to holes, being below the 
Fermi energy (i.e. being in the Fermi sea) with~: 
$E_{1p}(\p_+/2\pm\Delta \p)<E_F$ 
and the diagonal matrix elements in the effective sector Hamiltonian 
are $-[E_c(\p_1, \p_2) - 2E_F]>0$ (with $\p_{2,1}=\p_+/2\pm\Delta \p$
and $E_c$ given by (\ref{eq_Ec})) since it costs energy 
to excite holes and the interaction coupling 
matrix elements $U_p(\Delta\p',\Delta\p)$ are unchanged.
For convenience, we do not apply the sign change in the following energy 
landscape figures for holes (figures of style of Figs.~\ref{fig2}, \ref{fig5}) 
such that the forbidden white zones for holes correspond to positive 
values of $E_c(\p_1,\p_2)-2E_F>0$ (for the simple case $\p_+=0$). 

In this Section, we first consider the case of
static hole pairs with $\p_+=0$.
Examples of the energy landscape
with the frozen Fermi sea for hole dopings
$n_h = 1 -n$ at $n=0.74, 0.84$
are shown in top panels of Fig.~\ref{fig7}.
The ground states for these $n_h$ values (and $U=-1.5$) 
with attractive Hubbard interaction of holes
are also shown in this figure. The results show that
the pairs are very compact in the coordinate space
and in the momentum space they are located
at the (inside) vicinity of the Fermi surface with 
an effective width in momentum space being larger (smaller) 
if $\Delta p_x\approx \pm\pi$, $\Delta p_y\approx 0$ 
or $\Delta p_x\approx 0$, $\Delta p_y\approx \pm\pi$
($\Delta p_x\approx\Delta p_y$ respectively) in a similar way for 
electron pair states visible Fig. S2 of SupMat (located at the outside 
vicinity of the 
Fermi surface). The same approach also works 
for the case of attractive d-wave interaction
giving similar results for the ground state
energies and eigenstates but with an additional suppression 
of momentum wave function amplitudes in regions 
$\Delta p_x\approx\Delta p_y$ (not shown in figures here 
but similar to Fig. S3 of SupMat).

\begin{figure}
\begin{center}
\includegraphics[width=0.95\columnwidth]{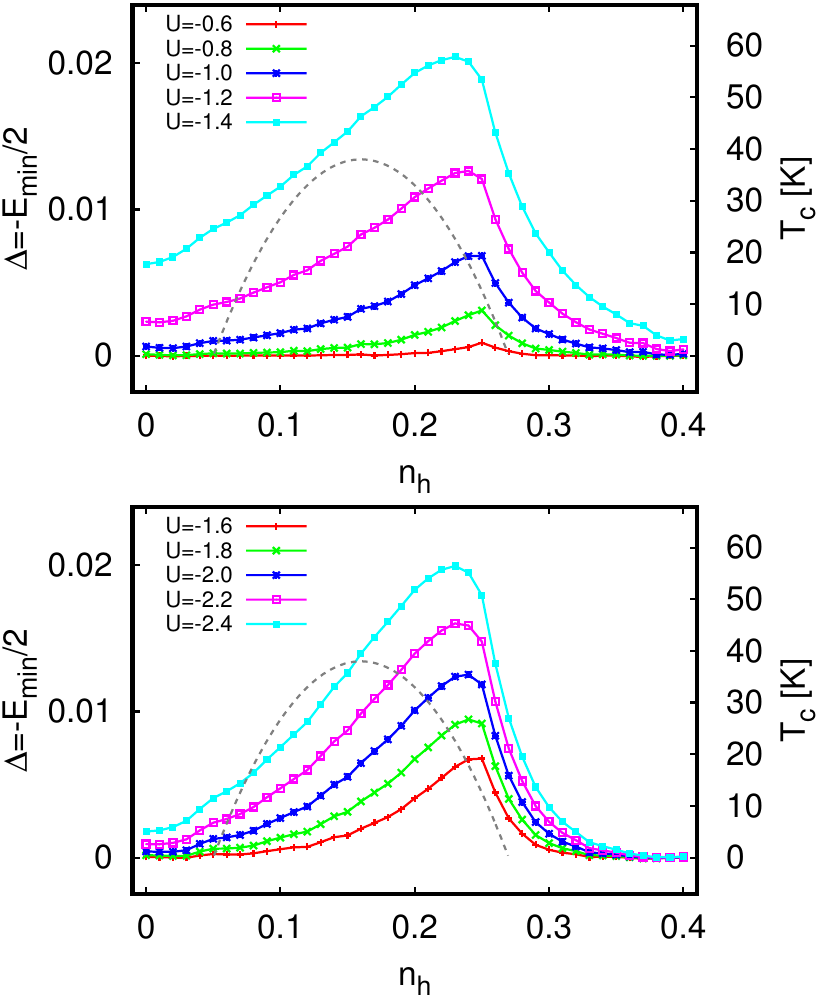}%
\end{center}
\caption{\label{fig8}
  \label{fig_nfill_gap_holes}
Gap dependence on hole doping $n_h$
in the HTC model of LSCO.
Shown is the gap energy $\Delta=-E_{\rm min}/2$ for hole excitations 
as a function of doping value $n_h=1-n$ for $N=1024$ and the 
sector $\p_+=0$.
The left vertical scale gives the energy values in units of the 
basic hopping matrix element $t$ and the right vertical scale 
gives the corresponding value of the critical temperature $T_c$ 
obtained from $\Delta=1.764\,k_B\,T_c$ and using $t=0.43\,$eV.
Top (bottom) panel corresponds to the Hubbard (d-wave) interaction 
with $U=-0.6, -0.8, -1, -1.2, -1.4$ ($U=-1.6, -1.8, -2, -2.2, -2.4$) 
for bottom to top curves. 
The parabolic grey dashed curve corresponds to the formula 
$T_c=T_{c0}[1-(\frac{n_0-n_h}{n_1})^2]$ 
with $T_{c0}=38\,K$, $n_0=0.16$, $n_1=0.11$ obtained 
from experimental data \cite{markiewicz}.
}
\end{figure}

We also computed the gap dependence on hole doping in LSCO 
for the attractive Hubbard and d-wave interactions at 
different interaction values $U$ and lattice size $N=1024$ (more
than a million lattice sites). 
The results are shown in Fig.~\ref{fig8} and 
the convergence of gap values with increasing lattice size from
$N=128$ to $N=1024$ is shown in Fig.~S8 of SupMat for an 
intermediate interaction value for both interaction cases. 
The curves exhibit still strong fluctuations at $N=128$ but the two curves 
at $N=512$ and $N=1024$ are nearly identical showing that $N=1024$ 
is sufficient to have gap values in the limit of infinite 
lattice size.

\begin{figure}
\begin{center}
\includegraphics[width=0.95\columnwidth]{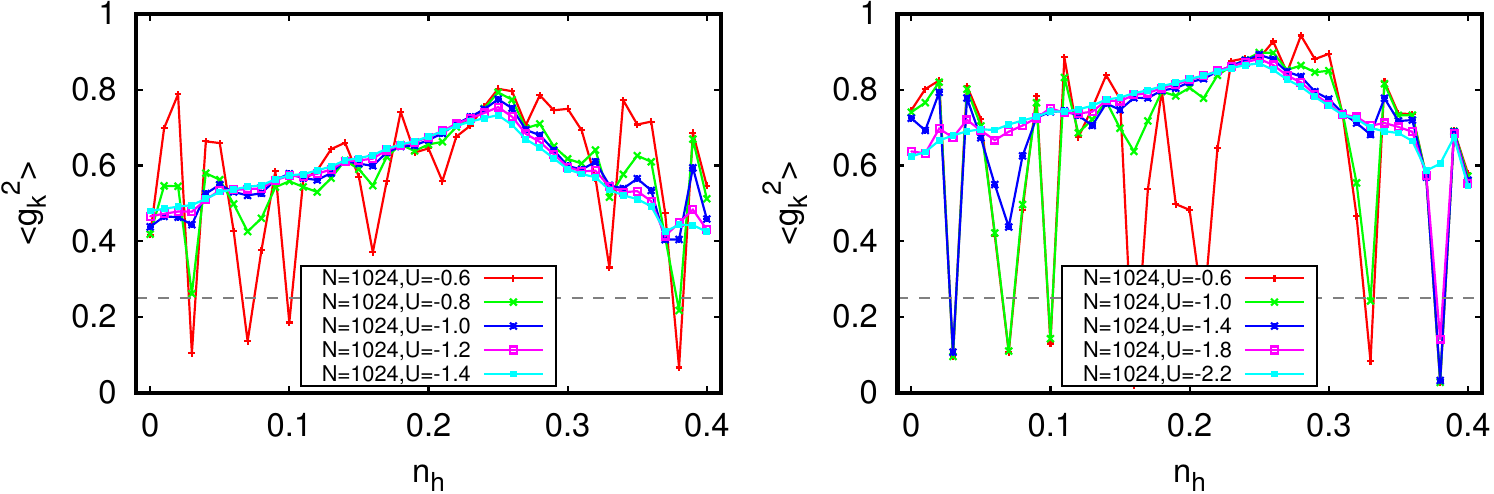}%
\end{center}
\caption{\label{fig9}
\label{fig_nfill_gk2_holes}
Ground state quantum average $\langle g_k^2\rangle$ 
for hole pairs 
as a function of doping  $n_h=1-n$ for $N=1024$ and the 
sector $\p_+=0$, with 
$g_k=(\cos(k_x)-\cos(k_y))/2$ and
$\k=\Delta \p = (k_x, k_y)$ being the 
quantum momentum space of the ground state at given doping $n_h$
($\p_+ =\k_+$ at $\hbar=1$) .
Left (right) panel corresponds to the Hubbard (d-wave) interaction 
with $U=-0.6, -0.8, -1, -1.2, -1.4$ ($U=-0.6, -1, -1.4, -1.8, -2.2$) 
for red, green, blue, pink, cyan curves respectively. 
The grey dashed line at the value $\langle g_k^2\rangle = 0.25$ corresponds 
to the uniform average over all values  $\k=(k_x,k_y)$ with probability
homogeneously distributed in the plane $-\pi < k_{x,y} < \pi$.}
\end{figure}

The gap values allow to obtain the critical
temperature $T_c$ of superconductivity
using the standard relation $\Delta = 1.764 k_B T_c$
(here $k_B$ is the Boltzmann constant and temperature $T_c$
is measured in Kelvin) \cite{tinkham}. 
In Fig.~\ref{fig8}, we also present the dependence of $T_c$ on
hole doping $n_h$ in LSCO.
For the Hubbard case at $U = -1.2$ we obtain the maximal
$T_c \approx 36 K$  (at the hopping $t=0.43\,$eV \cite{markiewicz})
being rather similar
to the maximal $T_{c0} = 38\,K$ obtained
experimentally (see Fig.11 in \cite{markiewicz}
and experimental Refs. therein).
The LSCO experimental results are satisfactorily
described by the doping dependence 
$T_c=T_{c0}[1-(\frac{n_0-n_h}{n_1})^2]$
with the optimal doping $n_0 = 0.16$
and $n_1=0.11$ \cite{markiewicz}.
The Hubbard results at $U=-1.2$
(at $t=0.43\,$eV this corresponds to $U =0.516\,$eV)
give the closest similarity of the $T_c$
dependence on hole doping $n_h$.
Still the numerical data at $U=-1.2$
give a somewhat different shape of the curve $T_c(n_h)$ as compared to
experimental data. Thus, the optimal doping
is at $n_h = 0.24$ for $U=-1.2$
(it slightly changes with $U$).
It is slightly below the doping value $n_{hs} =1-n_s \approx 0.26$ 
corresponding to the separatrix (see Fig.~\ref{fig1}).
Indeed, the density of states is maximal
at the van Hove singularity which significantly
contributes to the gap increase if the Fermi surface of holes
is located slightly below the separatrix value $n_{hs} \approx 0.26$. 
In this case we have $E_F>E_s$ ($E_s$ being the separatrix energy) and 
the accessible hole states include the region of $E_s$ that contributes 
to increase of the (sector) density of states. 
Our numerical data provides a dependence $T_c(n_h)$ on $n_h$ which seems 
to be rather close to the experimental data. 
We attribute certain differences (shift of the maximum position) 
to the fact that for LSCO three-dimensional effects significantly affect
the hopping parameters and the separatrix position 
as discussed in \cite{fresard}. 
In particular, Fig. 15 of \cite{fresard} indicates a separatrix position 
closer to $n=0.84$ ($n_h=0.16$) due to 3D and multiple band effects where 
the $k_z$ quantum number also plays a role. 
Furthermore, our computations are based on the simple 
Hubbard interaction which may be different from the real
effective interaction between holes. 

We also show the dependence $T_c(n_h)$ 
for the attractive d-wave interaction,
in the bottom panel of Fig.~\ref{fig8},
with curves being rather similar to the Hubbard case.
However, a somewhat stronger attractive interaction strength
$U=-2$ ($U = -0.86\,$eV for $t=0.43\,$eV)
is required to have a maximal $T_{c}$ value
close the experimental value $T_c =38\,K$
while the shape of the curves $T_c(n_h)$
remains rather similar to the Hubbard case.
Thus the comparison of $T_c(n_h)$ curves
for Hubbard and d-wave interactions
indicates that the shapes of the Fermi surface curves
is mainly at the origin of gap dependence on doping
in the HTC model.

In Fig.~S9 of SupMat, we also show for completeness
the dependence of $T_c(n)$ on $n_h=1-n$ for Cooper pairs of
electrons which have a rather similar structure as the 
hole case but in both cases there is certain a asymmetry around 
the maximum which is different between holes and electrons.
Thus at  doping
$n_h=0.2$ and $U=-1.2$ the gap for electron pairs
is about 50\% smaller than for hole pairs.

\begin{figure}
\begin{center}
\includegraphics[width=0.95\columnwidth]{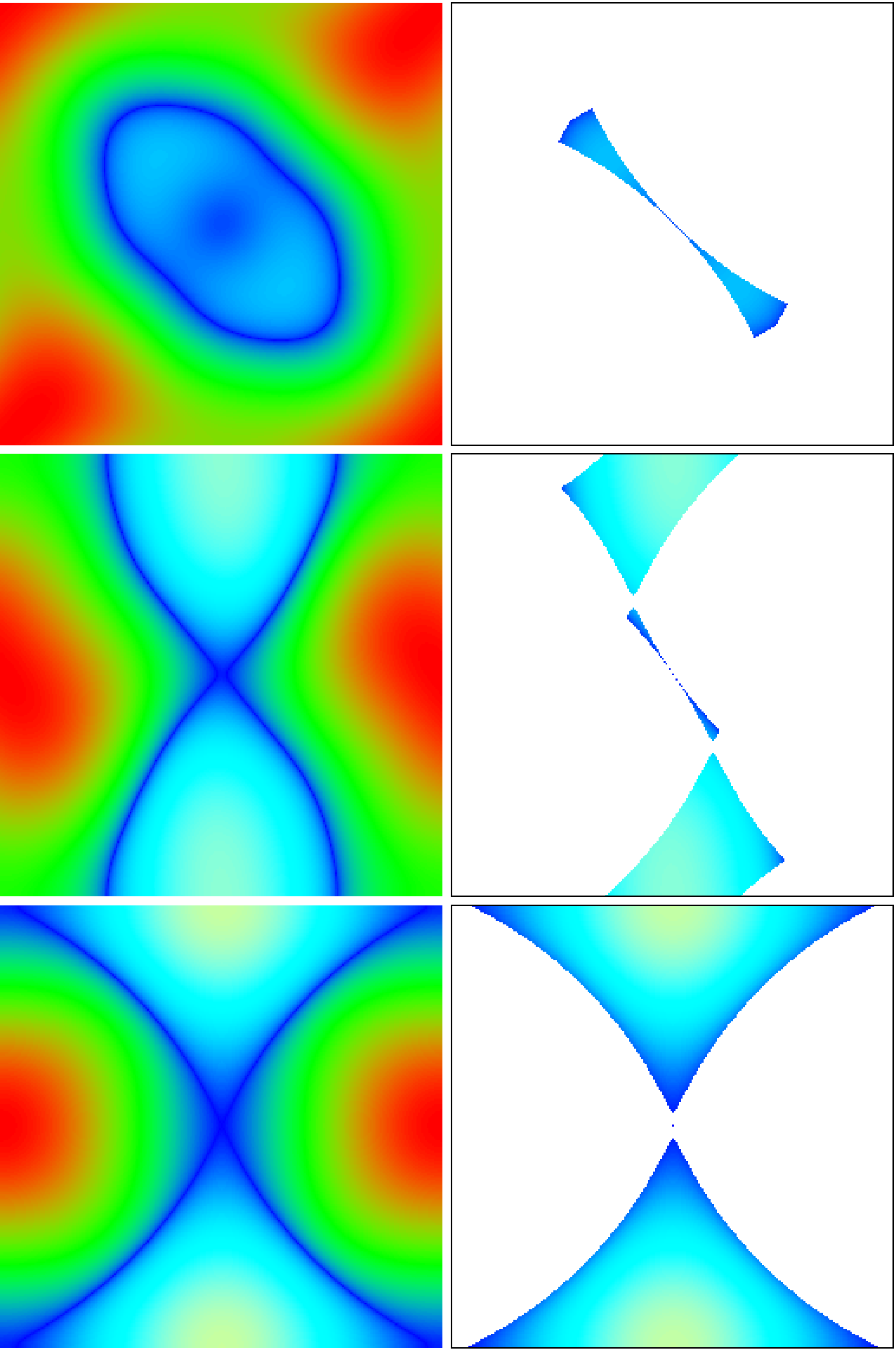}%
\end{center}
\caption{\label{fig10}
  \label{fig_energy_Holes0p74}
Energy landscape for mobile hole pairs.
Left panels show color plots 
of $E_c(\p_+/2-\Delta \p,\p_+/2+\Delta \p)-2E_F$ 
for the HTC model in the $\Delta p_x$-$\Delta p_y$ plane for 
$-\pi\le \Delta p_{x,y}<\pi$. The Fermi energy $E_F$ 
corresponds to the filling factor $n=0.74$ (in all panels). 
Top (center, bottom) panel corresponds to 
the sector $\p_+=2\pi(103,103)/256$ 
($\p_+=2\pi(46,172)/256$, $\p_+=2\pi(0,248)/256$).
The three values of $\p_+$ are chosen such that the center of 
mass momentum $\p_+/2$ is very close to the Fermi surface 
of virtual filling factor $n_v=0.74$ with three cases 
of $p_{+x}=p_{+y}$, $p_{+x}\approx p_{+y}/4$ and $p_{+x}$ ($p_{+y}$) 
minimal (maximal). 
The choice of discrete values is motivated by subsequent 
quantum computations at $N=256$ with these exact identical parameters.
The colors red (green) correspond to 
positive maximum (intermediate), blue to zero value
and yellow (cyan) to strongest (intermediate) negative values. 
Right panels are as the left panels but the forbidden 
zones of $\Delta \p$ (for hole excitations) 
such that each one-particle energy is above 
the Fermi energy, i.e. 
$E_{1p}(\p_+/2-\Delta \p)>E_F$ and 
$E_{1p}(\p_+/2+\Delta \p)>E_F$,  
are replaced by white color. Note that here 
the white zones include not only 
the positive value zones (red/green) in left panels but 
also additional zones of negative values due to $\p_{+}\neq 0$ and 
the more complicated selection rule using individual 
one particle energies.
}
\end{figure}

To characterize the d-wave structure of the ground state
we compute the value of the quantum average $\langle g_k^2\rangle$ 
over the ground state in momentum representation 
(with $k$ being $\Delta\p$ and $g_k=(\cos(k_x)-\cos(k_y))/2$).
The dependence  of $\langle g_k^2\rangle$
on hole doping $n_h$ is shown in Fig.~\ref{fig9}
for different values of $U$ for Hubbard and d-wave interactions.
At small $|U|$ the interactions and gap are too weak
and the discreteness of momentum values at finite lattice size
leads to strong fluctuations of $\langle g_k^2\rangle$
with $n_h$. This happens because at small $|U|$
only few specific $k$ values, closest to the Fermi surface, 
contribute to the ground state (a similar effect is discussed in detail
for rough billiards in \cite{roughbil}).
However, for moderate interactions
($|U| \geq 1$ for  Hubbard  and 
$|U| \geq 1.4 $ for d-wave cases), corresponding to 
$T_c$ values close to experimental ones
(see Fig.~\ref{fig8}),
the system size $N=1024$ 
is sufficiently close to the infinite $N$ limit
with a  smooth dependence of  $\langle g_k^2\rangle$
on $n_h$. 
As for $\Delta(n_h)$ shown in Fig.~\ref{fig8} the average 
$\langle g_k^2\rangle$ has also a maximum 
close to the optimal doping $n_h \approx 0.26$ corresponding
to the separatrix (van Hove singularity).
However, in contrast to $\Delta(n_h)$ the maximum is not 
very smooth and the lowest values of $\langle g_k^2\rangle$ 
(in the interval $0\le n_h\le 0.4$) are quite large, about $\sim 65$ \% 
of the maximal value. The maximal 
values themselves $\langle g_k^2\rangle \approx 0.75$ (Hubbard case)
and $0.87$  (d-wave case) are rather high 
and close to unity which corresponds to a strong 
concentration of the wavefunction in the vicinity 
of the antinode $k_x \approx 0, k_y \approx \pm\pi$
(or inverse). 

Such a concentration is indeed visible
for the ground state in momentum space
shown in Fig.~\ref{fig7}. We note that for the whole
considered range of dopings $0 \le n_h \le 0.4$
the obtained values of  $\langle g_k^2\rangle$
are significantly larger than the 
value $0.25$ corresponding to a homogeneous distribution
of probability over all $k_x, k_y$ values in the interval $[-\pi,\pi]$.
Using the classical local density $\rho_g(g_k)\sim 1/(1-|g_k|)$ 
with a cutoff $|g_k|\le g_{\rm max}$, where $g_{\rm max}<1$ 
is the maximal possible value of $|g_k|$ (for Fermi curves close to the 
separatrix curve; see Appendix \ref{appA2}), one can expect for the Hubbard case 
the analytical estimate~: 
$\langle g_k^2\rangle_{\rm cl.}\approx 1-3/(2|\ln(1-g_{\rm max})|)$ 
which provides theoretically unity for the exact separatrix curve but 
with a rather strong logarithmic correction even if $1-g_{\rm max}\ll 1$ 
which explains the rather larger values in Fig.~\ref{fig9} 
(significantly above $0.25$) but still somewhat smaller than unity. 

For the d-wave interaction, we remind that the momentum wave function 
amplitudes are essentially 
multiplied with $g_k$ (in comparison to the Hubbard wave function amplitudes 
at same gap value) and we expect that 
$\langle g_k^2\rangle_{q,{\rm d-wave}}\approx \langle g_k^4\rangle_{\rm cl.}
/\langle g_k^2\rangle_{\rm cl.}\approx 1-7/(12|\ln(1-g_{\rm max})|)$, with 
a reduced logarithmic correction explaining the somewhat larger values 
(closer to unity) in the right panel of Fig.~\ref{fig9}. 

We note that similar results are obtained for the dependence of
$\langle g_k^2\rangle$ on $n_h$ 
for electron pairs (see Fig.~S10 of SupMat).

The fact, that for both interactions the average $\langle g_k^2\rangle$, 
is significantly above the uniform average $0.25$, confirms the findings 
of Section~\ref{sec3} that the HTC-band structure alone induces a kind 
of d-wave preference in classical phase space (larger distance between 
two neighbor Fermi curves if $|g_k|\approx 1$) or for quantum states 
(with more occupied grid points in the regions close to the 
Fermi surface if $|g_k|\approx 1$). 
Therefore, to observe a d-wave dependence it is not necessary to inject 
a d-wave dependence in the interaction as such, as can be seen 
in the results of Fig.~\ref{fig8} and Fig.~\ref{fig9}
for the (s-wave) Hubbard interaction. For the d-wave 
interaction, the ``d-wave'' effect 
is somewhat enhanced but this enhancement is not the dominant part. 
Furthermore, the HTC-band structure also breaks the central symmetry
in the vicinity of optimal doping values.

\begin{figure}
\begin{center}
\includegraphics[width=0.95\columnwidth]{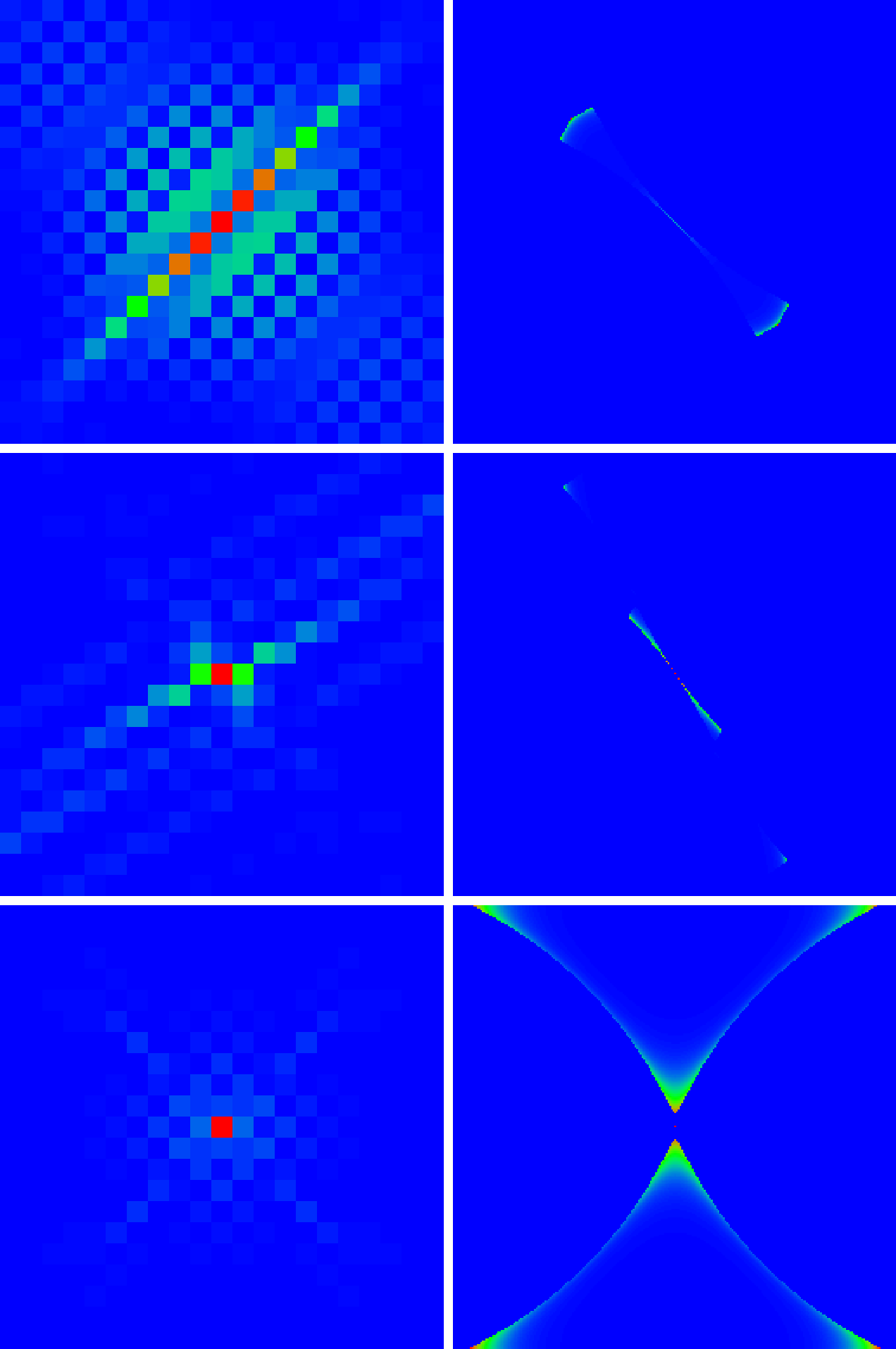}%
\end{center}
\caption{\label{fig11}
\label{fig_states_Holes0p74}
Ground state density plots for the Hubbard interaction, system size $N=256$, 
hole pairs, 
filling factor $n=0.74, n_h = 1 -n$ and three sectors $\p_+\neq 0$ (same 
values as in Fig.~\ref{fig_energy_P0p74}). Top (center, bottom) panels 
correspond to $U=-8$, $\p_+=2\pi(103,103)/256$ 
($U=-6$, $\p_+=2\pi(46,172)/256$; $U=-4$, $\p_+=2\pi(0,248)/256$).
Left panels show the ground state 
in $\Delta \r$-representation in a zoomed region with 
$-10\le \Delta x,\Delta y\le 10$ 
(color values outside the zoomed regions are blue) 
and right panels show the state 
in $\Delta \p$-representation (with $-\pi\le \Delta p_{x,y}<\pi$).
The two hole ground state energies $E_{\rm min}$ 
(of the sector Hamiltonian and 
in units of the basic hopping matrix element) 
 are $-0.03424$ ($-0.07144$, 
$-0.1884$) for top (center, bottom) panels.
}
\end{figure}

\section{Gap for  mobile  Cooper pairs of holes}
\label{sec6}

In this Section, we discuss the case of mobile pairs of holes 
with $\p_+\neq 0$. 
Similarly as in Section~\ref{sec4}, we use a virtual filling 
$n_v=0.74$ (and $n_v=0.84$ for SupMat figures) corresponding 
to certain center of mass values $\p_+/2$ being (very close) to 
the Fermi surface with filling $n_v$. 

\begin{figure}
\begin{center}
\includegraphics[width=0.95\columnwidth]{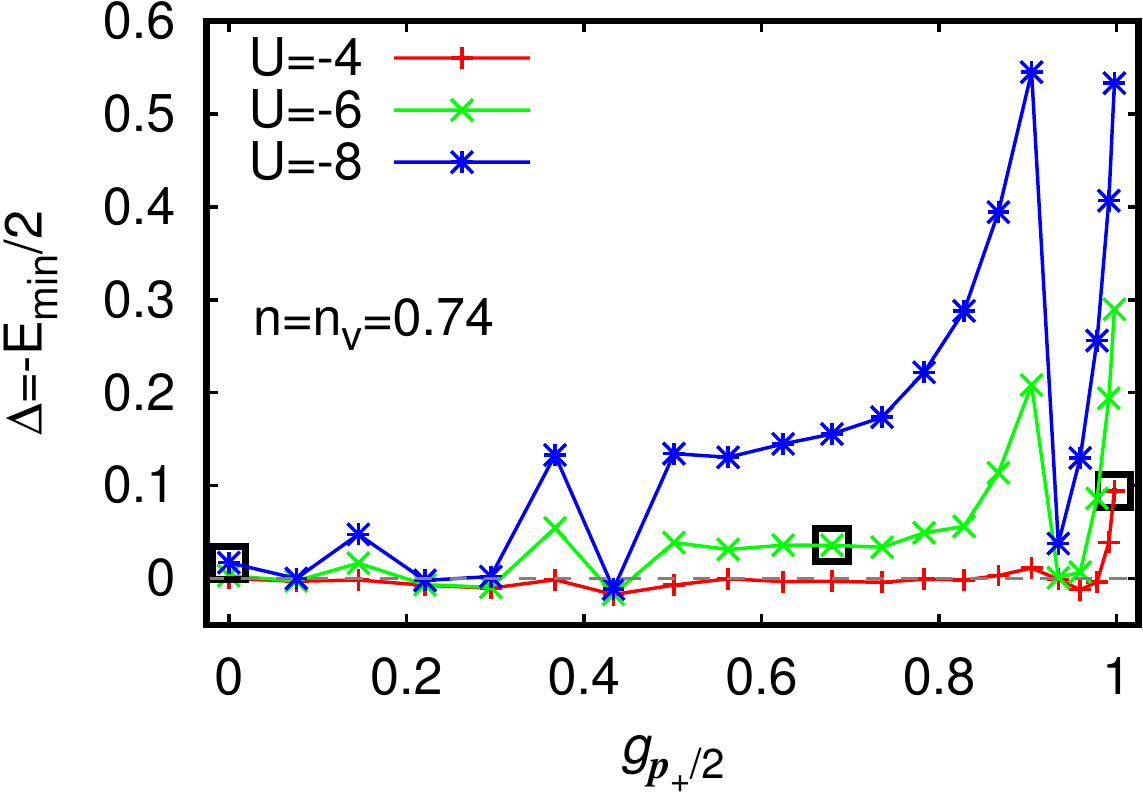}%
\end{center}
\caption{\label{fig12}
\label{fig_gap_0p74}
Gap energy $\Delta=-E_{\rm min}/2$ for hole pairs
as a function of 
$g_{\p_+/2}=[\cos(p_{+x}/2)-\cos(p_{+y}/2)]/2$ for $n=0.74, n_h = 1-n$ 
and 21 sector values of $\p_+$ such that the values
of the center of mass $\p_+/2$ lie uniformly on the line of virtual 
filling factor $n_v=0.74$ between the positions 
of $p_{+x}=p_{+y}$, with $g_{\p_+/2}=0$, 
and $p_{+x}\approx 0$, $p_{+y}\approx 2\pi$, 
with $g_{\p_+/2}\approx 1$. 
The three curves correspond to the three interaction values 
used in Fig. \ref{fig_states_Holes0p74} 
and the data points with black squares correspond to the 
three states shown in Fig. \ref{fig_states_Holes0p74} 
with $g_{\p_+/2}=0$ ($\approx 0.7$, $\approx 1$)
for top (center, bottom) row therein.
Data are obtained at $N=256$.
}
\end{figure}

An example of the energy landscape for mobile pairs
is shown in Fig.~\ref{fig10}
for the filling factor $n=0.74, n_h=1-n$
and the virtual filling factor being very
close to this value $n_v =0.74$
(up to discreteness lattice effects).
We see that the energy landscape changes 
significantly depending on the value of 
$\p_+/2$ on the virtual Fermi surface at $n_v$.
The landscape is shown for three cases 
of $p_{+x}=p_{+y}$, $p_{+x}\approx p_{+y}/4$ and $p_{+x}$ ($p_{+y}$) 
minimal (maximal). 
Even more striking are the changes of the zones 
of accessible $\Delta\p$ values 
shown in the right column of Fig.~\ref{fig10} 
due to the condition $E_{1p}(\p_+/2\pm\Delta \p)<E_F$ 
(see also discussion at the beginning of Section~\ref{sec5}). 
For $p_{+x}=p_{+y}$ these zones are composed of a quite small 
island with a dumbbell form. For $p_{+x}\approx p_{+y}/4$ this 
island is strongly reduced but two extra pieces around 
$\Delta\p=(0,\pm\pi)$ have been added. Finally, for $p_{+x}=0$ and $p_{+y}$ 
maximal the island has (nearly) disappeared and the extra pieces have 
increased in size with curved boundaries.

Examples of ground states of hole pairs
for parameters of Fig.~\ref{fig10}
are shown in Fig.~\ref{fig11}. 
Similarly, as in Fig.~\ref{fig6}, 
the ground states correspond
to compact pairs in $\Delta\r$-representation (left column) 
with a size decreasing with the increase of the gap $\Delta$.
Their densities in $\Delta\p$-representation (right column) 
are again concentrated at certain borders of the frozen 
Fermi sea (``blue'' Fermi sea borders with energies close to the Fermi 
surface; see right column of Fig.~\ref{fig10}). In particular, 
the (momentum) ground state for $p_{+x}=p_{+y}$, is concentrated on the 
outside borders of the dumbbell island. 

The important feature of these ground states
is that the gap values $\Delta = -E_{\rm min}/2$
are rather modest even if the Hubbard interaction strength 
is by a factor $4$ or even more higher
as compared to the case of static pairs of Fig.~\ref{fig8}. 
Similarly as with mobile electron pairs (see Section~\ref{sec4}) it 
is necessary to consider rather large interaction amplitudes $|U|$ 
between -4 and -8 to find nice pair states. 

It is convenient to express the gap dependence on
$p_{+x}$, $p_{+y}$ via the quantity 
$g_{\p_+/2}=[\cos(p_{+x}/2)-\cos(p_{+y}/2)]/2$
which characterizes the position of the center of mass 
$\p_+/2$ on the (virtual) Fermi surface 
(we note that this quantity is different
from $g_k$ used in the previous Sections since now $k$ 
corresponds to the center of mass $\p_+/2$ while previously it 
was given by the relative momentum $\Delta\p$).
The dependence of the gap $\Delta$ on this quantity 
is shown in Fig.~\ref{fig12} for three interactions values $U=-4,-6,-8$ 
and for 21 uniformly distributed data points on the virtual Fermi surface. 
The main observations from Fig.~\ref{fig12} 
can be listed as follows: 
the gap is very small at $g_{\p_+/2} \approx 0$ 
(symmetry point $p_{+x} = p_{+y}$) and is highest at
$g_{\p_+/2} \approx 1$ (asymmetry point $p_{+x} = 0$, $p_{+y}$ maximal,
or inverse). This can be understood from the fact that the
number of accessible states is significantly larger for 
$g_{\p_+/2} \approx 1$ than for
$g_{\p_+/2} \approx 0$ (small dumbbell island) or for 
other intermediate states (with intermediate $g_{\p_+/2}$) 
as it is well seen in the right column panels of Figs.~\ref{fig10} 
and \ref{fig11}.
The gap appears at rather large values $|U|$ for the Hubbard interaction 
as compared to the case of static pairs (with $\p_+=0$; 
see Section~\ref{sec5}). 
Similar results for another case,  $n=n_v = 0.84$, are 
shown in Figs.~S11,~S12,~S13 of SupMat.

We also considered the case of small values of $|\p_+|\to 0$ 
(nearly static pairs) at modest interaction strength $|U|=1$ 
(case of presence of a modest 
gap $\Delta\approx 0.06$ for holes and $\Delta\approx 0.1$ for electrons 
at $n=0.74$ and $\p_+=0$ and zero gap at $n_v=0.74$ with non-zero 
$\p_+$ values of Figs.~\ref{fig10} and \ref{fig11}).
It turns that for $N=512$ the gap rapidly disappears with increasing 
value of $|\p_+|$ at $|\p_+|\gtrsim (2\pi l)/512 $ with $l\approx 7-10$ 
for particles and $l\approx 3-5$ for holes. These borders correspond to 
very small virtual filling values $n_v\sim 10^{-4}-10^{-3}$.

In global, the results of this Section show that
it is possible to have coupled mobile pairs 
with an energy gap but the required (attractive) interaction amplitude 
should be $4-8$ times larger as compared to the case of static pairs.

\section{Pairs with Coulomb repulsion}
\label{sec7}

\begin{figure}
\begin{center}
\includegraphics[width=0.95\columnwidth]{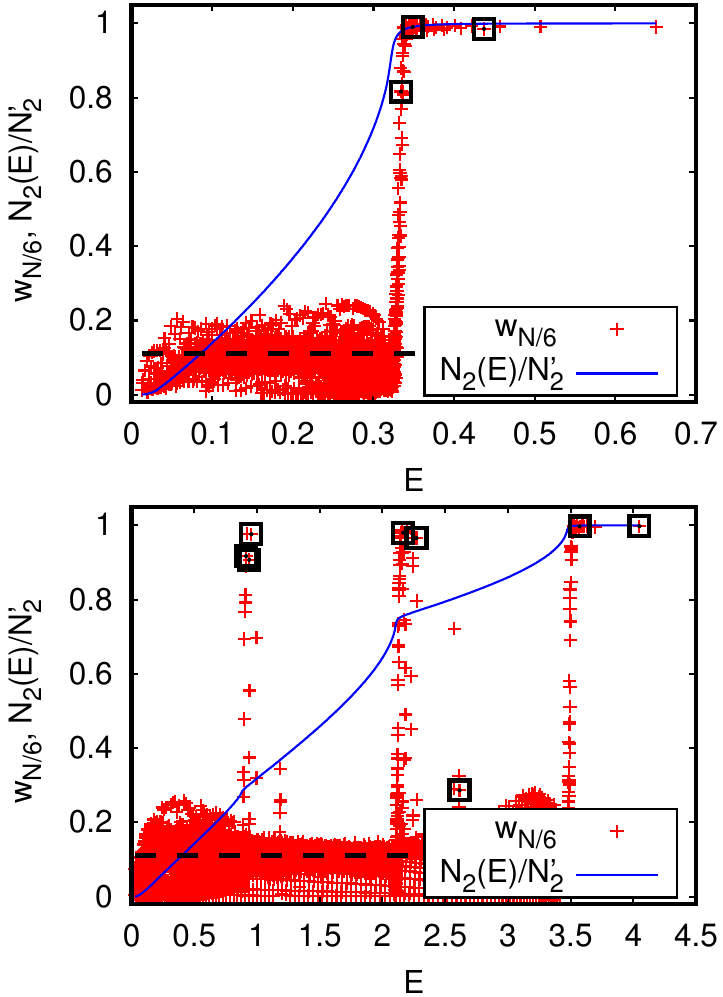}%
\end{center}
\caption{\label{fig13}
\label{fig_wN6}
The pair weight $w_{N/6}$ defined as the quantum probability for 
$|\Delta x|\le N/6$ and $|\Delta y|\le N/6$ is shown as a function of the 
pair excitation energy $E$ (eigenvalue of the sector Hamiltonian and with 
$E=0$ corresponding to the Fermi level; see Section~\ref{sec2} for details) 
for the case of repulsive Coulomb interaction 
with $U=2$, HTC model and filling factor $n=0.74$.
Top (bottom) panel corresponds to hole excitations with 
$\p_+=2\pi(207,\,207)/512$, sector dimension $N_2'=3040$ 
and $N=512$ with center of mass $\p_+/2$ 
being on the Fermi surface for virtual filling factor $n_v=0.74$ 
(particle excitations with $\p_+=2\pi(113,\,113)/256$, sector dimension $N_2'=8737$, 
$N=256$ and $n_v=1$).
The blue line shows $N_2(E)/N_2'$ with $N_2(E)$ being the number of levels 
below $E$. The energy values with strong energy derivative of this 
quantity, corresponding to strong peaks of sector density of states, 
coincide with the main peaks of $w_{N/6}$ for well defined close pair 
states. The dashed black line indicates the value $w_{N/6}=1/9$ 
for uniform ergodic states on the whole lattice. The energy landscape 
for these parameters is shown in Fig.~\ref{fig14}. 
The data points with black squares correspond to the 
states shown in Figs. \ref{fig15}, \ref{fig16}
(and in related Figs.~S14,~S15 of SupMat).
Additional data, especially raw png figures for pair states 
with $w_{N/6}>0.4$, for these 
two cases and also for $N=512$ for the parameters of the bottom panel 
are available at \cite{ourwebpage}. 
}
\end{figure}

In this Section, we present results of pair eigenstates for the repulsive 
Coulomb interaction (see case (i) in the discussion of Section~\ref{sec2})
combined with a frozen Fermi sea. 
In previous works \cite{prr2020,htcepjb}, the time evolution of electron 
pairs in NN and HTC lattices was studied for free pairs (in absence of 
a frozen Fermi sea) showing that the Coulomb repulsion can lead to 
Coulomb pair formation due to the appearance of an effective narrow or 
flat band when the total pair momentum is $\p_+\approx (\pm\pi, \pm\pi)$.
Such a mechanism is rather interesting but it is important to understand
if such states can have a gap ($E_{\rm min} <0$) or not and if such 
Coulomb pairs can exist in presence of a frozen Fermi surface and 
at which energies. 

\begin{figure}
\begin{center}
\includegraphics[width=0.95\columnwidth]{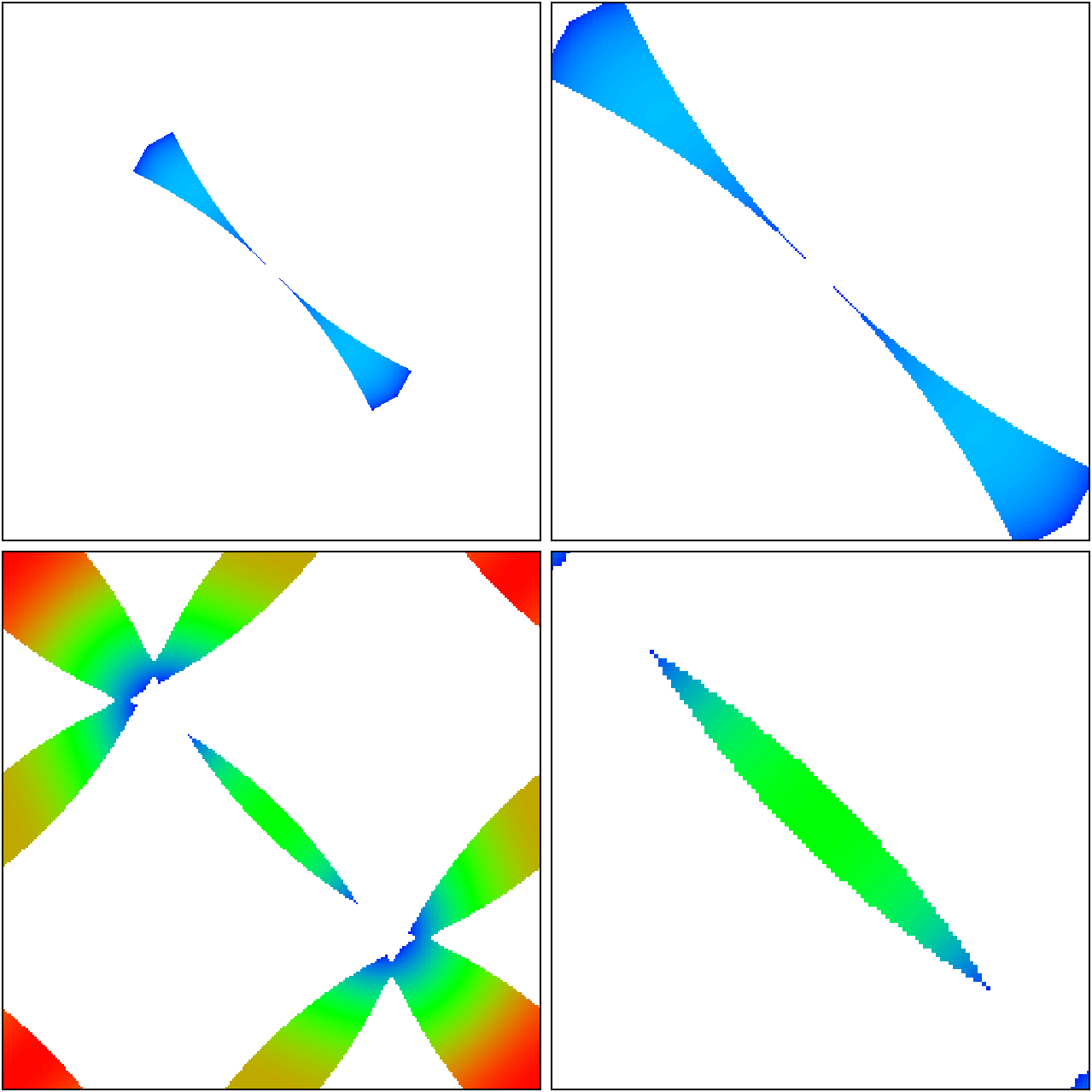}%
\end{center}
\caption{\label{fig14}
\label{fig_Coul_energy}
Colored allowed (forbidden white) zones in $\Delta \p$-plane 
(see captions of Figs.~\ref{fig5},~\ref{fig10} for details). 
Top (bottom) panels correspond to the parameters of top (bottom) panel 
of Fig.~\ref{fig13} with filling factor $n=0.74$, hole (electron) 
excitations, $\p_+=2\pi(207,\,207)/512$ and $N=512$ 
($\p_+=2\pi(113,\,113)/256$
and $N=256$). 
Right (left) panels show the full momentum cell: $-\pi\le \Delta p_{x,y}<\pi$ 
(zoomed center square: $-\pi/2\le \Delta p_{x,y}<\pi/2$) 
in $\Delta \p$-plane. 
Note that top left panel here 
corresponds nearly exactly to the top right panel 
of Fig.~\ref{fig10} (with a slight difference 
in $\p_+$ due to different choices of $N$).
}
\end{figure}

Due to a more complicated structure of coupling matrix elements 
the effective method of Appendix \ref{appA1} is not suitable and we determine 
the eigenstates and energies of the sector Hamiltonian (see Section~\ref{sec2} 
for details of its definition) by numerical full diagonalization. 
For this, we consider two particular cases: 

(i) Hole excitations at $n=0.74$ with one single value of $\p_+$ such that 
$p_{+x}=p_{+y}=2\pi(207/512)$ corresponding to $n_v=0.74$ and sector 
dimension $N_2'=3040$ at $N=512$. Note that the sector dimension is strongly 
reduced with respect to $N_2=N^2=512^2$ due to the small fraction of 
available states (case of dumbbell island visible in top right panel of 
Fig.~\ref{fig10}) allowing to choose the rather large system size $N=512$.

(ii) Electron excitations at $n=0.74$, also with one single value of 
$\p_+$ such that $p_{+x}=p_{+y}=2\pi(113/256)$ corresponding 
to $n_v=1$ and sector dimension $N_2'=8737$ at $N=256$. 
We have also computed the eigenstates and energies for the larger 
case $N=512$ with $p_{+x}=p_{+y}=2\pi(225/512)$, $N_2'=35030$, 
and verified that all physical conclusions remain valid. However, 
for practical reasons, we present here figures and the discussion only 
for the case $N=256$ (reduced number of data points and better visible 
eigenfunction figures at $N=256$, especially in momentum space). 
The choice of $\p_+$ corresponding 
to $n_v=1\neq n=0.74$ is motivated by its proximity to the ``optimal'' value 
$(\pi,\pi)$ found in \cite{prr2020,htcepjb} and the fact that the zone 
of allowed $\Delta\p$ values covers the vicinity of $\Delta\p=0$ which is 
actually a point of ``negative mass'' as we will see below (at $n_v=0.74$ and 
$p_{+x}=p_{+y}$ the region $\Delta\p=0$ would be in the forbidden zone; 
see top right panel of Fig.~\ref{fig5}).

Additional data, especially raw png figures for pair states, for these 
 cases (i), (ii) and (ii) for $N=512$ are available at 
\cite{ourwebpage}.

We also studied many other parameters (other choices of $n$, $n_v$, 
$\p_+$ with $\p_+/2$ being on different points of the virtual Fermi surface 
etc.) and in all cases the ground state energy $E_{\rm min}$ (of the 
sector Hamiltonian) was found to be positive such that there is no gap 
(in the framework of this approach) for repulsive Coulomb interaction. 
However, we discovered different mechanisms of pair formation at 
different excitation energies (sometimes close to the Fermi energy, sometimes 
at quite high excitation energies). The two specific examples above 
provide  eigenstates for all interesting cases which we will discuss 
below. 

\begin{figure}
\begin{center}
\includegraphics[width=0.95\columnwidth]{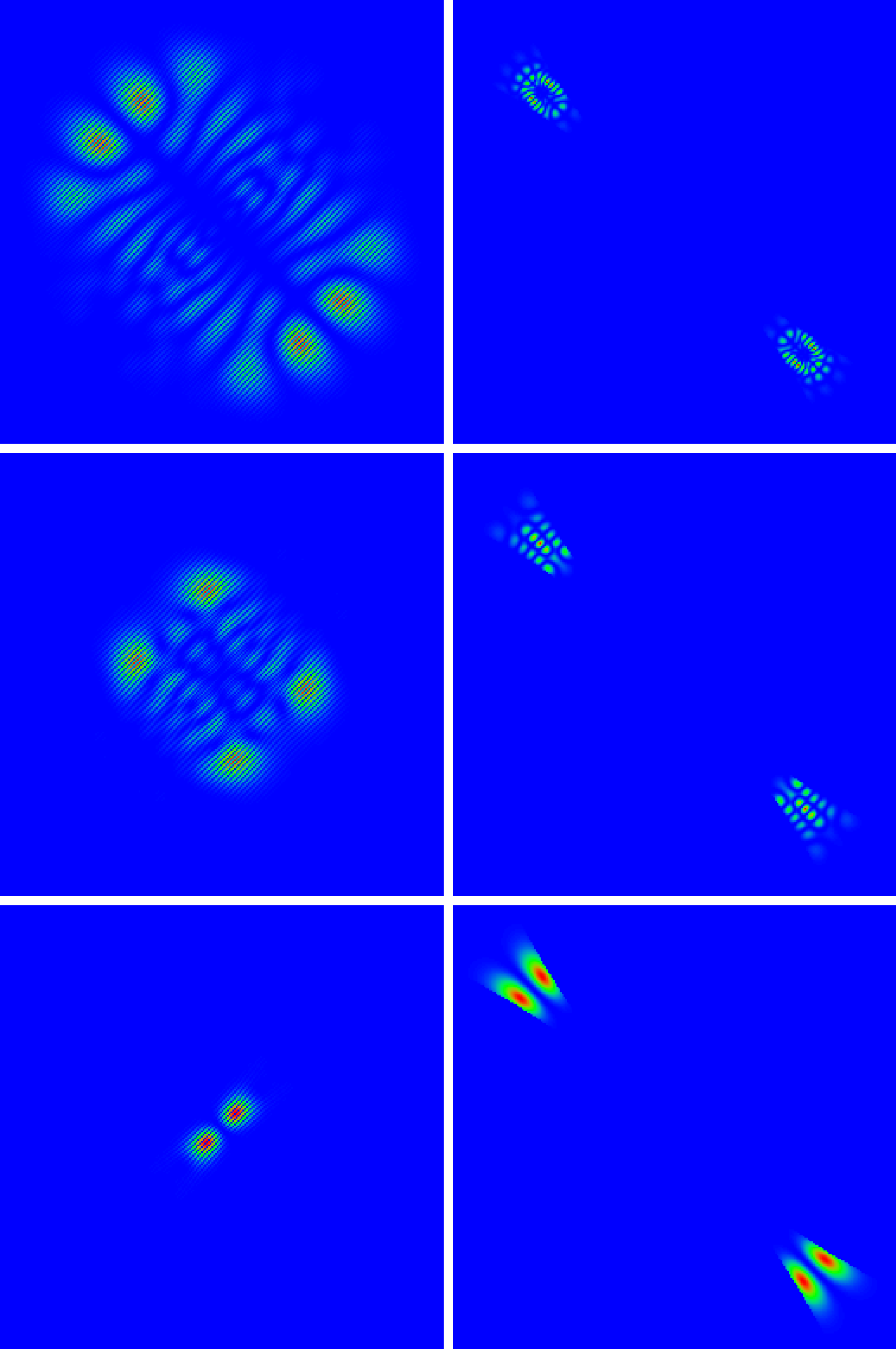}%
\end{center}
\caption{\label{fig15}
\label{fig_Cstates_holes}
Three strong pair eigenstates for the parameters of top panel of 
Fig.~\ref{fig13} (holes, $n=n_v=0.74$, $N=512$) 
for energies close to $0.3$ and marked by black squares 
therein. 
Left (right) columns correspond to the $\Delta \r$-
($\Delta \p$-) representation showing the two times zoomed center square 
for both cases: 
$-N/4\le \Delta x,\Delta y<N/4$ ($-\pi/2\le \Delta p_{x,y}<\pi/2$). 
The panels in $\Delta \p$-representation 
correspond to the top right panel 
of Fig.~\ref{fig14} concerning the identification of 
allowed and forbidden zones. 
Top (center, bottom) row corresponds to the eigenstates 
with level number $2974$ ($3011$, $3037$), 
energy $0.3342$ ($0.3485$, $0.4371$) and pair weight $w_{N/6}=0.8164$ 
($0.9900$, $0.9853$). Note that $N_2'=3040$ is the maximal possible 
level number for the largest energy (in the corresponding 
$\p_+$-sector).
}
\end{figure}

\begin{figure}
\begin{center}
\includegraphics[width=0.95\columnwidth]{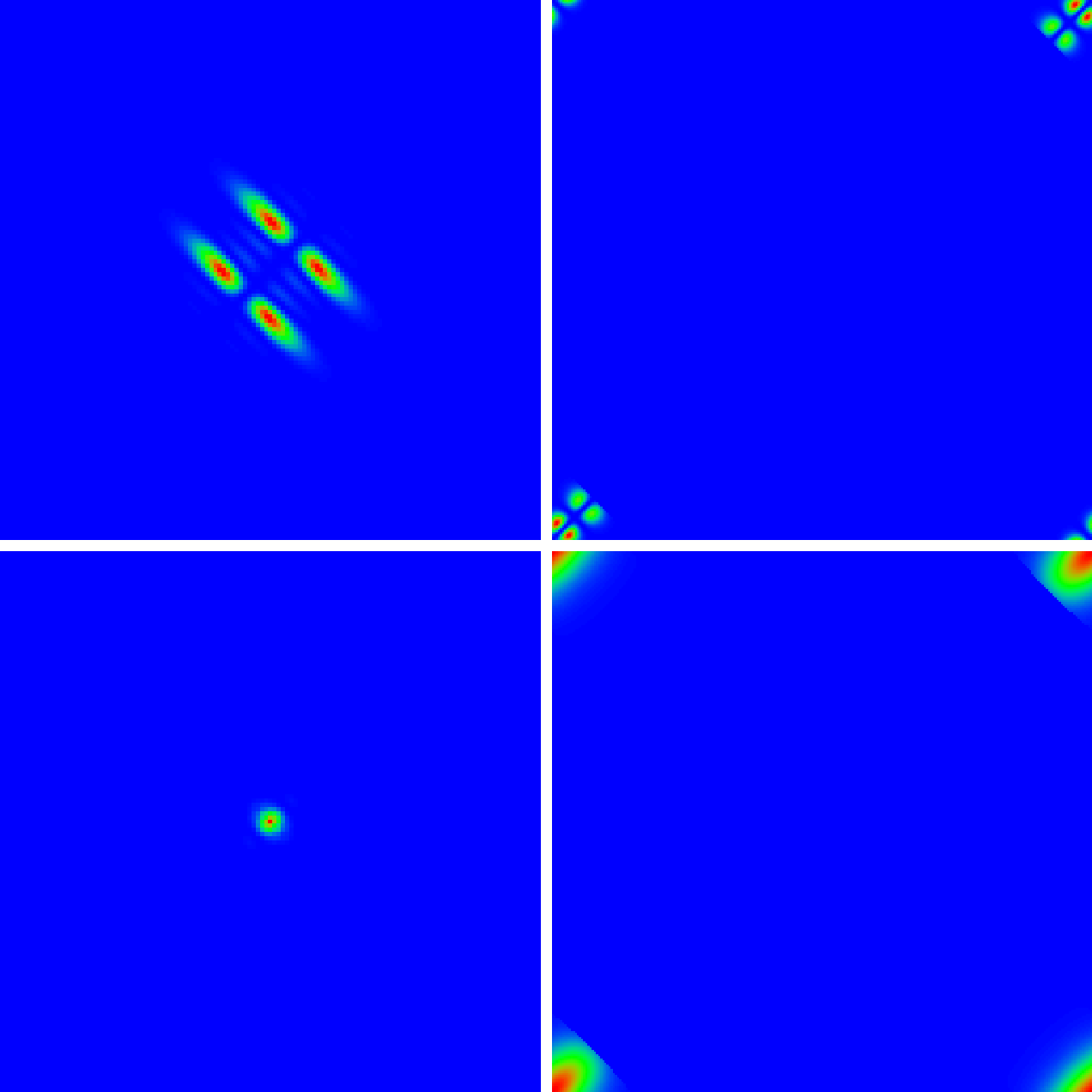}%
\end{center}
\caption{\label{fig16}
\label{fig_Cstates_particles3}
Two strong pair eigenstates for the parameters of bottom panel of 
Fig.~\ref{fig13} (particles, $n=0.74$, $n_v=1$, $N=256$) 
of the third strong peak of large 
$w_{N/6}$-values for energies close to $3.5$-$4$ and marked by black squares 
therein. 
Left (right) columns correspond to the $\Delta \r$-
($\Delta \p$-) representation showing the two times zoomed center square: 
$-N/4\le \Delta x,\Delta y<N/4$ (full momentum cell: 
$-\pi\le \Delta p_{x,y}<\pi$). 
The panels in $\Delta \p$-representation 
correspond to the bottom left panel 
of Fig.~\ref{fig14} concerning the identification of 
allowed and forbidden zones. 
Top (bottom) row corresponds to the eigenstates 
with level number $8732$ ($8737$), 
energy $3.571$ ($4.048$) 
and pair weight $w_{N/6}=0.9970$ ($0.9976$). 
Here $N_2'=8737$ is the maximal possible 
level number for the largest energy (in the corresponding 
$\p_+$-sector).
}
\end{figure}

To identify interesting Coulomb pairs 
at excited energies we compute for each eigenstate 
the quantity $w_{N/6}(E)$ defined as the 
quantum probability of $|\Delta x|\le N/6$ and $|\Delta y|\le N/6$
(obtained by summing $|\psi(\Delta\r)|^2$  over $\Delta\r$ satisfying 
this condition with $\psi(\Delta\r)$ being an eigenstate in 
$\Delta\r$-representation; see also Eq. (9) and text below of 
\cite{htcepjb} for the definition of the similar quantity $w_{10}$). 
In this work, we replace the width 
$10$ with $N/6$ since many nice pair states are still quite extended. 
Values of $w_{N/6}(E)$ significantly above the ergodic value 
$1/9$ (i.e. close to $1$) indicate pair states (at certain energies) 
which are quite well localized around $\Delta\r\approx 0$. 
We have also computed other quantities such as 
the quantum averages $\langle |\Delta\r|\rangle$, 
$e^{\langle \ln|\Delta\r|\rangle}$, $\langle \bar U(\Delta\r)\rangle$ 
or the inverse participation ratio in $\Delta\r$-representation, 
providing the same typical energies at which nice pair states appear.

Fig.~\ref{fig13} shows $w_{N/6}(E)$ (red data points) 
as a function of the 
excitation energy $E$ (eigenvalues of the sector Hamiltonian 
with diagonal matrix elements $\pm(E_c(\p_1,\p_2)-2E_F$; see 
discussion of Section~\ref{sec2} for details) for both above examples 
(i) in top panel and (ii) in bottom panel. In addition, also the 
normalized integrated (sector) density of states $N_2(E)/N_2'$, 
(fraction of states with energies below $E$; blue curves) are shown. 

For the case (i), there is one peak of strong pair states, 
with $w_{N/6}(E)$ close to 1, at the top of the energy spectrum, 
mostly at $E\approx 0.3$ and with a few states going up $E=0.4-0.7$ 
(energy measured in units of the basic hopping matrix element $t$). 
For the case (ii), there are three main peaks at energies 
$E\approx 0.9,\, 2.1-2.2$ and $3-3.5$ (top of the spectrum 
for the third peak). In addition there also two secondary peaks 
behind the first two peaks at $E\approx 1.2,\,2.6$. 
We observe at all main peak positions for both cases an enhanced 
slope of $N_2(E)/N_2'$ just before the energy corresponding to the peak 
indicating a strongly enhanced density of states at these energy values. 
For the case (ii) at $E\approx 0.9$ this effect is bit less strong, but 
still visible, as compared to the other peaks. 

To understand these observations and the physical nature of the 
pair states at these energies, we show in Fig.~\ref{fig14} the energy 
landscape of allowed $\Delta\p$ values together with the forbidden zones 
and in Figs.~\ref{fig15}, \ref{fig16} and Figs.~S14, S15 of SupMat examples of 
pair states at the peak energies marked by black squares in Fig.~\ref{fig13}. 
Top panels of Fig.~\ref{fig14} shows again the dumbbell island for the 
case (i) (see also top right panel of Fig.~\ref{fig10}) with a 50\% zoom 
in the right panel and bottom panels correspond to the case (ii) which 
is somewhat similar to the right panel of Fig.~\ref{fig5} but with an 
additional cigar-shape island in the region around $\Delta\p=0$ which 
appears due to the modified virtual filling $n_v=1$ with respect to 
$n_v=0.74$ in Fig.~\ref{fig5} (both with $p_{+x}=p_{+y}$). 

The panels of Fig.~\ref{fig14}, have to be viewed together with the 
eigenstate figures in $\Delta\p$-representation (right panels 
of Figs.~\ref{fig15}, \ref{fig16} and Figs.~S14, S15 of SupMat). 
For example, for the case (i) we see that the eigenstate densities in 
$\Delta\p$-space of the three pairs shown in Fig.~\ref{fig15} are 
concentrated a the outer border of the dumbbell island. The densities 
in $\Delta\r$-space are localized around $\Delta\r=0$ with 
a width of about 33\% (state of top panels), 25\% (state of center panels) 
and 5\% (state of top panels) of the available lattice showing 
that the width decreases when the energies approaches the top of the spectrum. 
These pair states are created by a combined mechanism of top spectrum, 
narrow band and island structure 
because at the top of the spectrum the repulsive Coulomb interaction behaves 
like an attractive interaction (at the bottom of the spectrum) confining 
the particles to a well defined pair. 

The eigenstates shown in Fig.~\ref{fig16} correspond to the case (ii) 
at the top of spectrum (third main peak at $E\approx 3.5-4.5$) with 
$\Delta\p$-densities concentrated 
at the regions $\Delta\p\approx (\pm\pi,\pm\pi)$ corresponding 
to red maximum regions in bottom panels of Fig.~\ref{fig14}. Here the 
pair creation mechanism is essentially due to the top spectrum 
(also negative mass; see below) and there is 
no strong island effect. Also the effective band is not very narrow 
(one may argue that the red zone region in bottom panels of 
Fig.~\ref{fig14} constitutes an effective narrow band). 
The width in $\Delta\r$ space (around $\Delta\r=0$) decreases 
very strongly when approaching the top of the spectrum. 

The  states shown in Fig. S14 of SupMat for the first main energy peak 
($E\approx 0.9$) of the case (ii) are very interesting. Their 
$\Delta\p$-densities are localized around $\Delta\p=0$ which constitutes 
a local energy maximum (center green zone of cigar shape island in 
bottom panels of Fig.~\ref{fig14}). This point is actually characterized 
by two negative eigenvalues of the Hessian matrix obtained by 
expanding $E_c$ given in (\ref{eq_Ec}) up to second order in $\Delta\p$. As 
can be seen in (bottom right panel) of Fig. S16 of SupMat, 
the symmetric value of $\p_+/2$ (i.e. with 
$p_{+x}=p_{+y}$) falls for $n_v=1$ clearly in one of the droplet 
regions where both eigenvalues are negative providing a point of negative 
mass with a clear local maximum in $\Delta\p$ space. Therefore, the 
pairs at $E\approx 0.9$ are created by the mechanism of negative mass which 
is similar to the mechanism of top spectrum where the repulsive Coulomb 
interaction confines particles. In $\Delta\r$-space the densities are 
again localized around $\Delta\r$ with width values 
between 15-25\% of the lattice. 

Example states of the second main energy peak ($E\approx 2.2-2.6$) of 
the case (ii) are shown in Fig. S15 of SupMat. These states 
correspond to the regions of olive color 
(in bottom panels of Fig.~\ref{fig14}) either 
at $\Delta\p\approx(0,\pm\pi)$, $\Delta\p\approx(\pm\pi,0)$ (for top and center panels 
with $E\approx 2.2$) or $\Delta\p\approx \pm 0.9(-\pi,\pi)$ (bottom panels 
with $E\approx 2.6$). Here the points $\Delta\p\approx(0,\pm\pi)$, 
$\Delta\p\approx(\pm\pi,0)$ correspond to regions with a local maximum 
in one direction and finite width in $\Delta\p$-space in the 
orthogonal direction due to the forbidden zone thus leading to a 
quasi-negative mass situation (note that Fig. S16 of SupMat does not apply 
to this case since $\Delta\p\neq 0$). 
The case of bottom panels is special, since there is no local maximum 
in $\Delta\p$-space at $\Delta\p\approx \pm 0.9(-\pi,\pi)$, and 
despite the optical appearance of a very close 
pair in $\Delta\r$-space the actual value of $w_{N/6}(E)\approx 0.3$ 
(see the black square data point at $E\approx 2.6$ in bottom panel of 
Fig.~\ref{fig13}) is quite small such that 
about 70\% of the quantum probability is still uniformly distributed over the 
full lattice. However, 
the other 30\% of probability produce a very strong peaked density around 
$\Delta\r=0$ with large maximum values such that the uniform background is not 
visible in the color plot. 

In all cases, we see that the region of allowed $\Delta\p$ values has a quasi 
1D-structure in $\Delta\p$-space (cigar or dumbbell shape island or 
finite width around the red or olive regions). Since these regions 
correspond (except for the special case of bottom panels of 
Fig. S15 of SupMat) to a global or local maximum of $E_c$ in 
$\Delta\p$-space this explains that, at the energies slightly below 
the maximum value, the density of states is strongly enhanced 
(see blue curves in Fig.~\ref{fig13}). 
In $1D$ the free momentum density of states is singular at a spectral border 
and in quasi-$1D$ with a finite width there should be still a strong 
enhancement. 

We mention that we have also studied a further case similar to (ii) 
but with $n_v=1.25$ 
(instead of $n_v=1$) such that the symmetric value $\p_+=(\pi,\pi)$ is 
exactly at the optimal point found in \cite{htcepjb}. In this case, 
the three main peaks visible in the bottom panel of Fig.~\ref{fig13} 
merge into one single peak at the (modified) top of the spectrum at 
$E\approx 2.2-2.7$ and both green/olive zones (for $n=1$) at $\Delta\p=0$ or 
$\Delta\p\approx(0,\pm\pi)$, $\Delta\p+\approx(\pm\pi,0)$ become red (for 
$n=1.25$) and all three maxima have now the same value. In particular, 
the negative mass point at $\Delta\p=0$ corresponds now to a global maximum
being degenerate with the other maxima at 
$\Delta\p\approx(0,\pm\pi)$, $\Delta\p+\approx(\pm\pi,0)$ 
and $\Delta\p\approx(\pm \pi,\pm\pi)$ and the $\Delta\p$-densities of pair 
states are concentrated at these points. 

These observations provide an additional explanation of the results 
of \cite{htcepjb} with optimal pair formation (in absence of a frozen 
Fermi sea) at $\p_+=(\pm\pi,\pm\pi)$. In Fig. S17, we show again the 
data of Fig.~4 of \cite{htcepjb} in a color plot superimposed with 
the Fermi surface for certain fillings $n=n_v$ by identifying 
$\p=\p_+/2$. The value of $\p_+/2\approx 0.4-0.45 (\pi,\pi)$ 
(i.e. $n_v=0.74$ or $n_v=1$) of the above two cases (i) and (ii) 
correspond to green zones of an enhanced pair formation probability. 
The optimal point $\p_+/2=0.5(\pi,\pi)$ corresponds to a red data point data 
(however, not very well visible in the figure).

Thus the three main mechanisms of pair formation by Cou\-lomb repulsion
are: narrow or flat band local spectrum structure
as discussed in \cite{prr2020,htcepjb};
negative effective mass for the pair energy
so that a repulsion works as an effective attraction;
restricted area (e.g. cigar shape or dumbbell islands) of 
states accessible for interaction
induced transitions above the frozen Fermi sea. At the same time 
a quasi-1D structure of allowed $\Delta\p$ zones leads also to an enhanced 
density of states for a small energy intervall slightly below typical 
pair energies (peak positions of $w_{N/6}$). 
We also point out that, in the framework of this 
approach, Coulomb repulsion does not lead to a gap with a ground state 
energy below the Fermi surface.

\section{Discussion}
\label{sec8}

In this work,  we apply the Cooper approach \cite{cooper}
to study the formation of coupled pairs of two interacting particles,
holes or electrons, in a tight-binding model of
La-based cuprate superconductors. The one-particle 
band structure of such systems is obtained from advanced numerical
analysis \cite{markiewicz,bansil,fresard} based on modern
methods of quantum chemistry.
We consider three types of interactions
being: attractive Hubbard and d-wave type interactions
and the standard repulsive Coulomb interaction. 
Following the Cooper approach \cite{cooper}, 
the interaction induced transitions are
taking place only over the pair states
where each particle (hole) is outside (inside) a frozen Fermi sea 
in a sector with a conserved fixed total momentum $\p_+$ of a pair 
at relative momentum $\Delta \p$.
Here, we do not discuss possible origins 
of the appearance of an attractive interaction and we 
simply assume that such interactions are given (for the 
cases of Hubbard and d-wave interactions). 


We establish that the energy landscape of the relative particle
motion in a pair strongly depends on the particular value of its 
center of mass $\p_+/2$ corresponding either to a static ($\p_+ = 0$)
or a mobile regime ($\p_+\neq 0$).
For the attractive Hubbard and d-wave interactions, 
we obtain a formation of static Cooper pairs ($\p_+ = 0$)
with a gap $\Delta$ depending  on the interaction
amplitude $U$  and hole (or electron) doping $n_h$
($n = 1 -n_h$). The gap and related $T_c$ dependence on doping
is compared with LSCO experimental 
results (see Fig.~\ref{fig8})
showing a satisfactory agreement.

We find the best agreement with the LSCO experimental data for the case of 
hole excitations at $|U| \approx 1.2 t \approx 0.5\,$eV (Hubbard 
interaction) or at $|U| \approx 2 t \approx 0.8\,$eV (d-wave interaction). 
The position of the optimal hole doping
is approximately located at $n_h \approx 0.24$
being influenced by the close van Hove singularity 
(separatrix Fermi curve) 
of  one-particle density of states. This value is higher 
as compared to the experimental optimal doping
$n_h \approx 0.16$. We attribute such a difference to
missing 3D corrections to the 2D band structure model
we used here for LSCO. 

Another important finding is that the ground state has pronounced 
d-wave features for {\em both} Hubbard and d-wave interactions which 
can be understood by the effective width of the Fermi surface 
in momentum space clearly breaking central symmetry (see Fig.~\ref{fig7}). 
Thus this width is smaller (larger) if the momentum is close to a node (anti-node). 
Therefore, we can conclude that the experimental observation of 
d-wave features does not necessarily imply that the interaction as such should
have d-wave symmetries. In our studies, the d-wave interaction model 
provided somewhat stronger d-wave effects but the latter were also 
clearly present, due to band-structure Fermi surface effects,
for the Hubbard interaction, 
which has only an s-wave symmetry.


For mobile pairs ($\p_+\neq 0$, with $\p_+/2$ being on a typical Fermi 
surface with $n_v=0.74$) the required attractive interaction 
strength $|U|$ to form a pair with a similar gap as at $\p_+=0$ 
is enhanced by a factor $3-4$ at same filling $n$. 
The gap value is minimal at the node region ($p_{+x} \approx p_{+y}$)
and maximal at the antinode region ($p_{+x}$  close to zero and $p_{+y}$ close
to maximum, or inverse). We point out that for mobile
pairs the region of accessible $\Delta\p$ values due to the frozen Fermi
sea has a very complex structure (see e.g. Figs.~\ref{fig5},~\ref{fig10}).
We expect that such mobile pairs can play a role for 
stripe formation in LSCO.

For the case of Coulomb repulsion we do not find gap
and coupled pairs at  the ground state. 
However, we find the formation of mobile Coulomb pairs
at excited energies provided by three different mechanisms
being: narrow or flat band as discussed in \cite{prr2020,htcepjb},
local effective negative mass of relative motion,
restrictions of motion due to island structures related to 
the restriction of interaction induced transitions
imposed by the frozen Fermi sea. 
We note that the quite complicated zones of accessible states 
(in $\Delta\p$-space) due to the 
frozen Fermi sea (see e.g. Figs.~\ref{fig5},~\ref{fig10})
could in principle  favor paring by the Kohn-Luttinger type
mechanism (see \cite{kohn,chubukov,guinea}) with emergence of an 
effective attraction in d-wave or higher-wave sectors due 
to the complexity of the accessible energy landscape.
However, we do not find signatures
of such an effective attraction nor pair formation
at the ground state by Coulomb repulsion in our studies.

We hope that the results
obtained in the framework of the Cooper approach \cite{cooper}
will lead to a better understanding of
unconventional superconductivity in cooper oxides.
\bigskip

\noindent {\bf Acknowledgments}
This work has been partially supported through the grant
NANOX $N^o$ ANR-17-EURE-0009 in the framework of 
the Programme Investissements d'Avenir (project MTDINA).
This work was granted access to the HPC resources of 
CALMIP (Toulouse) under the allocation 2022-P0110.
\bigskip

\noindent {\bf Data Availability Statement} This manuscript has no
associated data or the data will not be deposited. [Author's
comment: There are no external data associated with the
manuscript.]
\bigskip

\appendix
\section{Appendix}
\label{appA}
\subsection{Numerical Cooper pair method}
\label{appA1}

Let us consider the mathematical eigenvalue problem 
of a Hamiltonian matrix of the form~:
\begin{equation}
\label{eq_hamcooperdef}
H_{k,k'}=\delta_{k,k'}\eps_k-\frac{|U|}{N_2}\,g_k\,g_{k'}
\end{equation}
with diagonal unperturbed energies $\eps_k\ge 0$ and an ``interaction'' 
or ``coupling'' matrix of rank one. For the considerations in this 
appendix both $\eps_k (\ge 0)$ and $g_k$ may be rather arbitrary but for 
the physical applications in this work $\eps_k$ represents the excitation 
energy of two particles (holes) of the form: 
\begin{equation}
\label{eq_epsdef}
\eps_k=\pm\left[E_{1p}\left(\frac{\p_+}{2}-\Delta \p\right)
+E_{1p}\left(\frac{\p_+}{2}+\Delta \p\right)-2E_F
\right]
\end{equation}
with $k$ corresponding to $\Delta \p$, ``$+$'' (``$-$'') 
for particle (hole) excitations, $\p_+$ being the conserved 
total momentum of the particle (hole) pair and 
only the values of $k$ (or $\Delta \p$) are allowed that 
such $E_{1p}(\p_+/2\pm\Delta \p)-E_F>0$ for both particles 
(or $<0$ for both holes). The number $N_2$ corresponds to the dimension 
of the full unrestricted sector of $\p_+$ with all values of 
$\Delta \p$. For later use we note the dimension of the restricted 
sector (with allowed values of $\Delta\p$) as $N_2'$ (being 
a given fraction of $N_2$). 

The case $g_k=1$ corresponds to an attractive Hubbard interaction 
of interaction strength $U$ 
and $g_k=g_{\Delta \p}=[\cos(\Delta p_x)-\cos(\Delta p_y)]/2$ 
corresponds to an effective d-wave pairing attractive interaction 
used in typical mean field approaches (see for example \cite{bansil}). 
For $g_k=1$, $\p_+=0$ and a simpler energy band 
this model was already considered by Cooper in 1956 \cite{cooper}. 
His technical trick to compute the ground state energy (or gap) 
can be generalized to the more general model here and also be exploited 
for an efficient numerical method. 

Let $\psi_k$ be the $k$-component of an eigenvector 
of (\ref{eq_hamcooperdef}) of energy $E$. 
It satisfies obviously the equation~:
\begin{equation}
\label{eq_psicooper1}
(E-\eps_k)\psi_k=-\frac{|U|}{N_2}\,g_k\,S\quad,\quad
S=\sum_{k'} g_{k'}\psi_{k'}\ .
\end{equation}
There are two possibilities: either $S=0$ or $S\neq 0$. 
The case $S=0$ is possible if certain $\eps_k$ values are degenerate, 
e.g. due to symmetries (there are always $1$ to $3$ symmetries in 
our applications for the HTC model, depending on the value 
of $\p_+$, see \cite{htcepjb} for details), and corresponds 
to anti-symmetric wave functions with respect to those symmetries.
Also if $g_k=0$ for certain $k$-values, we may have $S=0$. 
For $S=0$, we have obviously $E=\eps_k$ for some $k$ (with degenerate 
$\eps_k$ or $g_k$=0) and $\psi_{k'}\neq 0$ (or $=0$) if $\eps_{k'}=\eps_{k}$ 
($\eps_{k'}\neq\eps_{k}$ respectively) 
and such states are not affected by the interaction.
For $S\neq 0$ (corresponding to totally symmetric states with 
respect to symmetries), we can insert 
\begin{equation}
\label{eq_psicooper2}
\psi_k=-\frac{|U|}{N_2}\,\frac{g_k\,S}{E-\eps_k}
\end{equation}
into the sum of $S$ and thus obtain an implicit equation 
for the energy $E$:
\begin{equation}
\label{eq_Eimplicit1}
1=-\frac{|U|}{N_2}\sum_k \frac{g_k^2}{E-\eps_k}\ .
\end{equation}
Due to the attractive interaction there is always exactly 
one (ground state) solution 
$E=E_{\rm min}$ with $E_{\rm min}<\eps_{\rm min}$ 
where $\eps_{\rm min}$ is the minimal value of $\eps_k$ (with $g_k\neq 0$ !). 

The implicit equation (\ref{eq_Eimplicit1}) allows for an efficient 
numerical method to compute the first energy $E_{\rm min}$ (and potentially 
also other eigenvalues) by standard algorithms to numerically determine 
function zeros. Once the energy is known, the eigenstate itself is obtained 
from (\ref{eq_psicooper2}) with $S$ being determined from the normalization. 
We have implemented this method and verified that it produces identical 
results to exact full numerical diagonalization (up to numerical precision). 

From (\ref{eq_Eimplicit1}) one can also obtain the limits of $E_{\rm min}$ 
for very small interaction (retaining in the sum 
only the $\eps_{\rm min}$-terms) and very large interaction 
(replacing in the sum all $\eps_k\to \eps_{\rm min}$)~: 
\begin{align}
\label{eq_limitsmallU}
E_{\rm min}&\approx \eps_{\rm min}-\frac{d_{\rm min}|U|}{N_2}
&\mbox{if}\quad
\frac{|U|}{N_2}&\ll \delta_\eps\sim \frac{\eps_{\rm max}}{N_2'}\ ,\\
\label{eq_limitlargeU}
E_{\rm min}&\approx \eps_{\rm min}-\frac{N_2'|U|}{N_2}
&\mbox{if}\quad
\frac{|U|}{N_2}&\gg \eps_{\rm max}\ .
\end{align}
In (\ref{eq_limitsmallU}) $\delta_\eps$ represents the 
typical spacing of $\eps_k$-levels (close to $\eps_{\rm min}$) and 
$d_{\rm min}$ is the degeneracy of the level $\eps_{\rm min}$ for the 
Hubbard case or the sum of $g_k^2$ over the $\eps_{\rm min}$ levels 
for the d-wave interaction case. Furthermore, $N_2'$ is the number 
of $\eps_k$-levels (dimension of the $\p_+$-sector of pair excitations).

Following Cooper \cite{cooper}, and for the simple Hubbard interaction 
case with $g_k=1$, one can also try a continuous limit 
if $N_2'\gg 1$:
\begin{equation}
\label{eq_Eimplicit2}
1=-\frac{|U|}{N_2}\int_{0}^{\eps_{\rm max}}
\rho_2(\eps) \frac{1}{E_{\rm min}-\eps}\,d\eps
\end{equation}
where $\rho_2(\eps)$ is the two-particle (two-hole) excitation density 
of states in the given $\p_+$-sector and normalized by 
$N_2'=\int_0^{\eps_{\rm max}} \rho_2(\eps)\,d\eps$. 
For simplicity, we have also replaced 
$\eps_{\rm min}\to 0$ by applying a uniform shift to all values 
of $\eps_k\to \eps_k-\eps_{\rm min}$ (actually in the 
limit $N_2'\to\infty$ we have anyway $\eps_{\rm min}\to 0$). We also 
assume that the ratio $N_2'/N_2$ remains finite in the limit $N_2'\to\infty$ 
(constant fraction of allowed states in the given $\p_+$-sector; 
see non-white zones in Figs.~\ref{fig2}, \ref{fig5}
and \ref{fig10}).

For the case of a constant density of the states 
$\rho_2(\eps)=N_2'/\eps_{\rm max}$ one obtains from (\ref{eq_Eimplicit2}) 
the expression 
\begin{equation}
\label{eq_Eminresult1}
E_{\rm min}=-\eps_{\rm max}\,\left[\exp\left(
\frac{\eps_{\rm max} N_2}{|U|N_2'}\right)-1\right]^{-1}\ .
\end{equation}
which is very similar to the well known result
of Cooper \cite{cooper} (only with different notations/parameters). 
One can note that in the limit of very strong interaction this expression 
reproduces (\ref{eq_limitlargeU}) plus a constant correction being 
``$+\eps_{\rm max}/2$'' (reduction of $|E_{\rm min}|$) 
which has to be added to (\ref{eq_limitlargeU}). 
On the other hand, for finite $N_2'$, (\ref{eq_Eminresult1}) is not valid 
in the regime where the very small interaction 
limit (\ref{eq_limitsmallU}) applies.

However, for the HTC-lattice, at filling factors close to the separatrix 
point, e.g. $n=0.74$, and for $\p_+=0$ the density of states is 
strongly enhanced for small energies due to the effect of the close van Hove 
singularity (separatrix) as can be seen in the right panel of Fig.~S1 of SupMat. 
To model this behavior we try the fit-ansatz~:
\begin{equation}
\label{eq_dosfit1}
\rho_2(\eps)=\frac{N_2'}{\eps_{\rm max}\log(1+\alpha)}\,
\frac{\alpha}{1+\alpha(\eps/\eps_{\rm max})}
\end{equation}
where $\alpha$ is a fit-parameter reproducing the constant DOS 
if $\alpha=0$ or providing a strongly enhanced DOS close to small energies 
if $\alpha\gg 1$ and a power law decay $\rho_2(\eps)\sim 1/\eps$ for 
larger energies. 
(Also negative values of $\alpha$ are potentially possible.) 
This form does not correspond exactly to the van Hove singularity but 
it is convenient for the subsequent analytical evaluation of 
(\ref{eq_Eimplicit2}) and in any case, we want to model the case 
close but still different from the van Hove singularity where the DOS 
at $\eps=0$ is still finite. The right panel of Fig.~S4 of SupMat 
shows that this ansatz produces an integrated DOS which fits very well 
the exact integrated DOS at $n=0.74$, particles for the sector 
$\p_+=0$. 
For $n=0.3$ the fit is of less quality but still provides an improvement. 

Using (\ref{eq_Eimplicit2}) and (\ref{eq_dosfit1}), we obtain:
\begin{equation}
\label{eq_dosfit2}
E_{\rm min}=-\eps_{\rm max} f^{-1}\left(
\frac{\eps_{\rm max} N_2}{|U|N_2'}\right)
\end{equation}
where $f^{-1}(\ldots)$ is the inverse function of:
\begin{equation}
\label{eq_ffunc}
f(x)=\frac{\alpha}{1-x\alpha}
\left(\frac{\log(x^{-1}+1)}{\log(1+\alpha)}-1\right)\ .
\end{equation}
(Here $x$ represents the ratio $-E_{\rm min}/\eps_{\rm max}$). 
In the limit $\alpha\to 0$ we recover from (\ref{eq_dosfit2}) 
the original Cooper type result (\ref{eq_Eminresult1}). 

The result (\ref{eq_dosfit2}) 
is shown as blue curves in Fig.~\ref{fig_coopertheory} and for 
$n=0.74$, with the fit value $\alpha=6.589$, the blue curve coincides very 
well with the numerical data points except for a very small shift while the 
green curve based on the assumption of a constant DOS (i.e. $\alpha=0$) 
provides much smaller gap values. 
For $n=0.3$ the situation is different. Here for modest values $|U|$ 
the green curve fits better the numerical data points. This is because in 
this case the uniform DOS (or linear integrated DOS) fits better the 
initial (integrated) DOS at small energies 
as can be seen in the left panel of 
Fig.~S4 of SupMat where the green line is closer to the 
red data points for $\eps<0.15\,\eps_{\rm max}$ than the blue curve 
corresponding to the ansatz (\ref{eq_dosfit1}). 
However, for larger values of $|U|\approx 8$ 
(not visible in Fig.~\ref{fig_coopertheory}) the blue curve is closer 
to the numerical data points since here the full range of energies 
$\eps\in[0,\eps_{\rm max}]$ is important. 

It is also possible to simplify (\ref{eq_dosfit2}) in the limit 
of very strong interaction, corresponding to $x\gg 1$ in 
(\ref{eq_ffunc}), which gives:
\begin{equation}
\label{eq_dosfit3}
E_{\rm min}=-\left[\frac{|U|N_2'}{N_2}-A_\alpha\eps_{\rm max}\right]
\ ,\ A_\alpha=\frac{1}{\log(1+\alpha)}-\frac{1}{\alpha}
\end{equation}
which is in agreement with the limit behavior of (\ref{eq_Eminresult1})
since $\lim_{\alpha\to 0} A_\alpha=1/2$. For larger values of $\alpha$ 
the coefficient $A_\alpha$ decreases 
with respect to this value, e.g $A_\alpha=0.3416$ for $\alpha=6.589$.
Even though, mathematically, the constant term with $A_\alpha$ provides 
``only a small'' correction to the first term $\sim |U|$, the fact that this 
coefficient decreases from $0.5$ (at $\alpha=0$) to $0.3416$ 
(at $\alpha=6.589$)
has a considerable impact on the quite significant 
difference between the blue and green curves in 
Fig.~\ref{fig_coopertheory} also for intermediate interaction values. 
For very large values of $|U|$ these 
curves are actually parallel with a constant shift due to different 
values of this coefficient. Furthermore, the third term in 
the large $|U|$-expansion of $E_{\rm min}$ would only provide 
an additional correction of the form $\sim |U|^{-1}$ in (\ref{eq_dosfit3}).

\subsection{Local $g_k$-density of states}
\label{appA2}

The density of states $\rho(E)$ for both lattices has a logarithmic 
van Hove singularity visible in Fig.~S1 of SupMat which is 
due to the vanishing value of $\bnabla E_{1p}(\k_s)=0$ 
at the separatrix points $\k_s=(0,\pm\pi)$ or $\k_s=(\pm\pi,0)$. 
Classically $\rho(E)\,dE$ can be obtained from the (relative) 
area in $\k$-space 
between the two Fermi curves at energies $E$ and $E+dE$. As can 
be seen in Fig.~\ref{fig1} this area is significantly enhanced 
in the region close to a separatrix point. To see this point 
more clearly, it is interesting to consider the angle resolved 
area between Fermi curves at energies $E$ and $E+dE$ and 
also angles $\varphi$ and $\varphi+d\varphi$ where $\varphi$ is 
the phase angle of the momentum vector 
$\k=k_E(\varphi)\e(\varphi)$ where 
$\e(\varphi)=(\cos\varphi,\,\sin\varphi)$ and $k_E(\varphi)$ is determined 
such that at given energy $E$ and angle $\varphi$ we have 
$E=E_{1p}[k_E\,\e(\varphi)]$.
This area (divided over $(2\pi)^2 dE\,d\varphi$) defines the 
local angle-density of states $\rho_\varphi(\varphi,E)$ which 
can be formally computed from the integral~:
\begin{eqnarray}
\label{eq_phidensdef}
\rho_{\varphi}(\varphi,E)&=&\frac{1}{\pi^2} \int_0^\pi dk_x
\int_0^\pi dk_x \delta[E-E_{1p}(\k)]\\
\nonumber
&&\times\delta[\varphi-\arctan(k_y/k_x)]\\
\label{eq_phidensres}
&=&\frac{k_E(\varphi)}{\pi^2
|\e(\varphi)\cdot \bnabla E_{1p}[k_E(\varphi)\e(\varphi)]|}\,
\end{eqnarray}
In (\ref{eq_phidensdef}) we limit ourselves to the first quadrant 
with $0\le \varphi\le\pi/2$ such that the normalization prefactor 
is $1/\pi^2$. The expression (\ref{eq_phidensres}) is obtained by 
computing the integral in polar coordinates for $\k$ and it is 
valid for angles $\varphi$ such that 
the equation $E=E_{1p}[k_E\,\e(\varphi)]$ has a solution for $k_E$. 
If this equation does not have a solution, we simply have 
$\rho_{\varphi}(\varphi,E)=0$. For example, for energies above 
the separatrix energy $E_s=E_{1p}(0,\pi)$ the local angle-density 
of states is limited to values $\varphi\le\varphi_{\rm max}<\pi/2$. 
Close to $\varphi_{\rm max}$ and for energies close to $E_s$ 
this density is not singular but has a 
strong peak value $\sim 1/(\pi/2-\varphi_{\rm max})^2$
(the exponent $2$ is due to a combination of small gradient and 
small scalar product in the denominator since the gradient 
and $\e(\varphi)$ are nearly orthogonal). For 
energies below $E_s$ there is a minimal angle $\varphi_{\rm min}>0$ 
with $\varphi\ge\varphi_{\rm min}$ and a density peak 
$\sim 1/\varphi_{\rm min}^2$. 

In this work, we prefer however to use the quantity 
$g_k=(\cos k_x-\cos k_y )/2$ instead of $\varphi$ 
with values $g_k\approx 1$ (or $-1$) if 
$\varphi\approx \pi/2$ ($\varphi\approx 0$) and $g_k=0$ if $\varphi=\pi/4$. 
This quantity allows also to characterize a position on a Fermi 
surface at given energy (in the first quadrant). 
Its local $g_k$-density of states 
is obtained by a similar expression as (\ref{eq_phidensdef}):
\begin{equation}
\label{eq_gkdensdef}
\rho_g(g,E)=\frac{1}{\pi^2}\int_0^\pi dk_x
\int_0^\pi dk_x \delta[E-E_{1p}(\k)]\,
\delta(g-g_k)
\end{equation}
and it satisfies the relation:
\begin{equation}
\label{eq_gkdensrel1}
\rho_g(g_k,E)=\rho_{\varphi}(\varphi,E)
\left(\frac{dg_k}{d\varphi}\right)^{-1}\ .
\end{equation}
We have used this relation together with (\ref{eq_phidensres}) 
(and a numerical evaluation of $dg_k/d\varphi$ by finite differences 
for a sufficiently dense set of data points) to compute 
numerically the $g_k$-density with results shown in Fig.~S5 of SupMat
and also in Fig.~\ref{fig4}. 

For energies $E$ close to $E_s$ and values $1-g_k\ll 1$, we can apply 
to $E_{1p}(\k)$ and $g_k$ a quadratic expansion for 
$\k$ close to the separatrix point $\k_s=(0,\pi)$ resulting in~:
\begin{equation}
\label{eq_Equad}
E_{1p}(\k)=E_{1p}(\k_s)+\frac12\left[a_x\,k_x^2-a_y(\pi-k_y)^2\right]
\end{equation}
with $a_x=a_y=2$ ($a_x=2.084$, $a_y=0.452$) for the NN-lattice 
(HTC-lattice) and
\begin{equation}
\label{eq_gkquad}
g_k=1-\frac14\left[k_x^2+(\pi-k_y)^2\right].
\end{equation}
Inserting (\ref{eq_Equad}) and (\ref{eq_gkquad}) in 
(\ref{eq_gkdensdef}) one obtains the following analytical result:
\begin{equation}
\label{eq_gkdensresult}
\rho_g(g_k,E)=\frac{C_1}{
\sqrt{(g_{\rm max}-g_k)(g_{\rm max}-g_k+C_2(1-g_{\rm max}))}}
\end{equation}
with constants $C_1=1/(2\pi^2\sqrt{a_x a_y})$ 
and $C_2=1+a_x/a_y$ ($C_2=1+a_y/a_x$) if $\Delta E=E-E_s\ge 0$ 
($\Delta E=E-E_s\le 0$), i.e. if the Fermi curve is above (below) 
the separatrix curve. Furthermore, $g_{\rm max}$ 
is the maximal possible value of $g$ given by~:
$g_{\rm max}=1-\Delta E/(2a_x)$ 
[$g_{\rm max}=1+\Delta E/(2a_y)=1-|\Delta E|(2a_y)$]
if $\Delta E\ge 0$ ($\Delta E\le 0$). 
We also note that (\ref{eq_gkdensresult}) is valid for $g_k>0$ 
because we have chosen the expansion around the separatrix point 
$\k_s=(0,\pi)$. 
Using that $\rho_g(g_k,E)=\rho_g(-g_k,E)$ due to the 
$x-y$ exchange symmetry it is sufficient to replace in 
(\ref{eq_gkdensresult}) $g_k\to|g_k|$ to obtain a more general expression 
for other separatrix points where $g_k$ is close to $-1$. 

For the separatrix case $\Delta E=0$ with $g_{\rm max}=1$ the expression 
(\ref{eq_gkdensresult}) simplifies to the simple power law 
\begin{equation}
\label{eq_gkdensseparatrix}
\rho_g(g_k,E)=\frac{C_1}{1-g_k}\ .
\end{equation}
This power law is also valid for the general case close to but outside the 
separatrix curve in the range of $g_k$ values sufficiently 
far away from the singularity at $g_{\rm max}$, i.e.: 
$1-g_{\rm max}\ll g_{\rm max}-g_k\ll 1$.
For values very close to the singularity 
$g_{\rm max}-g_k\ll 1-g_{\rm max}$ the expression (\ref{eq_gkdensresult}) 
becomes a power law with exponent $-1/2$. 
All these points are very nicely confirmed in Fig.~S5 of SupMat. 

Actually, the analytical expression (\ref{eq_gkdensresult})
based on the separatrix approximation is highly 
accurate (provided one uses for $g_{\rm max}$ the precise values 
for a given energy and not the approximate linear expressions 
in $|\Delta E|$ given above) 
even for filling factors not very close to the separatrix values 
and even in the interval $0\le g_k\le 0.8-0.9$ it is still 
rather close to the precise distribution obtained numerically.

Furthermore, from (\ref{eq_gkdensdef}) we immediately see that 
\begin{equation}
\label{eq_intgkdens}
\int_{-g_{\rm max}}^{g_{\rm max}} \rho_g(g_k,E)\,dE=
2\int_{0}^{g_{\rm max}} \rho_g(g_k,E)\,dE
=\rho(E)
\end{equation}
where $\rho(E)$ is the total density of states given by an 
expression similar to (\ref{eq_gkdensdef}) but without 
the delta-function factor $\delta(g-g_k)$. 
For $1-g_{\rm max}\ll 1$, we find that this integral 
behaves as $\log(1-g_{\rm max})\sim\log|\Delta E|=\log|E-E_s|$ 
(simply using (\ref{eq_gkdensseparatrix}) with a cut-off at 
$|g|<g_{\rm max}$)
thus confirming the logarithmic nature of the van Hove singularity 
in the density of states.



\begin{thebibliography}{00}

  \bibitem{hts1986}%
  K.A.~M\"uller, and J.G.~Bednorz,
  Z. Phys. B: Condens. Matter {\bf 64}, 189 (1986).

  \bibitem{dagotto}%
  E.~Dagotto, Rev. Mod. Phys.  {\bf 66}, 763 (1994).

  \bibitem{kivelson}%
  B.~Keimer, S.A.~Kivelson, M.R.~Norman, S.~Uchida, 
   and Z.~Zaanen, Nature {\bf 518}, 179 (2015).

  \bibitem{proust}%
  C.~Proust, and L.~Taillefer,
  Annu. Rev.  Condens. Matter Phys.  {\bf 10}, 409 (2019).

  \bibitem{anderson}%
  P.W.~Anderson, Science {\bf 235}, 1196 (1987).

  \bibitem{emery1}%
  V.J.~Emery, 
  Phys. Rev. Lett.  {\bf 58}, 2794 (1987).

  \bibitem{emery2}%
  V.J.~Emery, and G.~Reiter,
  Phys. Rev. B  {\bf 38}, 4547 (1988).

  \bibitem{varma}%
  C.M.~Varma, 
  Solid State Commun.  {\bf 62}, 681 (1987).

  \bibitem{loktev}%
  Y.B.~Gaididei, and V.M.~Loktev,
  Phys. Status Solidi  {\bf 147}, 307 (1988).

  \bibitem{markiewicz}%
  R.S.~Markiewicz, S.~Sahrakorpi,
  M.~Lindroos, H.~Lin, and A.~Bansil,
  Phys. Rev. B  {\bf 72}, 054519 (2005).

  \bibitem{bansil}%
    T.~Das,  R.S.~Markiewicz, and  A.~Bansil,
    Adv. Phys. {\bf 63}, 151 (2014)

    
  \bibitem{fresard}%
  R.~Photopoulos, and R.~Fresard,
  Ann. Phys. (Berlin) 1900177 (2019).

 \bibitem{cooper}%
  L.~Cooper, Phys. Rev.  {\bf 104}, 1189 (1956).

  
  \bibitem{prr2020}%
  K.M.~Frahm, and D.L.~Shepelyansky,
  Phys. Rev. Research  {\bf 2}, 023354 (2020).

  \bibitem{htcepjb} K.M.Frahm, and D.L.Shepelyansky,
    Eur. Phys. J. B {\bf 94}, 29 (2021).

  \bibitem{vishik1} I.M.Vishik, W.S.Lee, R.-H.He, M.Hashimoto,
    Z.Hussain, T.P.Devereaux, and Z.-X.Shen,
    New J. Phys. {\bf 12}, 105008 (2010),

  \bibitem{vishik2} M.Hashimoto, I.M.Vishik, R.-H.He, T.P.Devereaux,
    and Z.-H.Shen, Nature Phys. {\bf 10}, 483 (2014).
  
  \bibitem{roughbil}%
  K.M.~Frahm, and D.L.~Shepelyansky,
  Phys. Rev. Lett.  {\bf 79}, 1833 (1997).

  \bibitem{stripe1} R.Comin et al., Science {\bf 343}, 390 (2014).

  \bibitem{stripe2} K. von Arx et al., arXiv:2206.06695[cond-mat.supr-con] (2022).

  \bibitem{tinkham} M.~Tinkham, (1996). {\em Introduction to Superconductivity}, 
    Dover Publications. p. 63. 

  \bibitem{ourwebpage}%
  K.M.~Frahm, and D.L.~Shepelyansky,
   Available upon request:  \url{https://www.quantware.ups-tlse.fr/QWLIB/electronpairsforhtc/}; 
   Accessed September (2022)

  \bibitem{kohn} W.~Kohn, and J.M.~Luttinger,
    Phys. Rev. Lett. {\bf 15}, 524 (1976).

  \bibitem{chubukov} A.V.~Chubukov, Phys. Rev. B
    {\bf 48}, 1097 (1993).

  \bibitem{guinea} F.~Guinea, and B.~Uchoa,
    Phys. Rev. B {\bf 86}, 134521 (2012).


\end{thebibliography}

\newpage

\setcounter{figure}{0} \renewcommand{\thefigure}{S\arabic{figure}} 
\setcounter{equation}{0} \renewcommand{\theequation}{S\arabic{equation}} 
\setcounter{page}{1}

\noindent{{\bf Supplementary Material for\\
\vskip 0.2cm
\noindent{Cooper approach to pair formation 
in a tight-binding model of La-based cuprate superconductors}}\\
\noindent by
K.~M.~Frahm and D.~L.~Shepelyansky\\
\noindent Laboratoire de Physique Th\'eorique, 
Universit\'e de Toulouse, CNRS, UPS, 31062 Toulouse, France}

\vspace{0.5cm}
\noindent Here, we present additional Figures for the main part of the article.
\vspace{3cm}

\setcounter{figure}{0} \renewcommand{\thefigure}{S\arabic{figure}}

\begin{figure}[h]
\begin{center}
\includegraphics[width=0.95\columnwidth]{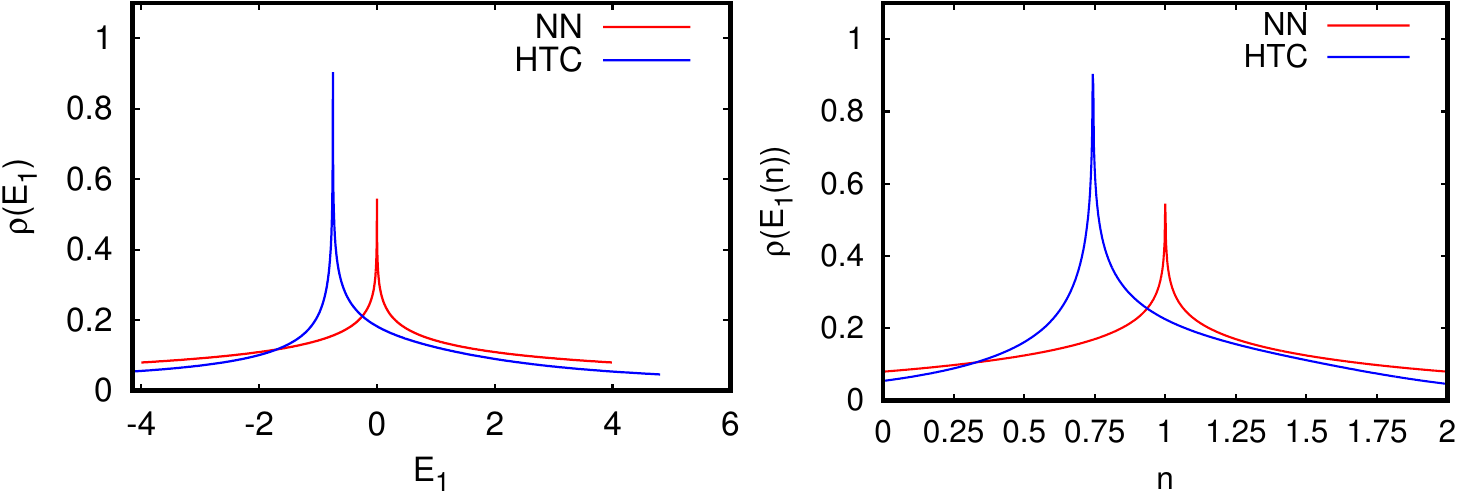}%
\end{center}
\caption{\label{figS1}
\label{fig_dos}
One-particle energy density of states $\rho(E_1)$ for both model types 
shown as a function of the one-particle energy $E_1$ (left panel) and filling 
factor $n$ (right panel). The van Hove singularities (or separatrix values) 
correspond to $E_1=-0.748$ ($E_1=0$) and $n=0.743465958$ ($n=1$)
for the HTC model (NN model).
Note that the right panel shows the identical quantity $\rho(E_1)$ 
as the left panel but as a function of $n$ and
without application of any Jacobian factor. 
In particular, this does not represent the density 
in the variable $n$ (obtained by taking into account the Jacobian factor) 
which has actually the simple uniform value $0.5$ for $0\le n\le 2$.}
\end{figure}

\begin{figure}
\begin{center}
\includegraphics[width=0.95\columnwidth]{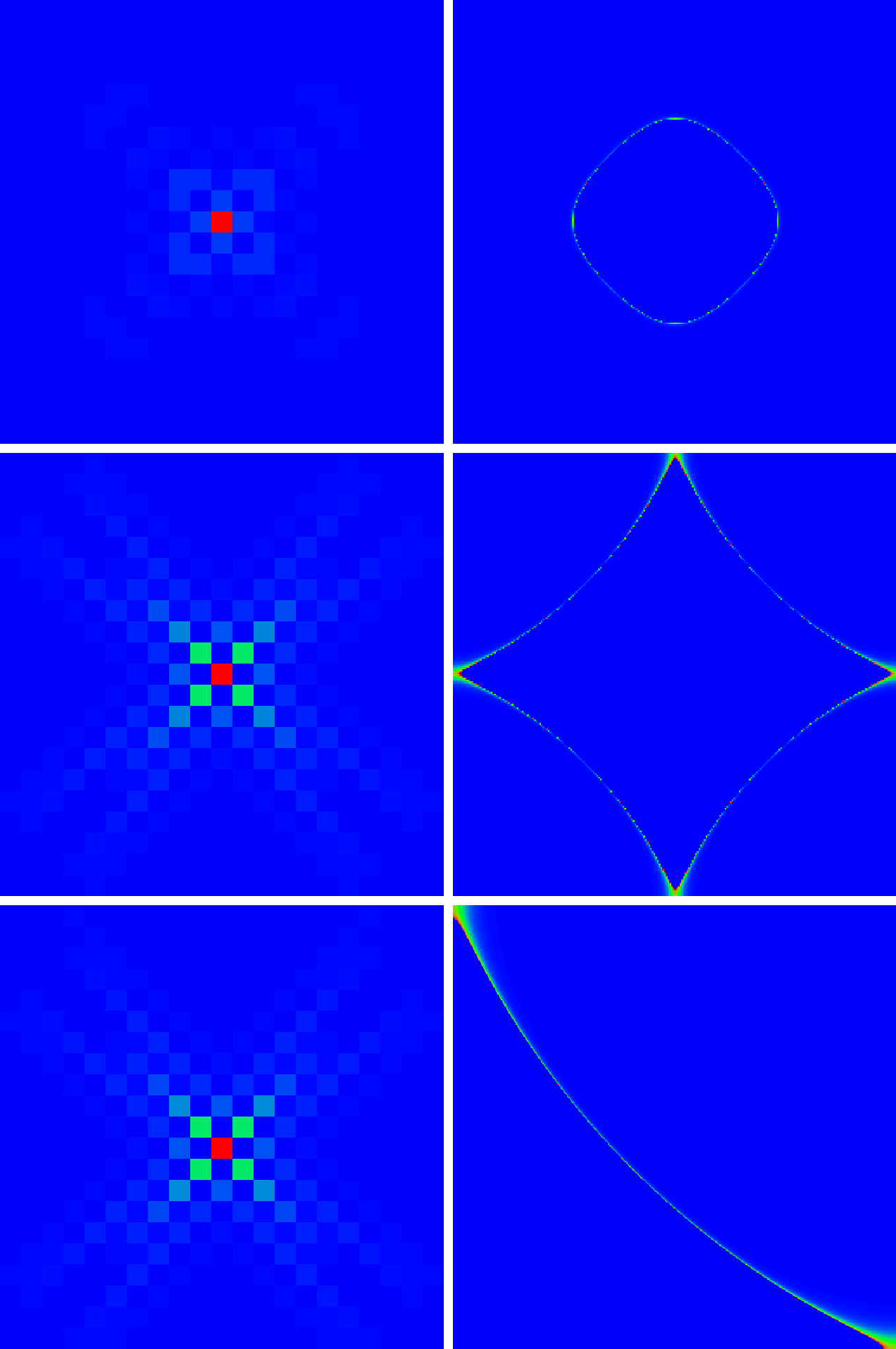}%
\end{center}
\caption{\label{figS2}
\label{fig_states_P0}
Ground state density plots for the Hubbard interaction, system size $N=256$ 
(top and center panels) and $N=1024$ (bottom panels), 
sector $\p_+=0$,  
and filling factor $n=0.3$, $U=-2.5$ (top) 
and $n=0.74$, $U=-1$ (center, bottom). Left panels show the ground state 
in $\Delta \r$-representation in a zoomed region with 
$-10\le \Delta x,\Delta y\le 10$ 
(color values outside the zoomed regions are blue) 
and right panels show the state 
in $\Delta \p$-representation (with $-\pi\le \Delta p_{x,y}<\pi$
for $N=256$ or zoomed top-right square $0\le \Delta p_{x,y}<\pi$ for $N=1024$).
The two particle ground state energies $E_{\rm min}$ 
(of the sector Hamiltonian) 
in units of the basic hopping matrix element are $-0.02656$ ($-0.01940$, 
$-0.01948$) 
for $n=0.3$, $N=256$, ($n=0.74$, $N=256$ or $n=0.74$, $N=1024$). 
The state for $N=1024$ in bottom panels is also used for the 
$g_k$-distribution shown in left panels of 
Fig.~4 (for $U=-1$).}
\end{figure}

\begin{figure}
\begin{center}
\includegraphics[width=0.95\columnwidth]{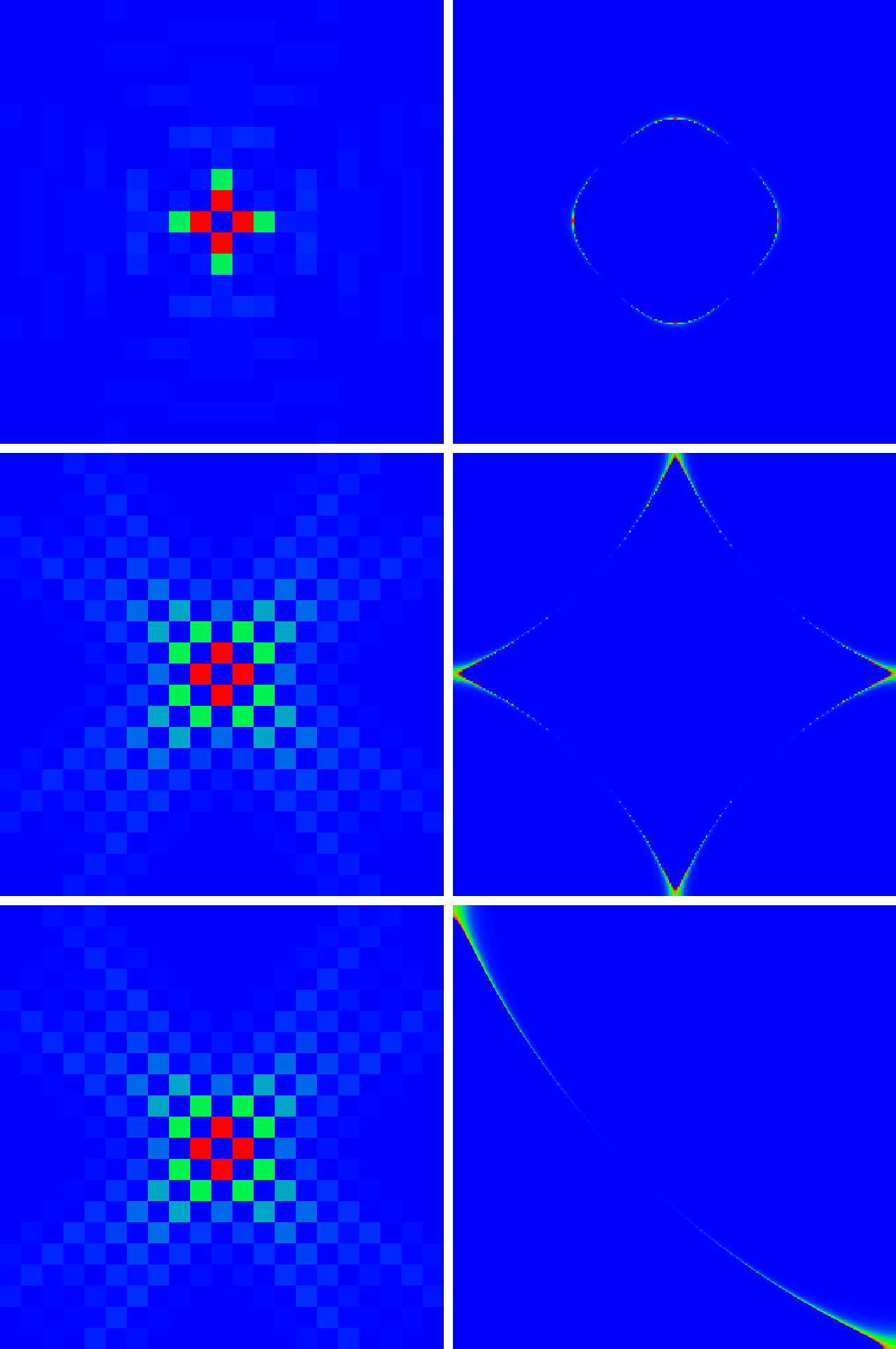}%
\end{center}
\caption{\label{figS3}
\label{fig_states_P0_dwave}
Ground state density plots for the d-wave interaction, system size $N=256$ 
(top and center panels) and $N=1024$ (bottom panels), 
sector $\p_+=0$,  
and filling factor $n=0.3$, $U=-5$ (top) 
and $n=0.74$, $U=-1.5$ (center, bottom). Left panels show the ground state 
in $\Delta \r$-representation in a zoomed region with 
$-10\le \Delta x,\Delta y\le 10$ 
(color values outside the zoomed regions are blue) 
and right panels show the state 
in $\Delta \p$-representation (with $-\pi\le \Delta p_{x,y}<\pi$
for $N=256$ or zoomed top-right square $0\le \Delta p_{x,y}<\pi$ for $N=1024$).
The two particle ground state energies $E_{\rm min}$ 
in units of the basic hopping matrix element are $-0.03140$ ($-0.01682$,
$-0.01681$)
for $n=0.3$, $N=256$, ($n=0.74$, $N=256$ or $n=0.74$, $N=1024$). 
The state for $N=1024$ in bottom panels is also used for the 
$g_k$-distribution shown in right panels of 
Fig.~4 (for $U=-1.5$).}
\end{figure}

\begin{figure}
\begin{center}
\includegraphics[width=0.95\columnwidth]{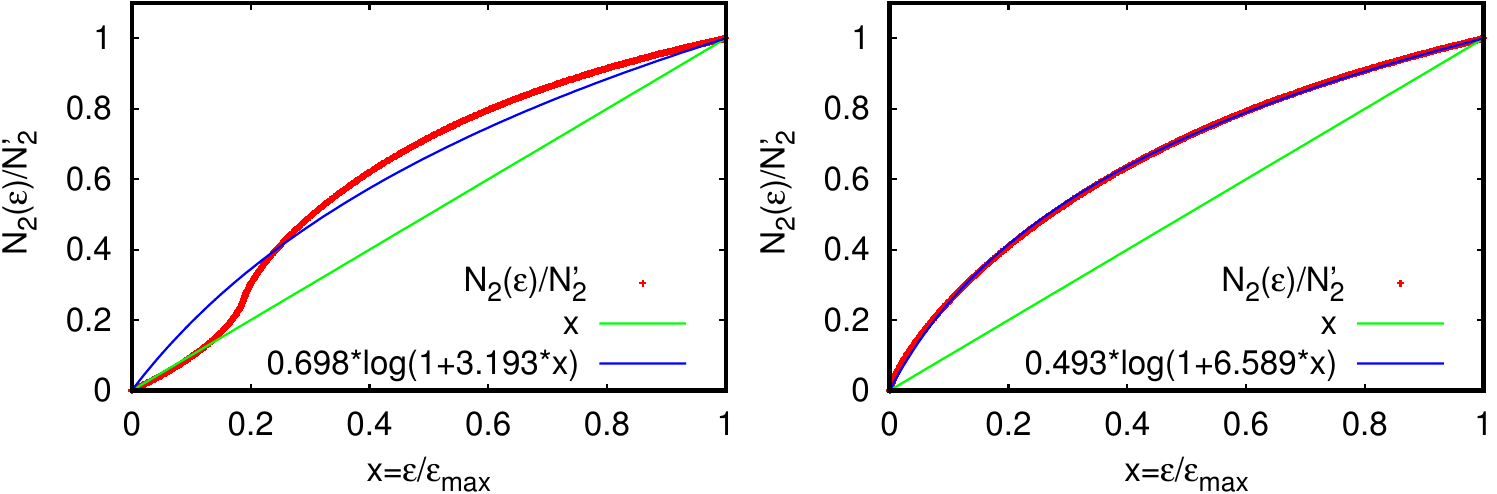}%
\end{center}
\caption{\label{figS4}
\label{fig_doslogfit}
Left (right) panel shows the rescaled integrated two-particle 
sector-density of states $N_2(\eps)/N_2'$ 
(red data points) for the sector $\p_+=0$, 
for $N=256$, $n=0.3$ ($n=0.74$) versus $\eps/\eps_{\rm max}$.
Here $N_2'$ represents the number of accessible levels in the 
given sector. The green line shows the linear behavior assuming 
a constant density of states and the blue line shows the 
fit $N_2(\eps)/N_2'=\log[1+\alpha(\eps/\eps_{\rm max})]/\log(1+\alpha)$ 
with $\alpha=3.193\pm 0.077$ ($\alpha=6.589\pm 0.017$).
These fits are used in Fig. \ref{fig3}.
}
\end{figure}

\begin{figure}
\begin{center}
\includegraphics[width=0.95\columnwidth]{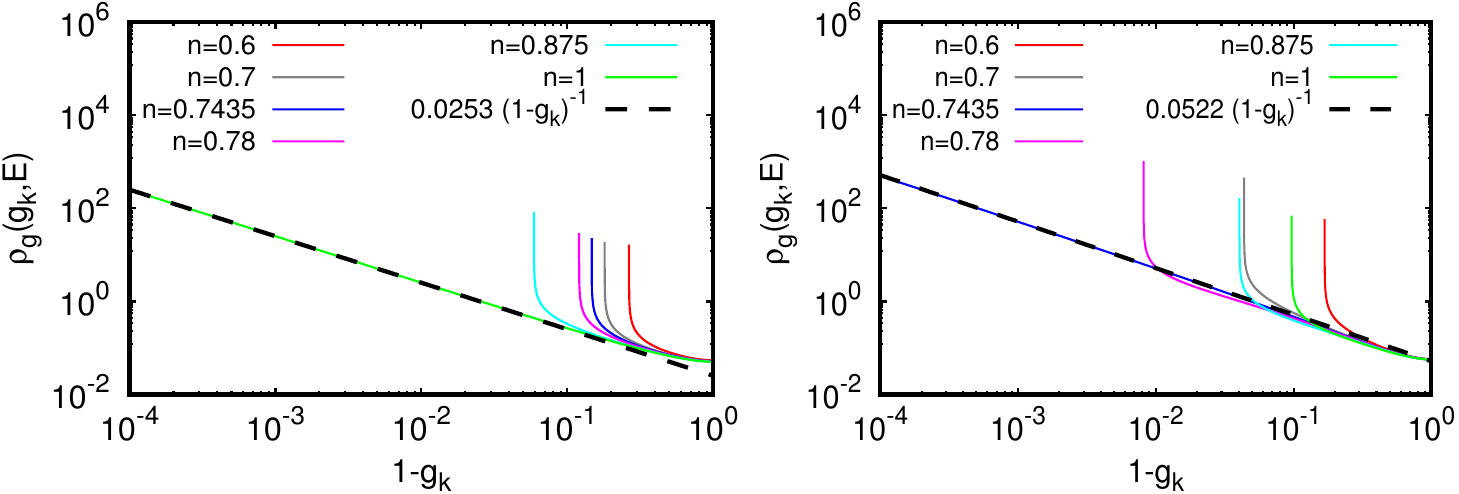}%
\end{center}
\caption{\label{figS5}
\label{fig_gkdens}
Local $g_k$-density of states $\rho_g(g_k,E)$ for 
the quantity $g_k=(\cos(k_x)-\cos(k_y))/2$ for different 
values of energies $E$/filling factors $n$ (see text 
for the definition). 
The dashed black line shows the analytical result 
$\rho_g(g_k,E)=C_1(1-g_k)^{-1}$ which is obtained for the exact 
separatrix case and if $1-g_k\ll 1$. 
Here the constant is given by $C_1=1/(2\pi^2\sqrt{a_x a_y})$
where $a_x=a_y=2$ for the NN model ($a_x=2.084$, $a_y=0.452$ 
for the HTC model) such that $C_1 = 0.0253303$ ($C_1 = 0.052198$). 
The values of $a_x$ and $a_y$ are obtained from the expansion 
$E_1(\k)=E_1(0,\pi)+\frac{1}{2}[a_x k_x^2 -a_y(\pi -k_y)^2]$
for $(k_x,k_y)$ being close to the separatrix point $(0,\pi)$ 
(see text for more details).
The constant $C_1$ for the HTC model is roughly twice as large 
than the constant $C_1$ for the NN model showing that for the HTC model 
$g_k$ values close to unity are more likely.
For the NN model (left panel) the green curve for the separatrix value $n=1$ 
extends numerically up to $(1-g_k)\approx 10^{-10}$. 
For the HTC model (right panel) the blue curve for $n=0.7435$
extends numerically to $(1-g_k)\approx 3\times 10^{-6}$; 
the curve for the precise separatrix value $n=0.743465958$ 
(not shown in the figure) 
extends numerically to very small values of $(1-g_k)\approx 10^{-12}$
(if computed properly).
The strong peak values at minimal values of $1-g_k=1-g_{\rm max}$ 
correspond to singularities of the type const.$/\sqrt{g_{\rm max}-g_k}$ 
and in this region the density coincides numerically very well 
with the analytical approximation 
(\ref{eq_gkdensresult}) 
showing a crossover from a power law with exponent $-1/2$ 
(for $g_{\rm max}-g_k\ll 1-g_{\rm max}$) 
to a different power law with exponent $-1$ corresponding to 
the black dashed line (for $1-g_{\rm max}\ll g_{\rm max}-g_k\ll 1$). 
Note that due to the exchange symmetry between $k_x$ and $k_y$
the local $g_k$-density of states is symmetric~: 
$\rho_g(g_k,E)=\rho_g(-g_k,E)$ and therefore this function 
is only shown for positive values of $g_k\ge 0$.}
\end{figure}

\begin{figure}
\begin{center}
\includegraphics[width=0.95\columnwidth]{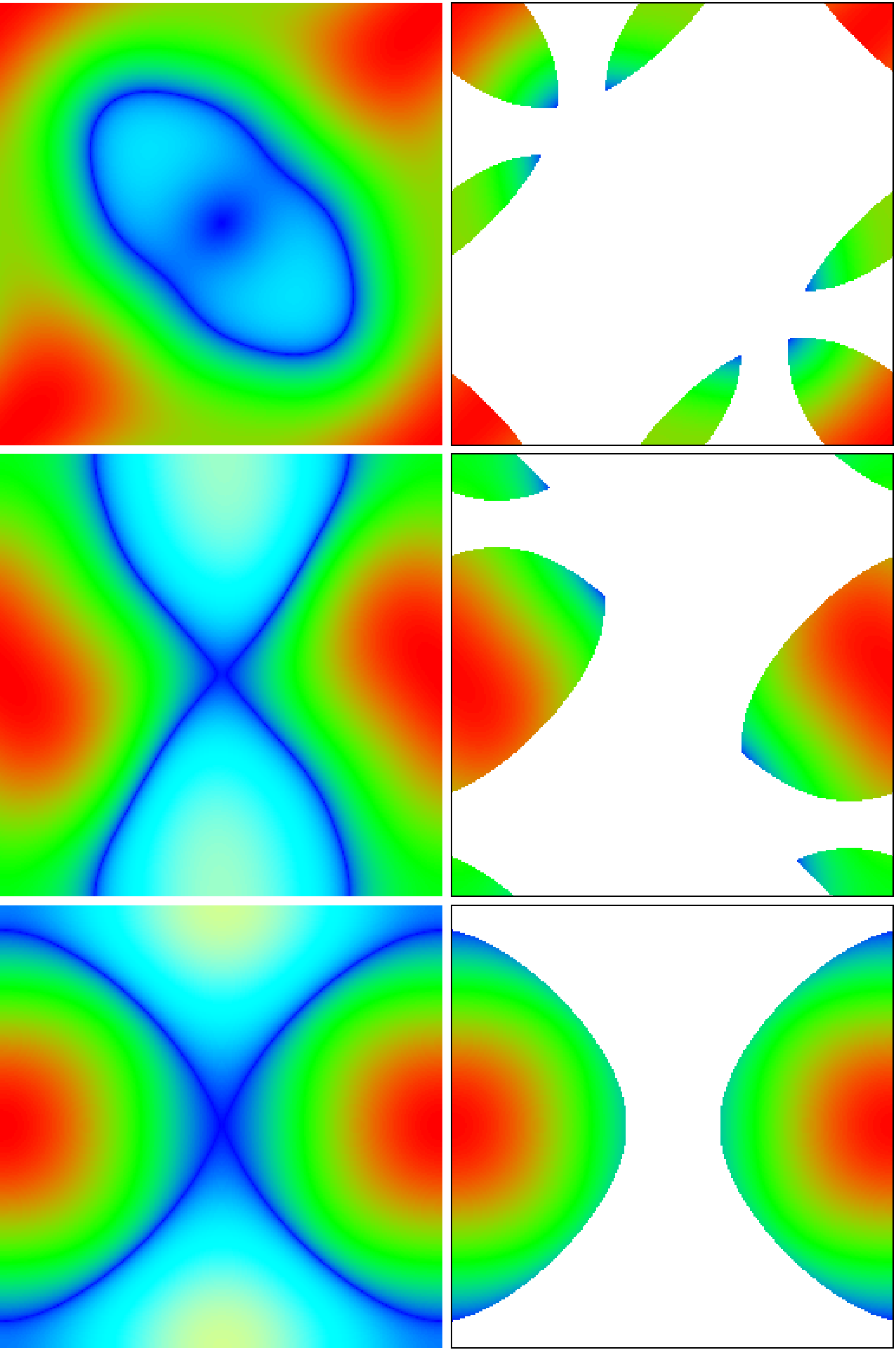}%
\end{center}
\caption{\label{figS6}
\label{fig_energy_P0p84}
Same as Fig.~5 with particle excitations 
but for the filling factor $n=0.84$ and the sectors 
$\p_+=2\pi(106,106)/256$ (top panels), 
$\p_+=2\pi(52,174)/256$ (center panels) and 
$\p_+=2\pi(27,256)/256$ (bottom panels) such that the center of 
mass momentum $\p_+/2$ is very close to the Fermi surface 
of virtual filling factor $n_v=0.84$ with three cases 
of $p_{+x}=p_{+y}$, $p_{+x}\approx p_{+y}/4$ and $p_{+x}$ ($p_{+y}$) 
minimal (maximal). 
}
\end{figure}

\begin{figure}
\begin{center}
\includegraphics[width=0.95\columnwidth]{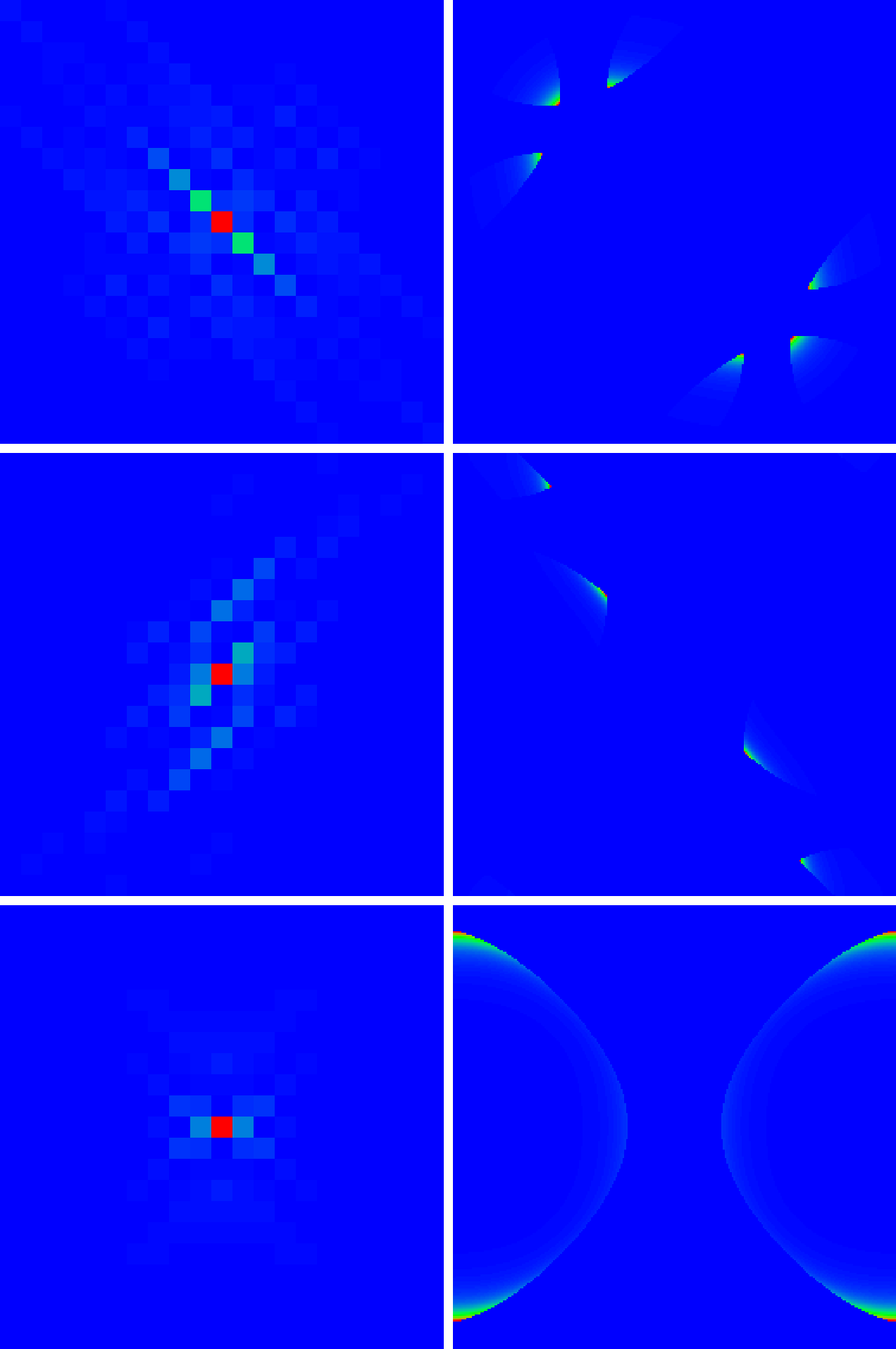}%
\end{center}
\caption{\label{figS7}
\label{fig_states_P0p84}
Ground state density plots for the Hubbard interaction, system size $N=256$, 
particle excitations, 
filling factor $n=0.84$ and three sectors $\p_+\neq 0$ (same 
values as in Fig.~\ref{fig_energy_P0p84}). Top (center, bottom) panels 
correspond to $U=-7$, $\p_+=2\pi(106,106)/256$ 
($U=-8$, $\p_+=2\pi(52,174)/256$; $U=-6$, $\p_+=2\pi(27,256)/256$).
Left panels show the ground state 
in $\Delta \r$-representation in a zoomed region with 
$-10\le \Delta x,\Delta y\le 10$ 
(color values outside the zoomed regions are blue) 
and right panels show the state 
in $\Delta \p$-representation (with $-\pi\le \Delta p_{x,y}<\pi$).
The two particle ground state energies $E_{\rm min}$ 
in units of the basic hopping matrix element are $-0.1337$ ($-0.1095$, 
$-0.2868$) for top (center, bottom) panels.
}
\end{figure}

\begin{figure}
\begin{center}
\includegraphics[width=0.95\columnwidth]{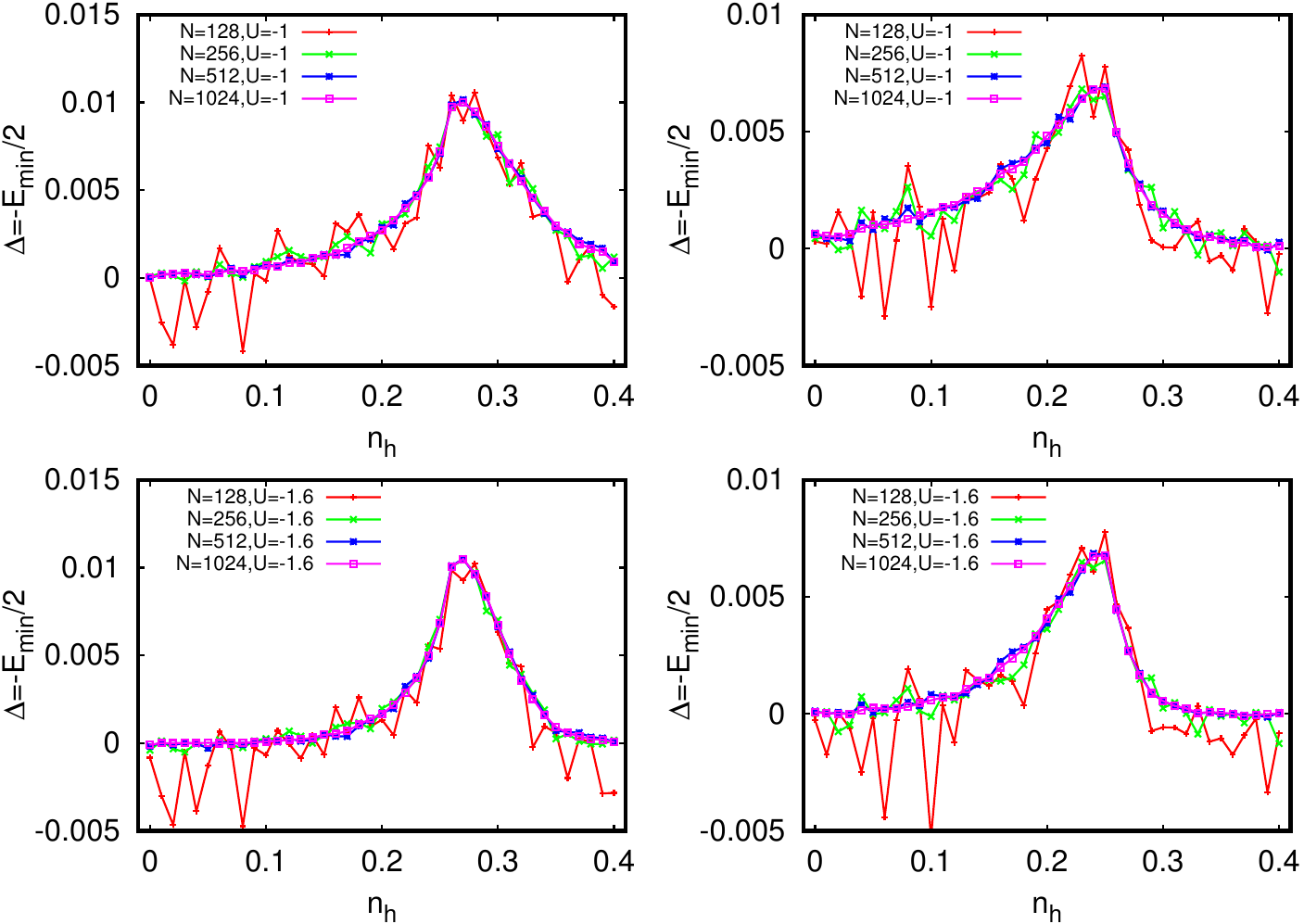}%
\end{center}
\caption{\label{figS8}
\label{fig_nfill_Nconv}
Convergence of gap energy $\Delta=-E_{\rm min}/2$ 
with increasing values of $N=128,\,256,\,512,\,1024$ and for the 
sector $\p_+=0$ as a function of doping value $n_h=1-n$.
The energy values are given in units of the 
basic hopping matrix element $t$.
Top (bottom) panels correspond to the Hubbard (d-wave) interaction 
with $U=-1$ ($U=-1.6$). Left (right) panels correspond 
to electron (hole) excitations. 
}
\end{figure}

\begin{figure}
\begin{center}
\includegraphics[width=0.95\columnwidth]{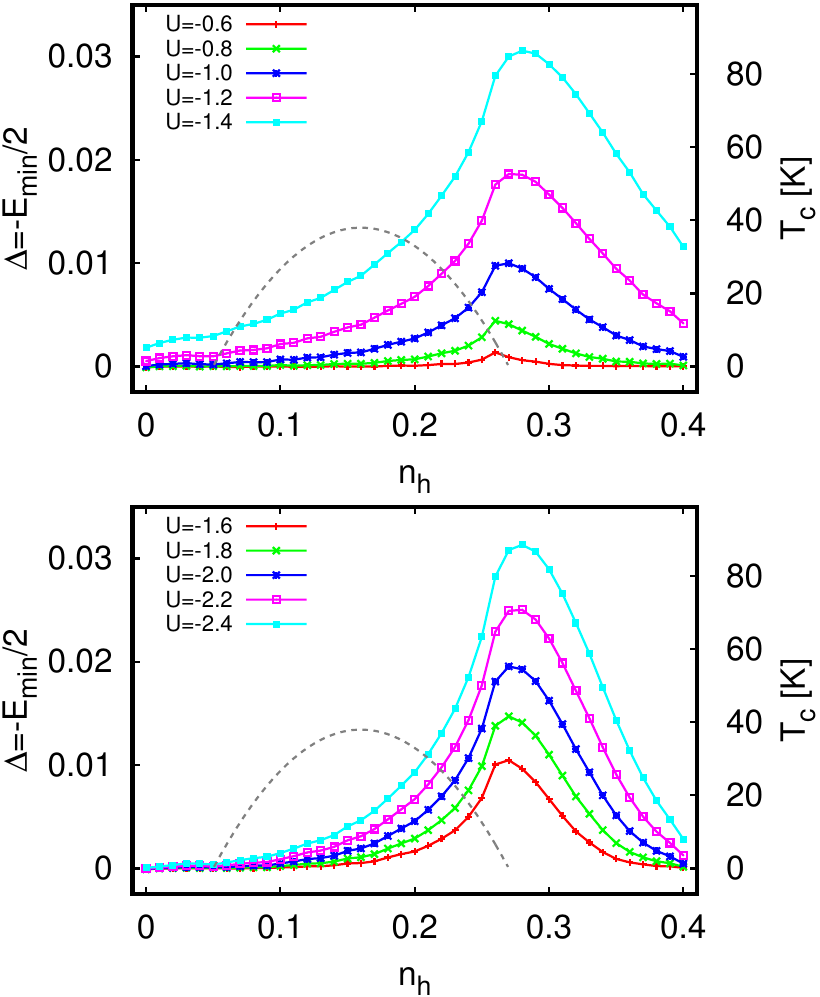}%
\end{center}
\caption{\label{figS9}
\label{fig_nfill_gap_particles}
As Fig.~8 but for electron excitations.}
\end{figure}

\begin{figure}
\begin{center}
\includegraphics[width=0.95\columnwidth]{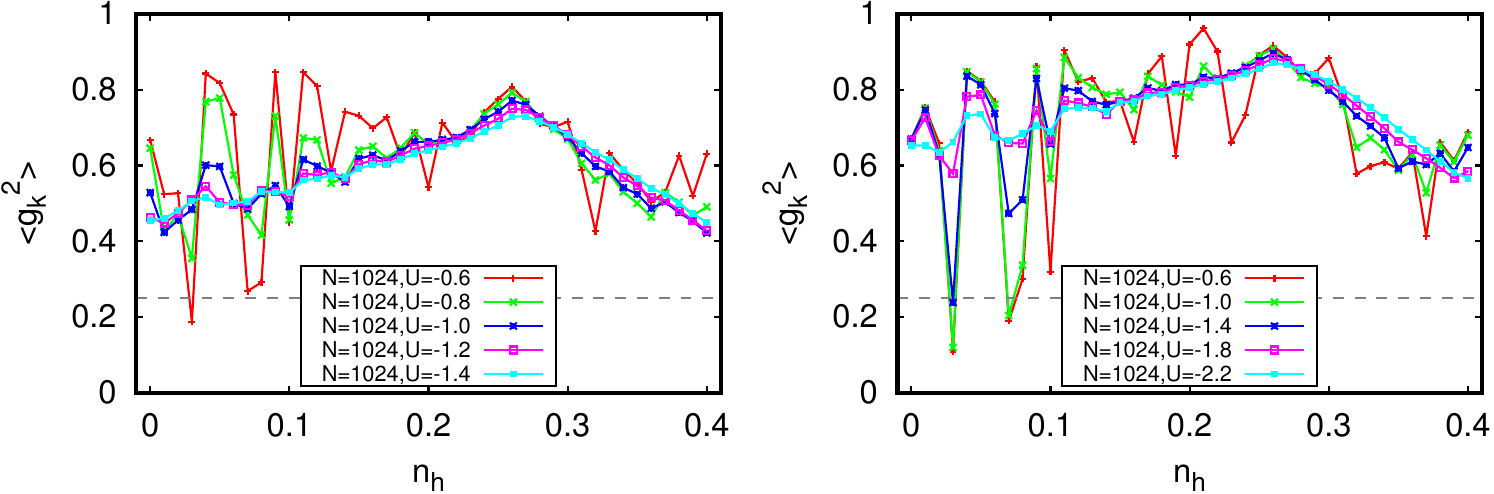}%
\end{center}
\caption{\label{figS10}
\label{fig_nfill_gk2_particles}
As Fig.~9 but for electron excitations.}
\end{figure}

\begin{figure}
\begin{center}
\includegraphics[width=0.95\columnwidth]{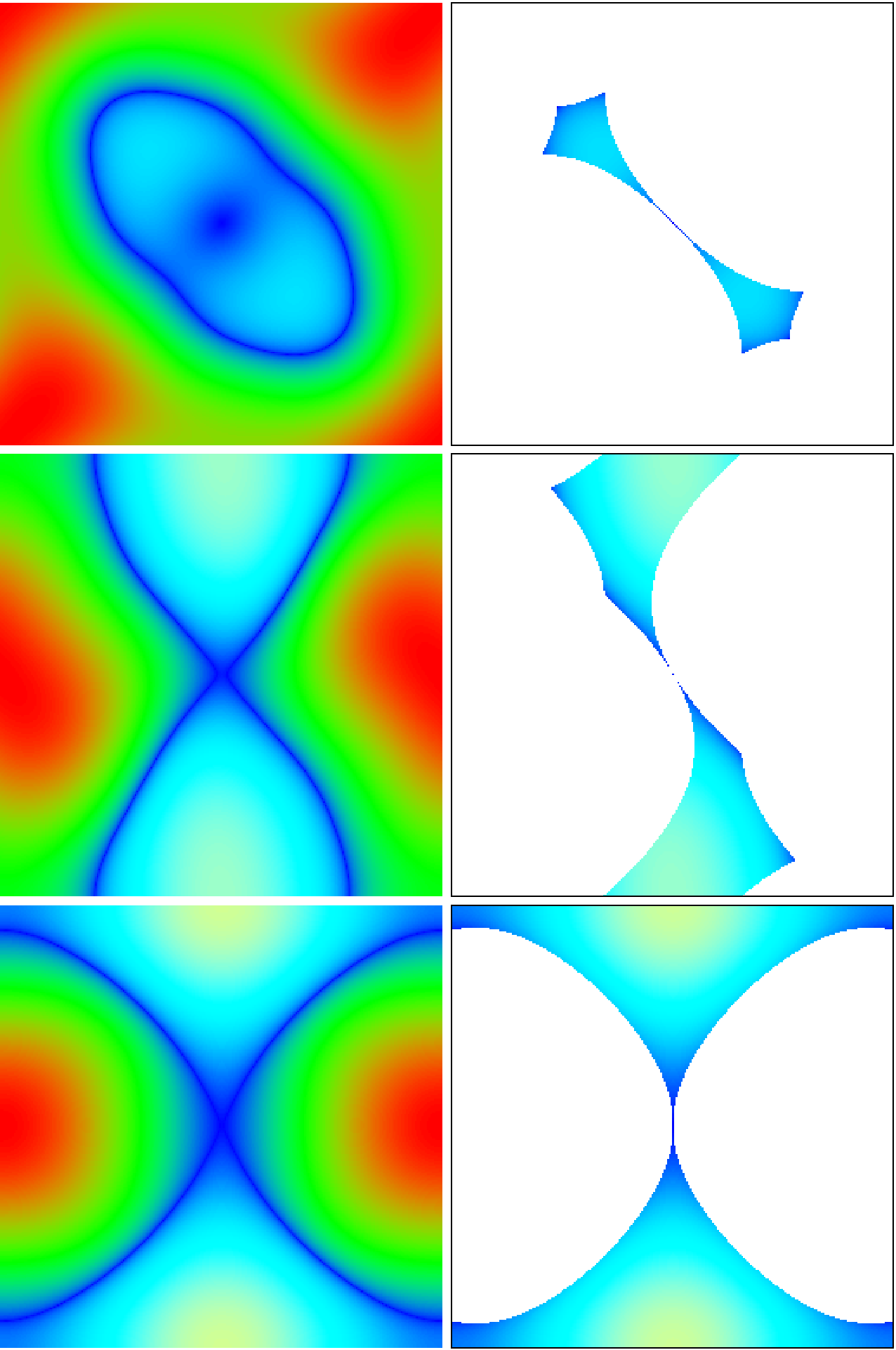}%
\end{center}
\caption{\label{figS11}
\label{fig_energy_Holes0p84}
Same as Fig.~10 with hole excitations 
but for the filling factor $n=0.84$ and the sectors 
$\p_+=2\pi(106,106)/256$ (top panels), 
$\p_+=2\pi(52,174)/256$ (center panels) and 
$\p_+=2\pi(27,256)/256$) (bottom panels) such that the center of 
mass momentum $\p_+/2$ is very close to the Fermi surface 
of virtual filling factor $n_v=0.84$ with three cases 
of $p_{+x}=p_{+y}$, $p_{+x}\approx p_{+y}/4$ and $p_{+x}$ ($p_{+y}$) 
minimal (maximal). 
}
\end{figure}

\begin{figure}
\begin{center}
\includegraphics[width=0.95\columnwidth]{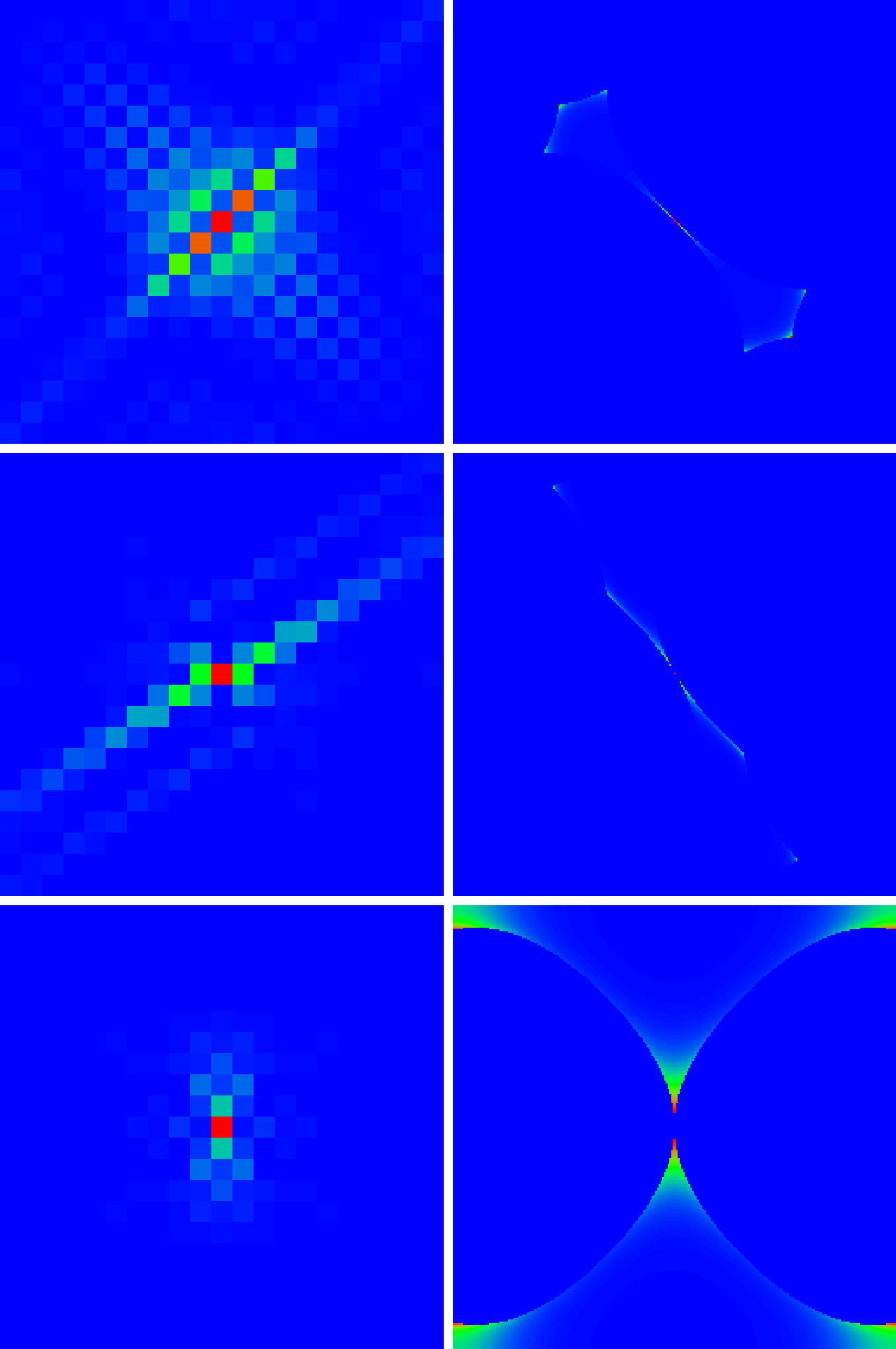}%
\end{center}
\caption{\label{figS12}
\label{fig_states_Holes0p84}
Ground state density plots for the Hubbard interaction, system size $N=256$, 
hole excitations, 
filling factor $n=0.84$ and three sectors $\p_+\neq 0$ (same 
values as in Fig.~\ref{fig_energy_Holes0p84}). Top (center, bottom) panels 
correspond to $U=-7$, $\p_+=2\pi(106,106)/256$ 
($U=-8$, $\p_+=2\pi(52,174)/256$; $U=-6$, $\p_+=2\pi(27,256)/256$).
Left panels show the ground state 
in $\Delta \r$-representation in a zoomed region with 
$-10\le \Delta x,\Delta y\le 10$ 
(color values outside the zoomed regions are blue) 
and right panels show the state 
in $\Delta \p$-representation (with $-\pi\le \Delta p_{x,y}<\pi$).
The two particle ground state energies $E_{\rm min})$ 
in units of the basic hopping matrix element $t$  are $-0.04761$ ($-0.05367$, 
$-0.3767$) for top (center, bottom) panels.
}
\end{figure}

\begin{figure}
\begin{center}
\includegraphics[width=0.95\columnwidth]{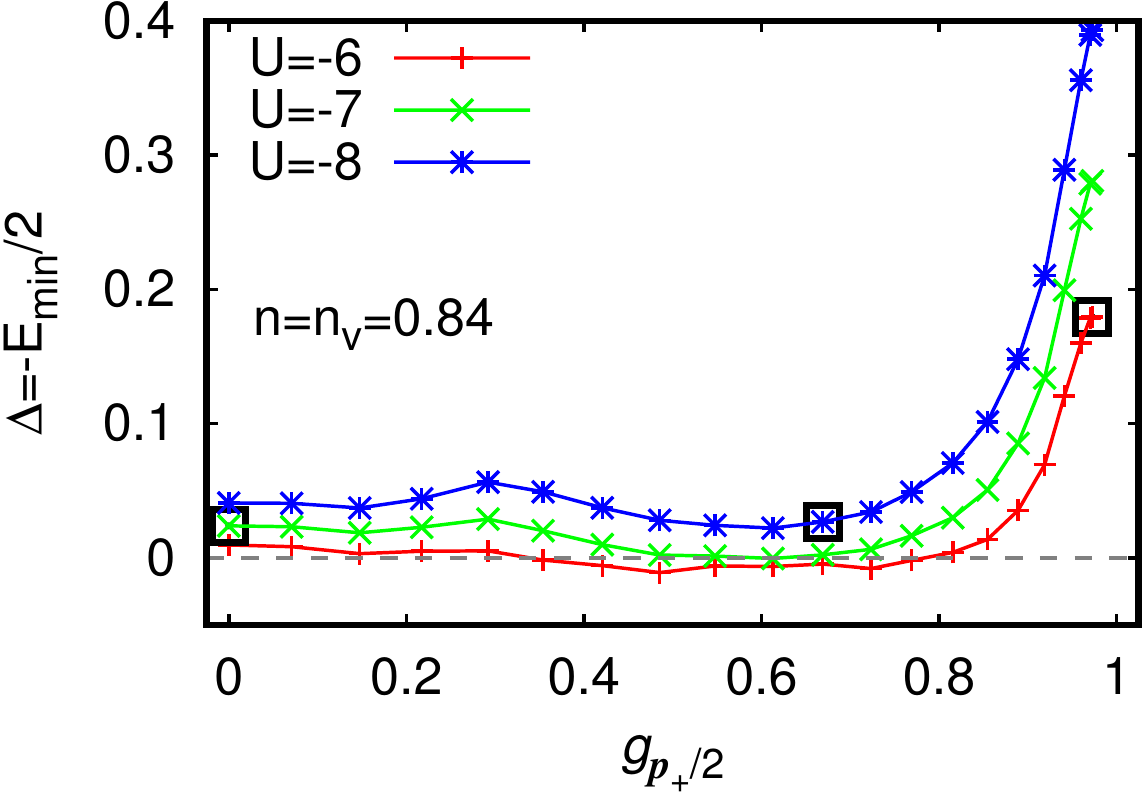}%
\end{center}
\caption{\label{figS13}
\label{fig_gap_0p84}
As Fig.~12 but for $n=n_v=0.84$ 
with three interaction values and the black square data points/states 
corresponding to Fig. \ref{fig_states_Holes0p84}. 
}
\end{figure}

\begin{figure}
\begin{center}
\includegraphics[width=0.95\columnwidth]{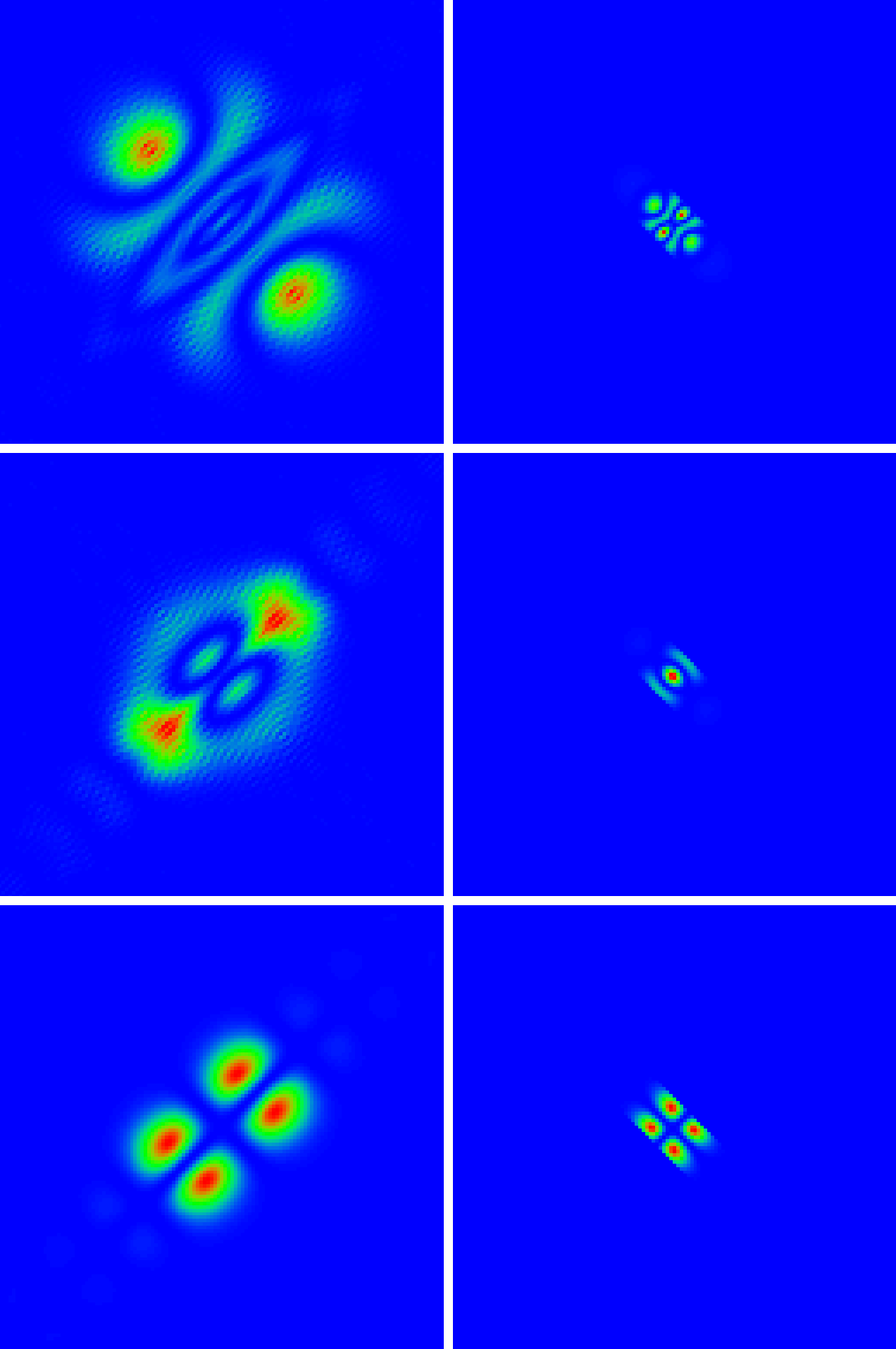}%
\end{center}
\caption{\label{figS14}
\label{fig_Cstates_particles1}
Three strong pair eigenstates for the parameters of bottom panel of 
Fig.~13 (electrons, $n=0.74$, $n_v=1$, $N=256$) 
of the first peak of large 
$w_{N/6}$-values for energies close to $0.9$ and marked by black squares 
therein. 
Left (right) columns correspond to the $\Delta \r$-
($\Delta \p$-) representation showing the 
two times zoomed center square for both cases: 
$-N/4\le \Delta x,\Delta y<N/4$ ($-\pi/2\le \Delta p_{x,y}<\pi/2$).
The panels in $\Delta \p$-representation 
correspond to the bottom right panel 
of Fig.~14 concerning the identification of 
allowed and forbidden zones. 
Top (center, bottom) row corresponds to the eigenstates 
with level number $2573$ ($2638$, $2679$), 
energy $0.9131$ ($0.9390$, $0.9550$)
and pair weight $w_{N/6}=0.9173$ ($0.9069$, $0.9761$). 
Here $N_2'=8737$ is the maximal possible 
level number for the largest energy (in the corresponding 
$\p_+$-sector).
}
\end{figure}

\begin{figure}
\begin{center}
\includegraphics[width=0.95\columnwidth]{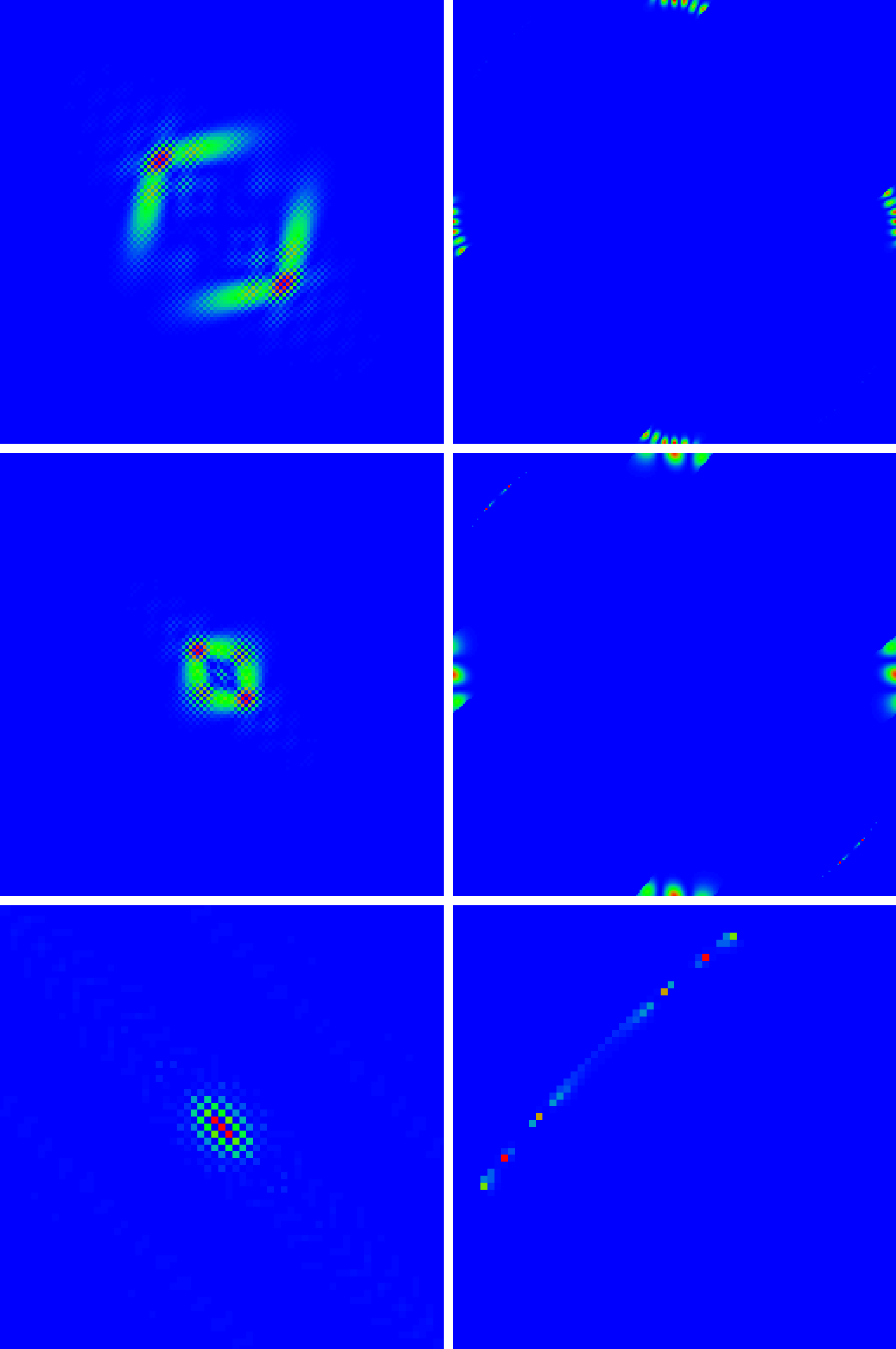}%
\end{center}
\caption{\label{figS15}
\label{fig_Cstates_particles2}
Three strong pair eigenstates for the parameters of bottom panel of 
Fig.13 (electrons, $n=0.74$, $n_v=1$, $N=256$) 
of the second peak (and the small peak behind it) of large 
$w_{N/6}$-values for energies close to $2.2$-$2.6$ and marked by black squares 
therein. 
Left (right) columns correspond to the $\Delta\r$-
($\Delta \p$-) representation.
Top and center rows show the two times zoomed center square: 
$-N/4\le \Delta x,\Delta y<N/4$ (full momentum cell: 
$-\pi\le \Delta p_{x,y}<\pi$). 
The bottom row shows the four times zoomed center square: 
$-N/8\le \Delta x,\Delta y<N/8$ (left panel) or the four 
times zoomed top left momentum corner: $-\pi\le \Delta p_{x,y}<-\pi/2$ 
(right panel) with other non-blue values in the (non-shown) 
bottom right momentum corner being the mirror image of the top left 
momentum corner 
(with $\Delta p_{x,y} \to \pi-\Delta p_{x,y}$).
The top and center panels in $\Delta \p$-representation 
correspond to the bottom left panel 
of Fig.14 concerning the identification of 
allowed and forbidden zones (for bottom $\Delta \p$-panel the top left 
corner has to be used). 
Top (center, bottom) row corresponds to the eigenstates 
with level number $6606$ ($6718$, $7062$), 
energy $2.165$ ($2.271$, $2.614$) 
and pair weight $w_{N/6}=0.9800$ ($0.9658$, $0.2862$).
Here $N_2'=8737$ is the maximal possible 
level number for the largest energy (in the corresponding 
$\p_+$-sector).
}
\end{figure}

\begin{figure}
\begin{center}
\includegraphics[width=0.95\columnwidth]{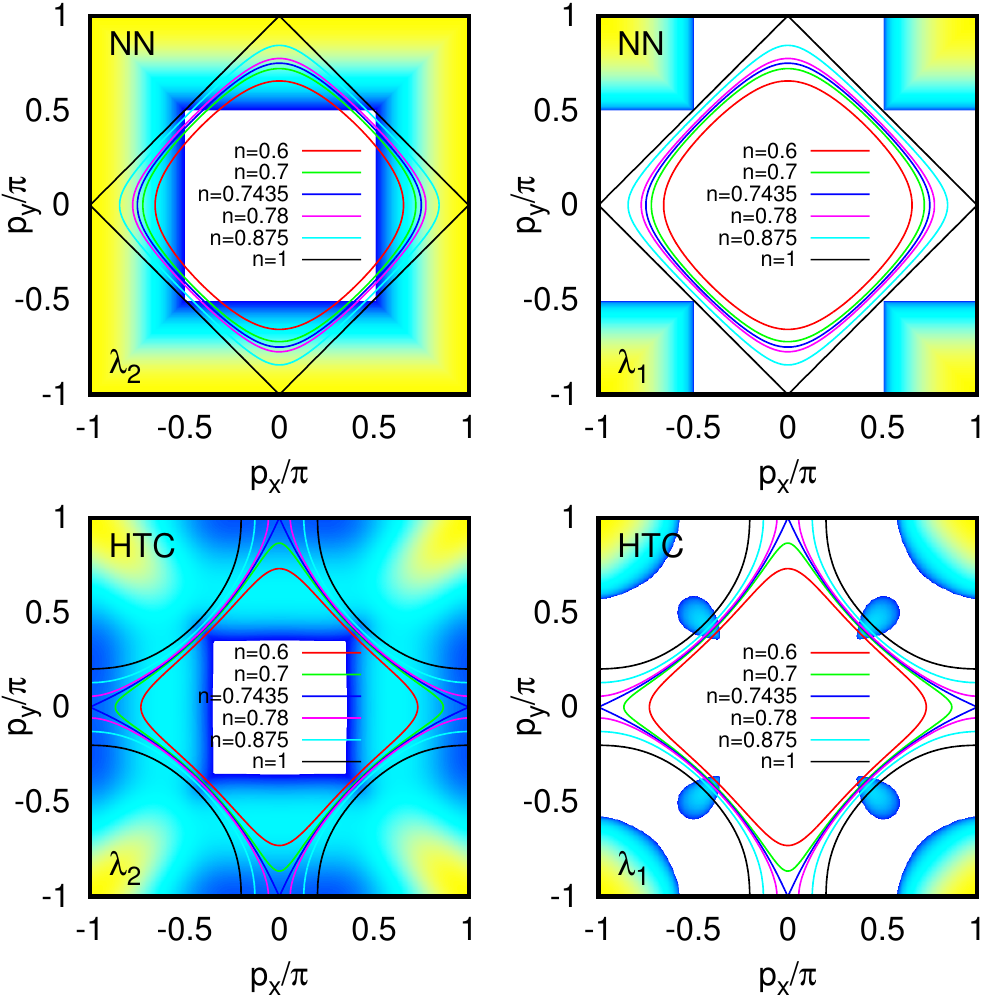}%
\end{center}
\caption{\label{figS16}
\label{fig_band_back}
Fermi surface for different filling factors as in Fig.~1
superimposed with color plots showing the regions of negative mass eigenvalues 
in classical phase space. Top (bottom) panels correspond to the NN model
(HTC model). Left (right) panels correspond the smaller eigenvalue 
$\lambda_2$ (larger eigenvalue $\lambda_1$). Shown are the regions 
of negative values for these eigenvalues with colors yellow (cyan, blue) 
for strong (intermediate, close to 0) negative values. The regions of 
white color correspond to positive $\lambda_{1,2}$. 
The eigenvalues $\lambda_1,\,\lambda_2$ as a function of the 
center of mass $\p_+/2=\p$ are computed as the eigenvalues
of the Hessian matrix obtained by expanding 
$E_{1p}(\p_+/2 - \Delta \p) + E_{1p}(\p_+/2 + \Delta \p)$
in $\Delta \p$ up to second order. 
Since $\lambda_1>\lambda_2$ the right panels show the regions where 
both eigenvalues are negative. The shown filling values $n$ in this 
figure actually correspond to the virtual filling $n_v=n$ as far as the 
superimposed color plot for the negative mass eigenvalues are concerned 
(since $\p_+/2=\p$). 
}
\end{figure}

\begin{figure}
\begin{center}
\includegraphics[width=0.95\columnwidth]{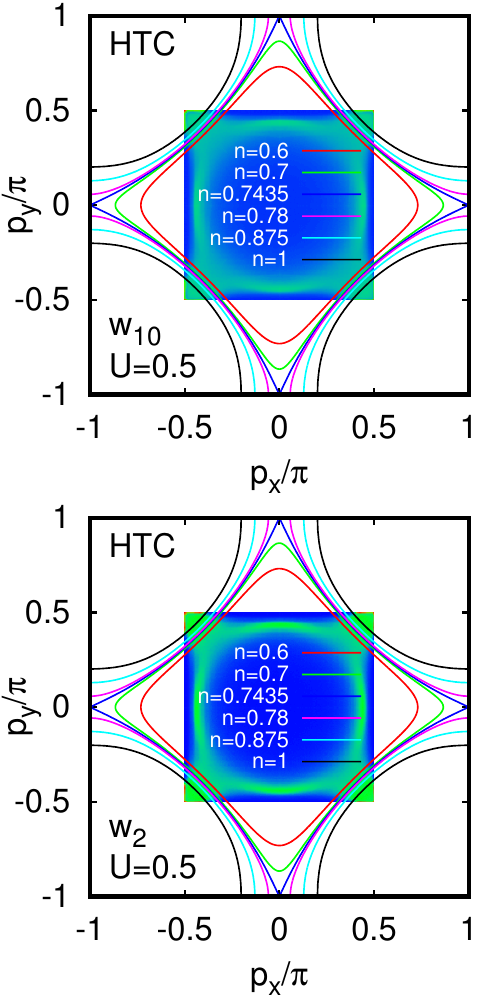}%
\end{center}
\caption{\label{figS17}
\label{fig_band_back_w2_w10}
Fermi surface for different filling factors as in Fig.~1
superimposed with color plots for the pair formation probabilities 
$w_{10}$ and $w_2$ of the HTC model computed in Ref.\cite{htcepjb} for $N=192$ 
from the long time evolution of an initially localized 
electron pair in relative coordinate propagating with the 
repulsive Coulomb interaction $U=0.5$. The color plots 
in the center of mass momentum $\p=\p_+/2$ 
are obtained by symmetric extension of the data of Fig.~4 in 
Ref.\cite{htcepjb}.
Note that the data of \cite{htcepjb} correspond to a free electron pair moving 
in an empty system without any other electrons (absence 
of frozen Fermi sea). 
The regions of strong pair formation probability close to 
$\p\approx (\pm \pi/2, \pm \pi/2)$ correspond also to regions 
of double negative mass eigenvalues shown in Fig.~\ref{fig_band_back}.
(Note that at $\p= (\pm \pi/2, \pm \pi/2)$ there are exact red data 
points for maximum values in the color plot. However, due the global figure
scale these data points are not well visible.) As in Fig.~\ref{figS16}, the 
filling values correspond to the virtual filling $n_v=n$. }
\end{figure}

\end{document}